\documentclass{aa}

\usepackage{natbib,afterpage}
\usepackage{graphicx,color}
\usepackage{amssymb,times,graphics, subfigure}
\usepackage[english]{babel}

\def\Msun{\hbox{$M_{\odot}$}}               
\def\Rstar{\hbox{$R_{\star}$}}              
\def\Mdot{\hbox{$\dot{M}$}}               
\def\Teff{\hbox{$\rm{T}_{\rm eff}$}}            
\def\twaalfco{\hbox{$^{12}$CO}}            
\def\dertienco{\hbox{$^{13}$CO}}            
\def\arcsec{\hbox{$^{\prime\prime}$}}

\bibpunct{(}{)}{;}{a}{}{,}

\begin{document}


\title{Circumstellar molecular composition of the oxygen-rich AGB star \object{IK~Tau}: II.\ In-depth non-LTE chemical abundance analysis}

\author{L.\ Decin \inst{1,2}\thanks{\emph{Postdoctoral Fellow of the
      Fund for Scientific Research, Flanders}} \and E.\ De Beck
  \inst{1} \and S.\ Br\"unken \inst{3,4} \and H.\ S.\ P.\ M\"uller \inst{4,5}
  \and K.\ M.\ Menten \inst{5} \and H.\ Kim \inst{5,6}  \and K.\ Willacy \inst{7} \and A.\ de Koter \inst{2,8} \and
  F.\ Wyrowski \inst{5}} \offprints{L.\ Decin, e-mail:
  Leen.Decin@ster.kuleuven.ac.be}

\institute{
  Department of Physics and Astronomy, Institute of Astronomy,
  K.U.Leuven, Celestijnenlaan 200D, B-3001 Leuven, Belgium
  \and Sterrenkundig Instituut Anton Pannekoek, University of
  Amsterdam, P.O.\ Box 9429, 1090 CE Amsterdam, The Netherlands
  \and Harvard-Smithsonian Center for Astrophysics, 60 Garden Street,
  Cambridge, MA 02138, USA
  \and I.\ Physikalisches Institut, Universit\"at zu K\"oln,
  Z\"ulpicher Street 77, 50937 K\"oln, Germany  
  \and Max-Planck Institut f\"ur Radioastronomie, Auf dem H\"ugel 69,
  53121 Bonn, Germany  
\and MPI f\"ur Gravitationsphysik, Calinstr.\ 38, 30167 Hannover, Germany
  \and  Jet Propulsion Laboratory, California Institute of Technology,
 Pasadena, CA 91109 
\and Astronomical Institute, Utrecht University, Princetonplein 5, 3584 CC Utrecht, The Netherlands
}

\date{Received January 26, 2010; accepted April 6, 2010}


\abstract
{The interstellar medium is enriched primarily by matter ejected from evolved low and intermediate mass stars. The outflow from these stars creates a circumstellar envelope in which a rich gas-phase chemistry takes place. Complex shock-induced non-equilibrium chemistry takes place in the inner wind envelope, dust-gas reactions and ion-molecule reactions alter the abundances in the intermediate wind zone, and the penetration of cosmic rays and ultraviolet photons dissociates the molecules in the outer wind region.} 
{Little observational information exists on the circumstellar molecular abundance stratifications of many molecules. Furthermore, our knowledge of oxygen-rich envelopes is not as profound as for the carbon-rich counterparts. The aim of this paper is therefore to study the circumstellar chemical abundance pattern of 11 molecules and isotopologs ($^{12}$CO, $^{13}$CO, SiS, $^{28}$SiO, $^{29}$SiO, $^{30}$SiO, HCN, CN, CS, SO, SO$_2$) in the oxygen-rich evolved star IK~Tau.}
{We have performed an in-depth analysis of a large number of molecular emission lines excited in the circumstellar envelope around IK~Tau. The analysis is done based on a non-local thermodynamic equilibrium (non-LTE) radiative transfer analysis, which calculates the temperature and velocity structure in a self-consistent way. The chemical abundance pattern is coupled to theoretical outer wind model predictions  including photodestruction and cosmic ray ionization. Not only the integrated line intensities, but also the line shapes, are used as diagnostic tool to study the envelope structure.}
{The deduced wind acceleration is much slower than predicted from classical theories. SiO and SiS are depleted in the envelope, possibly due to the adsorption onto dust grains. For HCN and CS a clear difference with respect to inner wind non-equilibrium predictions is found, either indicating uncertainties in the inner wind theoretical modeling or the possibility that HCN and CS (or the radical CN) participate in the dust formation. The low signal-to-noise profiles of SO and CN prohibit an accurate abundance determination; the modeling of high-excitation SO$_2$ lines is cumbersome, possibly related to line misidentifications or problems with the collisional rates. The SiO isotopic ratios ($^{29}$SiO/$^{28}$SiO and $^{30}$SiO/$^{28}$SiO) point toward an enhancement in $^{28}$SiO compared to results of classical stellar evolution codes. Predictions for H$_2$O emission lines in the spectral range of the Herschel/HIFI mission are performed.}
%
{}
\keywords{Line: profiles, Radiative transfer, Stars: AGB and post-AGB,
  (Stars): circumstellar matter, Stars: mass loss, Stars: individual:
  \object{IK Tau}}

\authorrunning{L.\ Decin et al.}
\titlerunning{The circumstellar chemistry in the O-rich AGB IK~Tau}

\maketitle


\section{Introduction}

Asymptotic Giant Branch (AGB) stars are well known to release significant amounts of gas and dust in the interstellar medium via (copious) mass loss. This mass loss dominates the evolution of the star and ultimately, when the stellar envelope is exhausted, causes the star to evolve off the AGB into the post-AGB phase.  The outflow from these evolved stars creates an envelope, which fosters gas-phase chemistry. The chemical complexity in circumstellar envelopes (CSEs)
is thought to be dominated by the elemental carbon to oxygen ratio: oxygen-rich M-stars have a C/O ratio less than unity, carbon-rich C-stars have C/O $>1$, and for S-stars C/O is $\sim$1.

Many have focused on the CSEs of carbon-rich stars in which a rich chemistry takes place. This is reflected by the detection of over 60
different chemical compounds, including unusual carbon chain radicals, in the CSE of \object{IRC~+10216}, the prototype of carbon stars \citep[e.g.][]{Cernicharo2000A&AS..142..181C}. In contrast, only 10--12 compounds have been identified in the chemically most interesting oxygen-rich evolved
stars, such as \object{IK~Tau} and \object{VY CMa} \citep[e.g.][]{Ziurys2007Natur.447.1094Z}. The first observations of carbon-bearing molecules (other than CO) in oxygen-rich AGBs were somewhat unexpected \citep[e.g.][]{Deguchi1985Natur.317..336D, Jewell1986Natur.323..311J}. Nowadays, the formation of carbon molecules is thought to be the result of shock-induced non-equilibrium chemistry in the inner circumstellar envelope \citep[e.g.][]{Duari1999AandA...341L..47D} and/or a complex chemistry in the outer envelope triggered by the penetration of cosmic rays and ultra-violet radiation \citep[e.g.][]{Willacy1997AandA...324..237W}. Recently, a new interstellar molecule, PO ($X\,^2\Pi_r$), has been detected toward the envelope of the oxygen-rich supergiant \object{VY~CMa} \citep{Tenenbaum2007ApJ...666L..29T}. Phosphorus monoxide is the first interstellar molecule detected that contains a P--O bond, a moiety essential in biochemical compounds. It is also the first new species identified in an oxygen-rich, as opposed to a carbon-rich, circumstellar envelope. These results suggest that oxygen-rich shells may be as chemically diverse as their carbon counterparts.

Circumstellar molecules have been extensively observed, both in the form of surveys of a single molecular species and in the form of searches for various molecular species in a limited number of carefully selected sources. The aim of these studies was to derive \emph{(i.)} the mass-loss rate (from CO rotational lines) or \emph{(ii.)} molecular abundances. For this latter purpose, several methods exist, each with varying degrees of complexity. \emph{(1.)}~For example, \citet{Bujrrabal1994AandA...285..247B} and \citet{Olofsson1998A&A...329.1059O} showed that simple molecular line intensity ratios, if properly chosen, may be used to study the chemical behaviour in CSEs. The use of line intensity ratios has the advantage of requiring no assumptions about a circumstellar model, but it also limits the type of conclusions that can be drawn. \emph{(2.)} Several authors have derived new constraints on chemical and circumstellar models based on the simplifying assumption of unresolved optically thin emission thermalized at one excitation temperature \citep[e.g.][]{Lindqvist1988AandA...205L..15L, Omont1993AandA...267..490O, Bujrrabal1994AandA...285..247B, Kim2009}. \emph{(3.)} Later on, observations were (re)-analyzed based on a non-LTE (non local thermodynamic equilibrium) radiative transfer model \citep[e.g.][]{Bieging2000ApJ...543..897B, Teyssier2006AandA...450..167T, Schoier2007AandA...473..871S}. In this study, we will go one step further and abandon or improve few of the assumptions still made in many non-LTE analyses. 
\begin{enumerate}
 \item Quite often, the temperature structure --- being the most important factor determining the molecular line excitation --- is approximated using a power-law  \citep[e.g.][]{Bieging2000ApJ...543..897B, Teyssier2006AandA...450..167T}. Effects of different heating and cooling mechanisms are hence not properly taken into account. For instance, in the outermost parts of the envelope the temperature profile deviates from a power law distribution once the influence of photoelectric heating by the external interstellar radiation field becomes important \citep[e.g.][]{Crosas1997ApJ...483..913C, Justtanont1994ApJ...435..852J, Decin2006A&A...456..549D}.
\item The shell is often assumed to expand at a constant velocity \citep[e.g.][]{Bieging2000ApJ...543..897B, Schoier2007AandA...473..871S}. However, for molecular lines primarily formed in the wind acceleration zone, the effect of a non-constant velocity structure on the derived molecular abundance may be significant.
\item The fractional abundances are estimated to follow an exponential or Gaussian distribution, assuming that the molecules are formed in the inner envelope, and photodissociated or absorbed onto dust grains further out \citep[e.g.][]{Bieging2000ApJ...543..897B, GonzalesDelgado2003AandA...411..123G, Schoier2007AandA...473..871S}. The effect of extra formation and/or depletion processes in the envelope can hence not be taken into account.
\item Often, a maximum of two molecules (CO and one other) is analyzed at once \citep[e.g.][]{GonzalesDelgado2003AandA...411..123G, Schoier2007AandA...473..871S}. 
\item Integrated line intensities are often used as a criterion to analyse the circumstellar chemical structure. However, line shapes provide us with strong diagnostic constraints to pinpoint the wind acceleration, which in turn has an influence on the deduced fractional abundances.
\end{enumerate}

In this paper, we will study the circumstellar chemical abundance fractions of eleven different molecules and isotopologs in the oxygen-rich AGB star \object{IK~Tau} based on the non-LTE radiative transfer code GASTRoNOoM \citep{Decin2006A&A...456..549D, Decin2007A&A...475..233D}, which computes the temperature and velocity structure in the envelope in a self-consistent way. Chemical abundance stratifications are coupled to theoretical non-equilibrium (non-TE) predictions in the outer envelope by \citet{Willacy1997AandA...324..237W} and compared to the shock-induced non-TE inner wind predictions by \citet{Duari1999AandA...341L..47D} and \citet{Cherchneff2006AandA...456.1001C}. \object{IK~Tau} has been chosen for study because of the wealth of observations which are available for this target and the fact that its envelope is thought to be (roughly) spherically symmetric \citep{Lane1987ApJ...323..756L, Marvel2005AJ....130..261M, Hale1997ApJ...490..407H, Kim2009}.

\object{IK~Tau}, also known as NML~Tau, was discovered in 1965 by \citet{Neugebauer1965ApJ...142..399N}. It is an extremely red Mira-type variable with spectral type ranging from M8.1 to M11.2 and a period around 470 days \citep{Wing1973ApJ...184..873W}. From dust shell motions detected at 11\,$\mu$m using the ISI interferometer, \citet{Hale1997ApJ...490..407H} deduced a distance of 265\,pc. This is in good agreement with the results of \citet{Olofsson1998A&A...329.1059O} who computed a distance of 250\,pc from integrated visual, near-infrared and IRAS data using a period-luminosity relation. Estimated mass-loss rates range from $3.8 \times 10^{-6}$ \citep{Neri1998AandAS..130....1N}  to $3 \times 10^{-5}$\,\Msun/yr \citep{GonzalesDelgado2003AandA...411..123G}. IK Tau's proximity and relatively high mass-loss rate (for a Mira) facilitates the observation of molecular emission lines.

In Sect.~\ref{data}, we present the molecular line observational data used in this paper. Sect.~\ref{analysis} describes the background of the excitation analysis: the radiative transfer model used, the molecular line data and the theoretical ideas on molecular abundance stratification in the envelope. Sect.~\ref{results} describes the results: we first focus on the velocity structure in the envelope with special attention to the acceleration zone, after which the derived stellar parameters are discussed. Thereafter, the abundance structure for each molecule is derived and compared to the theoretical inner and outer wind predictions and observational results found in the literature. The time variability and SiO isotopic ratios are discussed in Sect.~\ref{discussion} and water line predictions are performed in Sect.~\ref{H2O}. We end with some conclusions in Sect.~\ref{conclusion}.


\section{Observational data} \label{data}

Part of the observations were obtained from our own observing programs
scheduled at the JCMT, APEX and IRAM. These observations and the data
reduction are described in Sect.~\ref{obs_own}. Other
data are extracted from the literature and summarized briefly in
Sect.~\ref{obs_others}. An overview is given in Table~\ref{Table:data}.

\subsection{Observations and data reduction} \label{obs_own}

The \twaalfco(2--1), \twaalfco(3--2), \twaalfco(4--3) and the \dertienco(2--1) observations
were extracted from the JCMT\footnote{The James Clerk Maxwell Telescope (JCMT) is
  operated by The Joint Astronomy Centre on behalf of the Science and
  Technology Facilities Council of the United Kingdom, the Netherlands
  Organisation for Scientific Research, and the National Research
  Council of Canada} archive.  Additional data with the
APEX\footnote{APEX, the Atacama Pathfinder Experiment, is a
  collaboration between the Max-Planck-Institut fur Radioastronomie,
  the European Southern Observatory, and the Onsala Space
  Observatory. Program IDs are 077.D-0781 and 077.D-4004.} 12\,m telescope were
obtained for the \twaalfco(3--2), \twaalfco(4--3), \twaalfco(7--6),
and \dertienco(3--2) molecular line transitions. With the 30\,m telescope of the Institut de Radio
Astronomie Millimetrique (IRAM)\footnote{IRAM is supported by INSU/CNRS
(France), MPG (Germany), and IGN (Spain).} molecular line observations were
performed in December 2006. During this observing run, data on
the CO(2-1), SiS(8--7), SiS(12--11), SO$_2$($14_{3,11}-14_{2,12}$),
SO$_2$($4_{3,1}-4_{2,2}$), SO$_2$($3_{3,1}-3_{2,2}$),
SO$_2$($5_{3,3}-5_{2,4}$), and HCN(3--2)\footnote{When the isotopic 
  notation is not given, the molecular line transition is from the
  main isotopolog.}  line transitions were obtained. 

The JCMT and APEX observations were
carried out using a position-switching mode. The IRAM observations were done in the wobbler-switching mode  with a throw of ± 60". The frequency resolution for the JCMT-data equals 0.0305\,MHz, for
the APEX data it is  0.1221\,MHz. The resolution was 1.25\,MHz for the 3 and 2mm IRAM observations and 1 or/and 4\,MHz at 1.3 and 1\,mm, resulting in resolution slightly higher than 1 km/s for these observations. 

The JCMT data reduction was performed with the SPLAT devoted routines of STARLINK, the
APEX and IRAM data were reduced with CLASS.  A polynomial of first order was fitted
to an emission free region of the spectral baseline and subtracted.
To increase the signal-to-noise ratio, the data were rebinned to a
resolution of $\sim$1\,km/s  so that we have at
least 40 independent resolution elements per line profile.  The
antenna temperature, $T_A^*$, was converted to the main-beam
temperature ($T_{\rm mb} = T_A^*/\eta_{\rm mb}$), using a main-beam efficiency, $\eta_{\rm mb}$
as specified in Table~\ref{Table:data}. The absolute uncertainties are $\sim$20\,\%.

\subsection{Literature data} \label{obs_others} 

To have better constraints on the chemical abundance pattern in the
wind region around \object{IK~Tau}, additional data were taken from
the literature (see Table~\ref{Table:data}). 

High-quality observations were performed by
\citet{Kim2006} with the APEX telescope in Chile during observing
periods in November 2005, April 2006 and August 2006 \citep[see also][hereafter referred to as Paper~I]{Kim2009}. In total, 34
transitions from 12 molecular species, including a few maser lines, were
detected toward \object{IK~Tau}.

\citet{Schoier2007A&A...473..871S} published the observations of four SiS lines: the (5--4) and (6--5) rotational line transitions were obtained with the Onsala Space Observatory (OSO) telescope, the (12--11) and (19-18) rotational line observations were performed with the JCMT telescope.

SiO thermal radio line emission from a large sample of M-type AGB stars, including \object{IK~Tau}, was studied by \citet{GonzalesDelgado2003AandA...411..123G}. The SiO(2--1) line transition was obtained with the OSO telescope, the SiO(5--4) and (6--5) transitions with the Swedish ESO Submillimeterwave Telescope (SEST).

$^{12}$CO line data were obtained by \citet{Teyssier2006AandA...450..167T} with the IRAM and JCMT telescope. The CO(1--0) line was observed by \citet{Ramstedt2008AandA...487..645R} in December 2003 with the 20\,m OSO telescope.
 \citet{Olofsson1998A&A...329.1059O} reported on the detection of the CS(2--1) line at 98\,GHz with the OSO telescope with an integrated intensity of 0.5\,K\,km/s. 

\begin{table}[htp]
  \caption{Overview of the molecular line transitions used in this
    research, with indication of the frequency, the upper energy level, the telescope,  the main beam half
    power beam width (HPBW) and main 
    beam efficienciy ($\eta_{\rm mb}$).  } 
\vspace*{-1ex}
\label{Table:data}
\setlength{\tabcolsep}{1mm}
{\small{
\begin{tabular}{lcrcccc}
\hline
\hline
\rule[0mm]{0mm}{5mm}Transition & Frequency & E$_{\rm upper}$ & Telescope & HPBW  & $\eta_{\rm mb}$ & 
Ref. \\
                  & \multicolumn{1}{c}{[GHz]}     &  \multicolumn{1}{c}{[cm$^{-1}$]} &         & (\arcsec) & 
        & \\
\hline
\multicolumn{7}{c}{{\textbf{\rule[0mm]{0mm}{5mm}H$\mathbf{^{12}}$C$\mathbf{^{14}}$N}}} \\ 
3--2 &  265.886& 17.7382 & IRAM &   9.5 & 0.46    & 1\\
4--3 & 354.505 & 29.5633 & APEX & 18 & 0.73 & 2\\
\multicolumn{7}{c}{{\textbf{$\mathbf{^{12}}$C$\mathbf{^{14}}$N}}} \\
$3_{5/2}-2_{3/2}$ & 340.032 & 22.6784 & APEX & 18 & 0.73 & 2\\
$3_{7/2}-2_{5/2}$ & 340.248 & 22.7039  & APEX & 18 & 0.73 & 2 \\
\multicolumn{7}{c}{{\textbf{$\mathbf{^{12}}$C$\mathbf{^{16}}$O}}} \\
 1--0 &  115.271 & 3.8450 & IRAM &  21  &   0.45 &  5 \\
 1--0 &  115.271 & 3.8450 & OSO &   33 &  0.43  &  6  \\
 2--1 &   230.538 & 11.5350& JCMT & 20  & 0.69   & 1 \\
  2--1 &   230.538 & 11.5350& IRAM & 10.5 & 0.57  & 1 \\
  3--2 & 345.796 & 23.0695& JCMT &  14 & 0.63  & 1\\
  3--2  & 345.796 & 23.0695& APEX &  18 & 0.73  & 1\\
  3--2 & 345.796 & 23.0695 & APEX & 18 & 0.73 & 2 \\
 4--3 & 461.041 & 38.4481& JCMT &  11 & 0.52     &  1\\
  4--3 & 461.041 & 38.4481& APEX & 14 & 0.60 & 2 \\
  7--6 & 806.652 & 107.6424 & APEX & 8 & 0.43 & 2 \\
\multicolumn{7}{c}{{\textbf{$\mathbf{^{13}}$C$\mathbf{^{16}}$O}}}  \\
2--1 & 220.398 & 11.0276 & JCMT & 20 & 0.69 & 1 \\
3--2 & 330.588 & 22.0549 & APEX & 19 & 0.73 & 2 \\
\multicolumn{7}{c}{{\textbf{$\mathbf{^{12}}$C$\mathbf{^{32}}$S}}}  \\
6--5 & 293.912 & 34.3152& APEX & 21 & 0.80 & 2 \\
7--6 & 342.883 & 45.7525& APEX & 18 & 0.73 & 2 \\
\multicolumn{7}{c}{ {\textbf{$\mathbf{^{28}}$Si$\mathbf{^{16}}$O}}} \\
2--1 & 86.846 & 4.3454 & OSO &  42 & 0.55 &4 \\
 5--4 & 217.104 & 21.7261 & SEST & 25 & 0.55 &4 \\
 6--5 & 260.518 & 30.4161  & SEST & 21 & 0.45 &4 \\
 7--6 & 303.990 & 40.5540 & APEX & 20 & 0.80 &2 \\
 8--7 & 345.061 & 52.1397 & APEX & 18 & 0.73 & 2 \\
\multicolumn{7}{c}{ {\textbf{$\mathbf{^{29}}$Si$\mathbf{^{16}}$O}}} \\
7--6 & 300.120 & 40.0461 & APEX & 20 & 0.80 &2 \\
 8--7 & 342.980  & 51.4867 & APEX & 18 & 0.73 & 2 \\
\multicolumn{7}{c}{ {\textbf{$\mathbf{^{30}}$Si$\mathbf{^{16}}$O}}}\\
 7--6 & 296.575 &  39.5730 & APEX & 20 & 0.80 &2 \\
 8--7 & 339.930 & 50.8785 & APEX & 18 & 0.73 & 2 \\
\multicolumn{7}{c}{{\textbf{$\mathbf{^{28}}$Si$\mathbf{^{32}}$S}}} \\
5--4 & 90.771 & 9.0836 & OSO &  42 & 0.60 & 3 \\
 6--5 & 108.924 & 12.7169 & OSO & 35  & 0.50 & 3 \\
 8--7 & 145.226 & 21.7999 & IRAM & 17  & 0.69 & 1 \\
 12--11 & 217.817 & 47.2306 & IRAM &  10.5 & 0.57 & 1 \\
 12--11 & 217.817 & 47.2306 & JCMT & 22  & 0.70 & 3 \\
 16--15& 290.380 & 82.3445 & APEX & 21 & 0.80 & 2 \\
 17--16&308.516  & 92.6355  & APEX & 20 & 0.80 & 2 \\
 19--18 & 344.779 & 115.0319 & APEX & 18 & 0.73 & 2 \\
 19--18 & 344.779 & 115.0319  & JCMT & 14 &  0.63 & 3 \\
 20--19& 362.906 & 127.1372 & APEX & 18 & 0.73 & 2 \\
\multicolumn{7}{c}{ {\textbf{$\mathbf{^{32}}$S$\mathbf{^{16}}$O}}} \\
 $7_7-6_6$& 303.927 & 49.3181 & APEX & 20 & 0.80 & 2 \\
 $8_8-7_7$ & 344.310 & 60.8030 & APEX & 18 & 0.73 &2  \\
\multicolumn{7}{c}{{\textbf{$\mathbf{^{32}}$S$\mathbf{^{16}}$O$\mathbf{^{16}}$O}}} \\
 $14_{3,11}-14_{2,12}$ & 226.300 & 82.6982 & IRAM & 10.5 &0.57 & 1\\  
 $4_{3,1}-4_{2,2}$ & 255.553 & 21.7502 & IRAM & 9.5 & 0.46 & 1\\   
 $3_{3,1}-3_{2,2}$ & 255.958 & 19.1968 & IRAM & 9.5 & 0.46 & 1\\  
 $5_{3,3}-5_{2,4}$ & 256.247 & 24.9425 &IRAM & 9.5 & 0.46 & 1\\  
$3_{3,1}-2_{2,0}$ & 313.279 & 19.1968 & APEX & 20 & 0.80 & 2\\  
 $17_{1,17}-16_{0,16}$ & 313.660 & 94.5810 & APEX & 20 &   0.80 & 2\\  
 $4_{3,1}-3_{2,2}$ &  332.505& 21.7502 & APEX & 19 & 0.73 & 2\\  
 $13_{2,12}-12_{1,11}$ &  345.338 & 64.6273 & APEX  & 18 & 0.73 & 2\\  
 $5_{3,3}-4_{2,2}$ &   351.257& 24.9425 & APEX & 18 &  0.73 & 2\\  
 $14_{4,10}-14_{3,11}$ &   351.873 & 94.4354 & APEX & 18 & 0.73 & 2\\  
\hline\hline
\end{tabular}
\tablebib{(1)~new data (see Sect.~\ref{obs_own}); (2)~\citet{Kim2009};  
   (3)~\citet{Schoier2007A&A...473..871S};
    (4)~\citet{GonzalesDelgado2003AandA...411..123G};
    (5)~\citet{Teyssier2006AandA...450..167T}; (6)~\citet{Ramstedt2008AandA...487..645R}.
}
}}
\end{table}


\section{Excitation analysis} \label{analysis}

\subsection{Radiative transfer model} \label{rad_model}
The observed molecular line transitions provide information on the
thermodynamic and chemical structure in the envelope around \object{IK
  Tau}. The line profiles were modeled with our non-LTE (non-Local
Thermodynamic Equilibrium) radiative transfer code GASTRoNOoM
\citep{Decin2006A&A...456..549D}. The code \emph{(1)}  calculates the
kinetic temperature and velocity structure in the shell by solving the
equations of motion of gas and dust and the energy balance
simultaneously; then \emph{(2)} solves the radiative transfer
equation in the co-moving frame using the Approximate Newton-Raphson
operator as developed by \citet{Schonberg1986A&A...163..151S} and
computes the non-LTE level-populations; and finally
\emph{(3)} determines the observable line profile by ray-tracing. For a full description of the code we
refer to \citet{Decin2006A&A...456..549D}.

The main assumption of the code is a spherically symmetric wind. The
mass-loss rate is allowed to vary with radial distance from the
star. The local line width is assumed to be described by a Gaussian
and is made up of a microturbulent component with a Doppler width of
1.5\,km/s and a thermal component which is calculated from the derived
kinetic temperature structure.

Two major updates have been made since the original publication in
\citet{Decin2006A&A...456..549D}. 
\begin{itemize}
\item
The code now iterates on steps \emph{(1)} and \emph{(2)} to obtain the kinetic
temperature structure in a self-consistent manner from solving the
energy balance equation, where the CO and H$_2$O line cooling (or
heating) are directly obtained from the excitation analysis, i.e.
\begin{equation}
\Lambda = n_{\rm{H_2}} \sum_l \sum_{u>l} (n_l \gamma_{lu} - n_u
\gamma_{ul}) h \nu_{ul}\,,
\end{equation}
where $n_l$ and $n_u$ are the level populations in the lower and upper
levels participating in the transition at rest frequency $\nu_{ul}$,
and $\gamma_{ul}$ and $\gamma_{lu}$ are the CO-H$_2$ collisional rate
coefficients. The cooling rate $\Lambda$ (in erg s$^{-1}$ cm$^{-3}$) is
defined as positive for net cooling. In case of \object{IK Tau} the
water line cooling dominates the CO line cooling by more than an order
of magnitude in the inner wind region; for regions beyond
$10^{16}$\,cm the CO line cooling dominates over H$_2$O cooling with
the adiabatic cooling being the dominant coolant agent.  
\item While the original version of the code approximates the stellar
  atmosphere with a blackbody at the stellar effective temperature, an
  additional option is now implemented to use {\sc marcs} theoretical
  model atmospheres and theoretical spectra \citep{Gustafsson2008A&A...486..951G,
    Decin2007A&A...472.1041D} to estimate the stellar flux. Molecular species in the CSE, less abundant than CO and with larger dipole  moments are primarily excited by infrared radiation from the central star \citep[with the possible exception of  HCN,][and H$_2$O, see Sect.~\ref{H2O}]{Jura1983ApJ...267..647J}. For CO, the infrared radiation
  competes with rotational excitation by collisions and by trapped
  rotational line photons to determine the populations of the
  rotational levels \citep{Knapp1985ApJ...292..640K}. For those minor species, the blackbody approximation of the
  stellar flux may lead to inaccurate absolute intensity predictions
  in the order of 5 to 20\,\%. 
\end{itemize}

\citet{Decin2006A&A...456..549D, Decin2007A&A...475..233D} and \citet{Crosas1997ApJ...483..913C} demonstrated that not only the
integrated line intensities, but particularly the line shapes should
be taken into account for a proper determination of the envelope
structure. This is particularly true since \emph{(i.)} the line shapes contain valuable information on the line forming region, e.g.\ a Gaussian line profile points toward line formation partially in the inner wind where the stellar winds has not yet reached its full terminal velocity \citep{Bujarrabal1991A&A...251..536B}, and \emph{(ii.)} while the absolute uncertainty of the  intensity ranges between 20 to 50\,\%, the relative accuracy (of the line shapes) is of the order of a few percent.  In contrast to most other studies, we therefore will not use a
classical $\chi^2$-analysis using the integrated line intensity, but
will perform the model selection and the assessment of goodness-of-fit
using the log-likelihood function on the full line profiles as described in
\citet{Decin2007A&A...475..233D}.

\subsection{Molecular line data} \label{molecular_data}
In this paper, line transitions of CO, SiO, SiS, CS, CN, HCN, SO, and
SO$_2$ will be modeled, and H$_2$O line profile predictions for the
Herschel/HIFI instrument will be performed. The molecular line data used in this paper are described in Appendix~\ref{appa}.

\subsection{Molecular abundance stratification}

\begin{figure*}
\begin{center}
\includegraphics[height=.9\textwidth,angle=270]{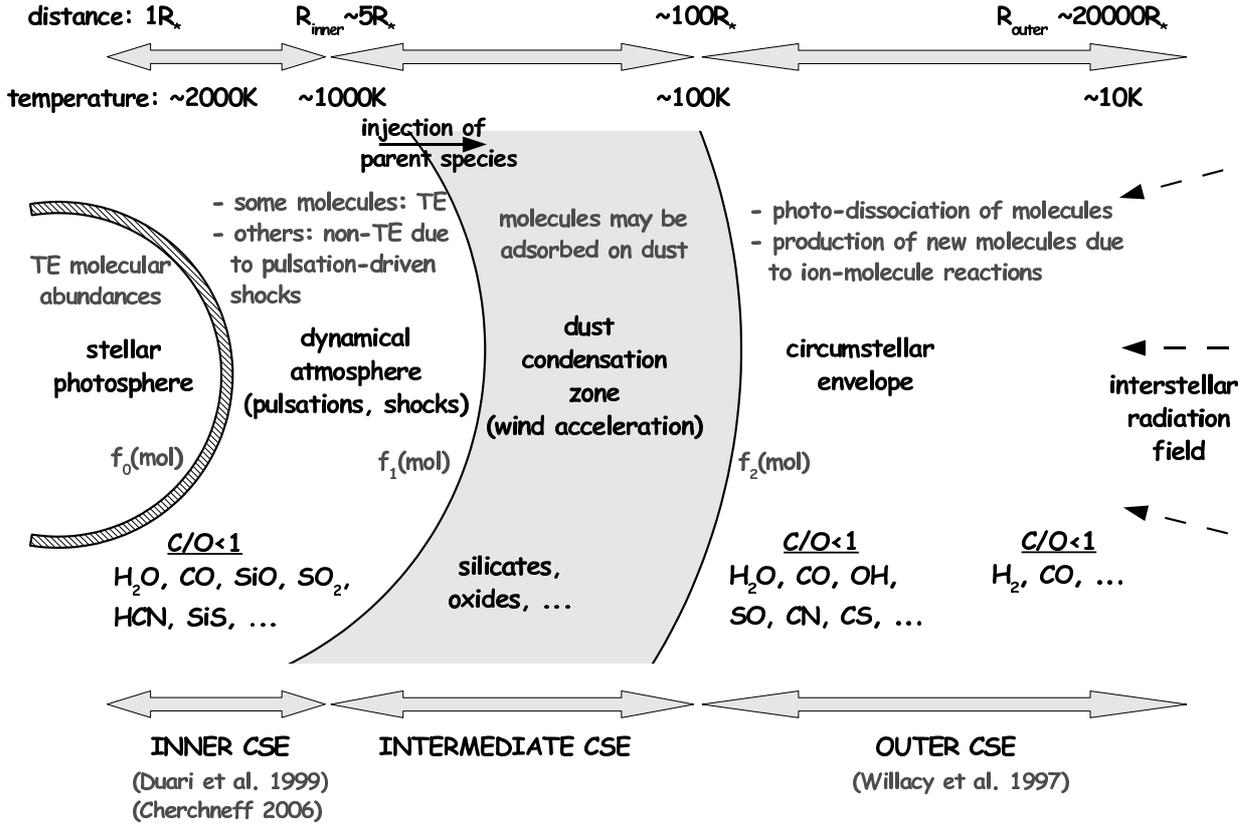}
\caption{Schematic overview (not to scale) of the
   circumstellar envelope (CSE) of an oxygen-rich AGB star. Several chemical processes are indicated at the typical temperature and radial distance from the star in the envelope where they occur. The nomenclature as used in this paper is given.}
\label{sketch}
\end{center}
\end{figure*}

Theoretical chemical calculations clearly show that the fractional
abundances (relative to H$_2$) vary throughout the envelope.  Chemical
processes responsible for the molecular content are dependent on the
position in the envelope (see Fig.~\ref{sketch}). In the stellar photosphere and at the inner
boundary of the envelope, the high gas density and temperature ensure
thermal equilibrium (TE). TE is suppressed very close to the
photosphere because of the action of pulsation-driven shocks
propagating outwards. Furthermore the regions of strong shock activity
correspond to the locus of grain formation and wind acceleration. This
region is referred to as the inner envelope (or inner wind) which
extends over a few stellar radii. At larger radii ($\sim$5 to
100\,R$_\star$) the newly formed dust grains interact with the cooler
gas. Depletion or formation of certain molecular/atomic species may
result from such an interaction and these layers are referred to as
the intermediate envelope. This is also the region where parent
molecules, injected in the envelope, may begin to break down, and
daughter molecules are formed.  At still larger radii
($>$100\,R$_\star$), the so-called outer envelope is penetrated by
ultraviolet interstellar photons and cosmic rays resulting in a
chemistry governed by photochemical processes.

Since our modeling results will be compared to chemical abundance predictions in the outer envelope by \citet{Willacy1997AandA...324..237W} and in the 
inner envelope by \citet{Duari1999AandA...341L..47D} and \citet{Cherchneff2006AandA...456.1001C}, we first briefly
describe these studies in Sect.~\ref{outer} and Sect.~\ref{inner} respectively.  In
Sect.~\ref{strategy} we discuss how we have implemented this knowledge
in the modeling of the molecular line transitions.

\subsubsection{Chemical stratification in the outer envelope} \label{outer}
\begin{figure}[htp]
\begin{center}
\includegraphics[width=.5\textwidth,angle=0]{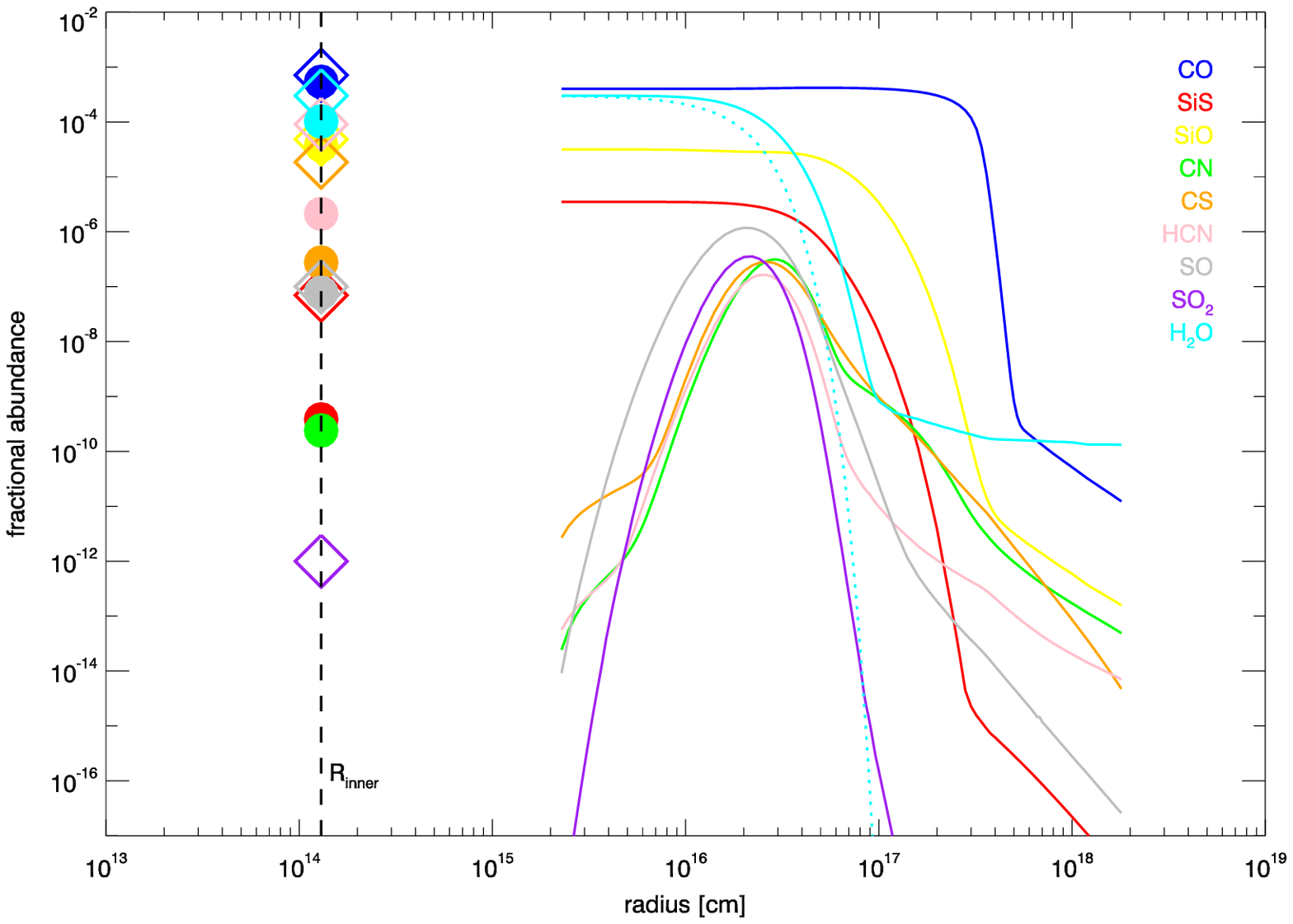}
\caption{Theoretically predicted molecular abundance stratification in the envelope of \object{IK~Tau}. The full lines represent the results as derived by \citet{Willacy1997AandA...324..237W} for the outer envelope of \object{IK~Tau} assuming  a mass-loss rate of $1 \times 10^{-5}$\,\Msun/yr. The diamonds represent the abundances derived
  by \citet{Cherchneff2006AandA...456.1001C} in the framework of
  non-equilibrium chemistry due to the passage of shocks in the inner
  envelope of \object{TX~Cam} at 2\,\Rstar; the filled circles indicate the results by
  \citet{Duari1999AandA...341L..47D} for \object{IK~Tau} at 2.2\,\Rstar.  The inner wind predictions are projected onto the dust-condensation radius derived in Sect.~\ref{velocity_structure}. The dotted line represents the
  Gaussian distribution of circumstellar H$_2$O as used by
  \citet{Maercker2008A&A...479..779M}.
}
\label{fig:Willacy_data}
\end{center}
\end{figure}

  The chemistry in the outer envelope of \object{IK Tau} has been
  modeled by \citet{Willacy1997AandA...324..237W}. This chemical kinetic model aims at deriving
  the abundance stratification in the outer envelope (between $2
  \times 10^{15}$ and $2 \times 10^{18}$\,cm). The chemistry is driven
  by a combination of cosmic-ray ionization and ultraviolet radiation
  and starts from nine parent species injected into the envelope (see
  Table~\ref{table:Willacy}). The CSE of IK~Tau was assumed to be
  spherically symmetric with a constant mass loss rate and a constant expansion velocity of
  19\,km/s. The temperature was described by a power law
  \begin{equation}
    T(r) = T(r_0) \left(\frac{r_0}{r}\right)^{0.6}
  \end{equation}
  with $r_0 = 2 \times 10^{15}$\,cm and 
$T(r_0)$ taken to be 100\,K for \Mdot$> 5 \times 10^{-6}$\,\Msun/yr and 300\,K otherwise. For \object{IK~Tau} $T(r_0)$  was assumed to be 100\,K. 
  The derived fractional abundances for the
  molecules studied in this paper are represented in
  Fig.~\ref{fig:Willacy_data}.

\begin{table}[htp]
\caption{The fractional abundance (relative to H$_2$) taken for the
  parent species by \citet{Willacy1997AandA...324..237W}.}
\label{table:Willacy}
\begin{center}
\begin{tabular}{llll}
\hline
Parent & Fractional & Parent & Fractional \\
species & abundance & species & abundance \\
\hline
\rule[3mm]{0mm}{0mm}CO     & $4.00 \times 10^{-4}$  & N$_2$ & $5.00 \times10^{-5}$ \\
H$_2$O & $3.00 \times 10^{-4}$  & H$_2$S & $1.27 \times 10^{-5}$ \\
CH$_4$ & $3.00 \times 10^{-5}$  & SiS & $3.50 \times 10^{-6}$ \\
NH$_3$ & $1.00 \times 10^{-6}$  & SiO & $3.15 \times 10^{-5}$ \\
He     & $1.00 \times 10^{-1}$  & PH$_3$& $3.00 \times 10^{-8}$\\
HCl    & $4.40 \times 10^{-7}$  &       &          \\
\hline
\end{tabular}
\end{center}
\end{table}

The model of \citet{Willacy1997AandA...324..237W} succeeded in reproducing the observed values of certain species but failed for some other molecular abundances: the calculated abundance of HCN was too low and the injected abundance of the parent species SiS was  about 10 times higher than observed.
\citet{Duari1999AandA...341L..47D} noted that the input molecular abundances of some parent species are sometimes questionably high since there exists no observational or theoretical evidence for the formation of these species in the inner and intermediate envelopes of O-rich Miras (see Sect.~\ref{inner}). More importantly, \citet{Duari1999AandA...341L..47D} showed that HCN should form in the inner envelope or extended stellar atmosphere due to non-equilibrium shock chemistry, and may be a parent species injected to the outer envelope. Recent observational studies also indicate that HCN must be formed in the inner envelope \citep{Bieging2000ApJ...543..897B, Marvel2005AJ....130..261M}. These results are in contrast to the modeling efforts of \citet{Willacy1997AandA...324..237W} where HCN was not yet considered as a parent species. 

\subsubsection{Chemical stratification in the inner envelope} \label{inner}

Carbon-bearing molecules have been identified in the envelopes of many oxygen-rich AGB stars \citep[e.g.,][]{Bujrrabal1994AandA...285..247B} and it was first thought that the observed carbon species were produced in the outer wind of O-rich stars via photochemical processes. However, \citet{Duari1999AandA...341L..47D} showed that shock-induced non-equilibrium chemistry models predict the formation of large amounts of a few carbon species, such as HCN, CS and CO$_2$, in the inner envelope of \object{IK~Tau}: these molecules are hence formed  in the post-shocked layers and are then ejected in the outer wind as 'parent' species. For some parent species, the non-equilibrium chemistry does not alter significantly the initial photospheric TE abundances. But other species, abundant in the TE photosphere, are quickly destroyed in the outflow by the non-equiblibrium chemistry generated by shocks (e.g., OH, SiS and HS). Again other species (such as SO) appear to be absent in the inner regions of the wind, and are thought to be produced by ion-molecule reactions in the photo-dissociation regions of the outer wind.

  \citet{Cherchneff2006AandA...456.1001C} continued on the study of shock-induced non-equilibrium chemistry in the inner wind of AGB stars. She demonstrated that whatever the enrichment in carbon of the star (i.e.\ the C/O ratio), the atomic and molecular content after the passage of the first shock in the gas layers just above the stellar photosphere is very much the same, and in many cases, totally different from what would be expected from thermodynamic equilibrium
  (TE) calculations. In the case of the oxygen-rich envelope around \object{TX~Cam} --- being almost a stellar twin of \object{IK~Tau}, but with slightly lower luminosity --- \citet{Cherchneff2006AandA...456.1001C} found that while,  e.g.,  the TE abundance of HCN (CS) is predicted to be $\sim$$1.9\times10^{-11}$ ($\sim$$2.5\times10^{-11}$), the non-TE fractional
  abundances at 2.5\,R$_\star$ are predicted to be $\sim$$9\times10^{-5}$ ($\sim$$1.85\times10^{-5}$). The
  fractional abundances derived by
  \citet{Cherchneff2006AandA...456.1001C} differ from the abundances of
  the injected parent molecules in the study of
  \citet{Willacy1997AandA...324..237W} (see Table~\ref{table:Willacy} and Fig.~\ref{fig:Willacy_data}):
  \citet{Willacy1997AandA...324..237W} did not consider CS and HCN as
  being parent molecules, and the (injected) abundance of SiS ($3.5\times10^{-6}$) is
  much higher than the abundance stratification derived by
  \citet{Cherchneff2006AandA...456.1001C} (see their Fig.~8).

\subsection{Modeling strategy} \label{strategy}

\subsubsection{Envelope structure as traced by the CO lines}

The physical properties of the circumstellar gas, such as the
temperature, velocity and density structure, are determined from the
radiative transfer modeling of the multi-transitional (sub)millimetre
CO line observations. Since higher-$J$ lines are formed at higher
temperature, different transitions offer the possibility to trace
different regions in the envelope. The highest CO energy level traced
is the CO $J^{\rm up}=7$ level at 154.8\,K. The available rotational CO lines
will hence be good tracers for the region beyond $\sim$100\,R$_\star$,
but they do not put strong constraints on the temperature in the inner
CSE. The upcoming Herschel/HIFI mission will be crucial in the study
of the temperature structure in this inner wind region.

An extensive grid has been calculated with parameters ranging from 2000 to 3000\,K for the stellar temperature $T_{\rm eff}$, from $1 \times 10^{13}$ to $6 \times 10^{13}$\, cm for the stellar radius \Rstar, an inner (dust condensation) radius between 2 and 30\,\Rstar, distance between 200 and 300\,pc, and a constant mass-loss rate between $1 \times 10^{-6}$ and $5 \times 10^{-5}$\,\Msun/yr.
As briefly explained in Sect.~\ref{rad_model}, a log-likelihood method \citep{Decin2007A&A...475..233D} is used to find the best-fit model and derive a 95\,\% confidence interval for the model parameters. The results will be presented in Sect.~\ref{results}.

\subsubsection{Abundance stratification through the envelope}

From the descriptions of theoretical abundance estimates in the inner and outer envelope (Sect.~\ref{outer} -- \ref{inner}) it is clear that there is still some debate about the  abundance structure in the envelope. SiS and HCN were already given as an example, but other molecules such as e.g.\ CS and SO also pose a problem. This is illustrated in Fig.~\ref{fig:Willacy_data}, where one notices for a few molecules a large difference between the theoretically predicted fractional abundance in the inner envelope by \citet{Duari1999AandA...341L..47D} and \citet{Cherchneff2006AandA...456.1001C} and the abundance of the parent molecules injected in the outer envelope by \citet{Willacy1997AandA...324..237W}. One of the big questions still existing concerns the modifications of the molecular abundances in the intermediate wind region due to gas-grain reactions. Currently, no theoretical efforts have been made to model this region in terms of molecular `leftovers' after the dust formation has occurred. In the case of O-rich envelopes, it is thought that CO, HCN and CS are quite stable and travel the entire envelope unaltered until they reach the photo-dissocation region of the outer wind as these molecules do not participate in the formation of dust grains such as silicates and corundum \citep{Duari1999AandA...341L..47D}. In contrast, SiO is a candidate molecule for depletion in the intermediate wind region due to the formation of SiO$_2$ (via a reaction with OH) whose condensation product, silica, is tentatively identified in post-AGB stars \citep{Molster2002A&A...382..222M} and is claimed to be the carrier of the 13\,$\mu$m  feature in low mass-loss rate AGB stars \citep[][but other studies argue that this feature is due to spinel]{Speck2000A&AS..146..437S}. The theoretical modeling of \citet{Duari1999AandA...341L..47D} and \citet{Cherchneff2006AandA...456.1001C} predict an SiS abundance 2 to 3 orders of magnitude lower than the observed value, indicating that SiS is produced in the outer envelope of \object{IK~Tau}. However, recent observational results by \citet{Decin2008A&A...480..431D} argue for a formation process in the inner envelope.

From the above arguments, it is clear that we should allow some variation in modeling the abundance structure in the envelope. However, one should also realize that we sometimes only have 2 rotational transitions of one isotopolog at our disposal with a restricted range in excitation temperature. The highest upper level energy traced is the SiS(20--19) transition at $\sim$183\,K; hence none of the studied transitions is sensitive to the abundance in the inner envelope (R$\la 5$\,\Rstar). In order to use some prior knowledge on the (theoretical) photo-dissociation rate in the outer regions and to allow for a depletion or an extra formation process in the intermediate/outer envelope, we therefore have opted to divide the envelope in different regimes (see also Table~\ref{ass:abundances} and Fig~\ref{sketch}): \emph{(i.)} in the dust-free zone ($R \le R_{\rm inner}$) the abundance is constant ($f_1({\rm mol})$),  \emph{(ii.)} between $R_{\rm inner}$ and $R_{\rm max}$ the abundance can decrease/increase from $f_1({\rm mol})(R_{\rm inner})$ to $f_2({\rm mol})(R_{\rm max})$ linearly on a log-log scale, where both $R_{\rm max}$ and $f_2({\rm mol})$ are free parameters, \emph{(iii.)} from $R_{\rm max}$ onwards, the abundance stratification follows the (photodissociation) results of \citet{Willacy1997AandA...324..237W} scaled to $f_2({\rm mol})$ at $R_{\rm max}$. In that way, 3 parameters ($f_1({\rm mol})$, $f_2({\rm mol})$, and $R_{\rm max}$) have to be estimated to determine the abundance stratification of a species.

\begin{table}[htp]
 \caption{Modeling assumptions of the abundance stratification.}
\label{ass:abundances}
\setlength{\tabcolsep}{1.mm}
\begin{tabular}{lll}
\hline
\hline
Region & Variable & Comment \\
\hline
 $R < R_{\rm{inner}}$ & $f_0$(mol)  & not sensitive \\
\hline
 $R =R_{\rm{inner}}$ &$f_1$(mol)  &  $f_0$(mol)$ \equiv f_1$(mol) \\
\hline
 $R =R_{\rm{2}}$ & $f_2$(mol)  &  allow for depletion {\footnotesize{(e.g. due to dust}} \\
                                 &    $R_{\rm{2}}$                &  formation) or extra formation process \\
\hline
\multicolumn{1}{l}{$R_{\rm{inner}} < R < R_{\rm{2}}$} & & linear interpolation in $\log$(R)-$\log$(f) \\
\hline
\multicolumn{1}{l}{$R > R_{\rm{2}}$}& & theor. pred. of Willacy scaled to $f_2$(mol) \\
\hline
\hline
 \end{tabular}
\end{table}

Most studies use the photodissociation results of \citet{Mamon1988ApJ...328..797M} to describe the CO spatial variation in the outer envelope. For other molecules, the abundance pattern is often assumed to be described by a simple Gaussian or expontential distribution \citep[e.g.][]{Bieging2000ApJ...543..897B, GonzalesDelgado2003AandA...411..123G, Schoier2007A&A...473..871S}. The $e$-folding radius then describes the photodissociation by ambient UV photons penetrating the dusty envelope or depletion of the molecules from the gas into dust grains in the outflowing stellar wind.
 That way, however, all molecules are assumed to be created in the extended atmosphere or inner wind region. Moreover, a combination of depletion and photodissociation or 
extra depletion/formation processes in the intermediate/outer region can not be captured,  and one can not use the results by  \citet{Willacy1997AandA...324..237W} describing extra formation of a few molecules by ion-ion reactions in the outer wind region. The methodology outlined above (Table~\ref{ass:abundances}) captures these flaws, and may serve to considerably strengthen our knowledge on the abundance stratification in the envelope.

As was already alluded to in the previous paragraph, the line profiles in this study are not sensitive to a change in abundance in the inner wind region ($R \le R_{\rm inner}$). To assess the abundance stratification in this region, one either needs high-resolution near-infrared  \citep[see, e.g.,][]{Decin2008A&A...484..401D} or far-infrared spectroscopy (as will be provided by the Herschel/HIFI instrument). Nonetheless, the derived abundance stratifications will be compared to the theoretical inner wind predictions by \citet{Duari1999AandA...341L..47D} and \citet{Cherchneff2006AandA...456.1001C}, since this comparison may yield hints on the (un)reactivity of the molecules in the dust forming region and on uncertainties in the inner wind predictions.


\section{Results} \label{results}

Using the log-likelihood method, the parameters for the model yielding the best-fit to the CO line profiles are derived (see parameters listed in the 2nd column in Table~\ref{param_IKTau}, `model 1'). The CO lines, however, only trace the envelope beyond $\sim$100\,\Rstar. One therefore should use other molecules to put constraints on the structure in the inner wind region.
HCN is the only molecule for which we have observational evidence that it is  formed in the inner wind region: using interferometric data \citet{Marvel2005AJ....130..261M} deduced a maximum size for the HCN distribution of 3.85\arcsec (in diameter), or a radius of $7.2 \times 10^{15}$\,cm at 250\,pc. They concluded that the deduced size indicates a shock origin for HCN close to the star and a radius for the HCN distribution limited by photodissociation.
The HCN line profiles (Fig.~\ref{HCN_model}) are clearly Gaussian indicating a line formation (at least partly) in the inner wind region, where the wind  has not yet reached its terminal velocity. That way, HCN observations yield important clues on the velocity structure in the inner wind region. Using the stellar parameters as given in the 2nd column in Table~\ref{param_IKTau} ('model 1'), we were unable to derive a HCN-abundance structure yielding a satisfactory fit to the line profiles. While the integrated intensities could be well predicted, the line profiles were flat-topped with a FWHM (full width at half maximum) being too large. The only way to reconcile this problem was \emph{(1)} concentrating the HCN abundance in the inner $2 \times 10^{15}$\,cm with [HCN/H]\,=\,$9\times10^{-6}$,  or \emph{(2)} allowing for a smoother velocity law. While the abundance in the former solution is within the predictions of \citet{Cherchneff2006AandA...456.1001C}, the angular distance is much smaller than the 3.85\arcsec diameter deduced by \citet{Marvel2005AJ....130..261M}.

\subsection{Velocity structure} \label{velocity_structure}
\begin{figure}
 \begin{center}
  \includegraphics[height=.48\textwidth,angle=90]{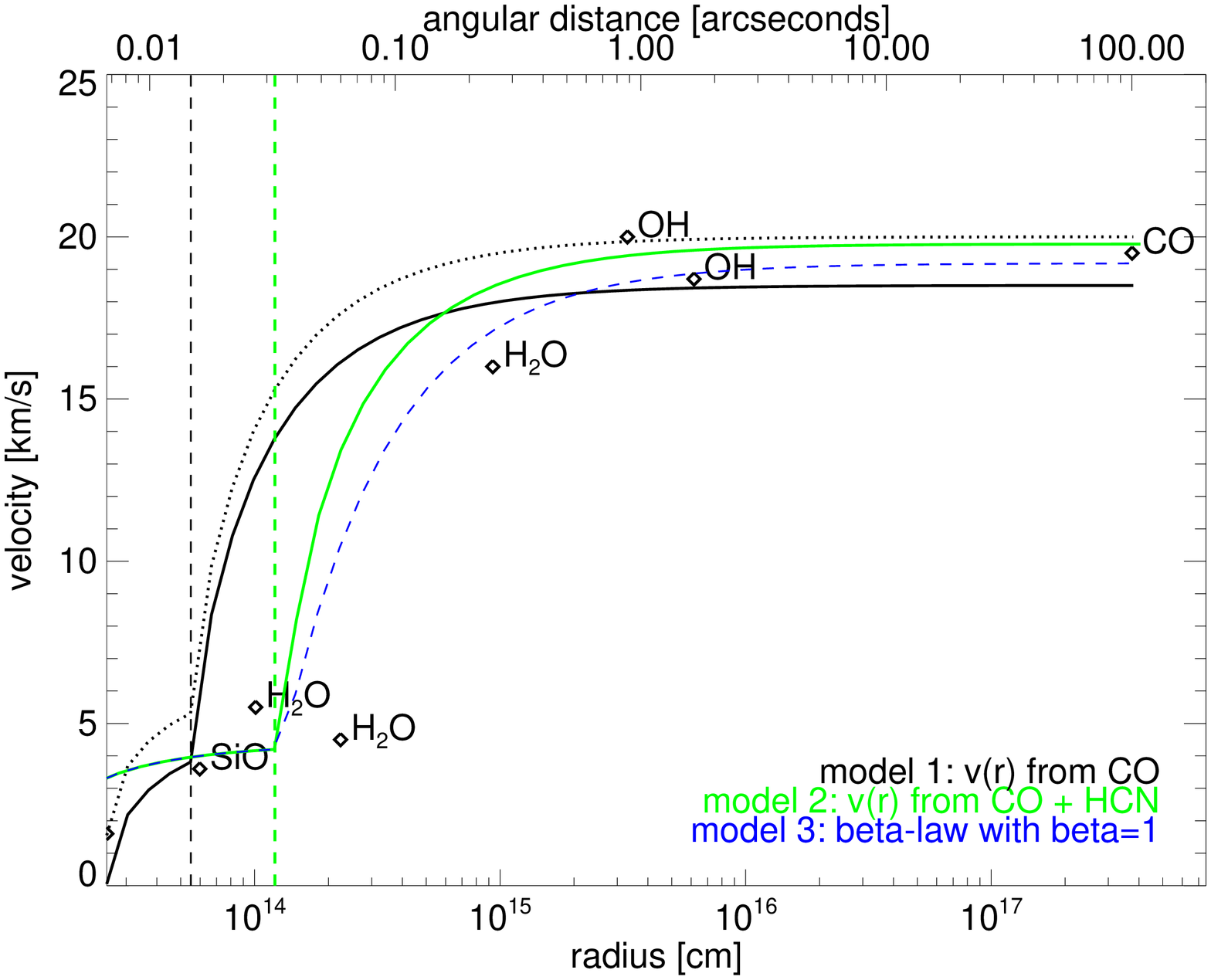}
\caption{Velocity profile of \object{IK~Tau}. Velocity data are obtained from mapping of maser emission: SiO \citep{Boboltz2005ApJ...625..978B}, H$_2$O \citep{Bains2003MNRAS.342....8B}, and OH \citep{Bowers1989ApJ...340..479B}. The CO expansion velocity derived from our CO data is also indicated. The expansion velocity deduced from the CO data alone (see `model 1' in Table~\ref{param_IKTau}) is plotted as full black line. The dotted black line indicates the velocity structure taking a turbulent velocity (of 1.5\,km/s) into account. The green line gives the expansion (+turbulent) velocity deduced from both the CO and HCN lines (model 2 in Table~\ref{param_IKTau}). The dashed blue line represents an even smoother expansion (+turbulent) velocity structure, applying Eq.~\ref{beta}, with $\beta$\,=\,1 (model 3 in Table~\ref{param_IKTau}). The vertical dashed black and green lines indicate the dust condensation radius R$_{\rm inner}$.  Note that $\beta$=0.5 in the inner wind region.}
\label{velocity}
 \end{center}

\end{figure}

The expansion velocity of SiO, H$_2$O, and OH masers can be used to put further constraints on the velocity structure, and specifically on the acceleration in the inner wind region (see Fig.~\ref{velocity}). It is clear that the velocity structure as derived from the parameters of the best-fit model only based on CO lines (model 1), is far too steep in the inner wind region as compared to the velocity indications of the maser lines. This problem can be solved by either increasing the  dust condensation radius or by allowing for a smoother velocity profile. This latter can be simulated using the classical $\beta$-law \citep{Lamers1999isw..book.....L} with $\beta > 0.5$ (see Fig.~\ref{velocity})
\begin{equation}
{v(r)\simeq v_0+(v_\infty-v_0)\left(1-\frac{R_\star}{r}\right)^\beta}
\label{beta}
\end{equation}
with $v_0$ the velocity at the dust condensation radius. 

One should realize that several assumptions are inherent to the velocity structure derived from solving the momentum equation: \emph{(i.)} All dust species at all different grain sizes are assumed to be directly formed at the dust condensation radius $R_{\rm inner}$. However, theoretical results from, e.g., \citet{Gail1999A&A...347..594G} show that formation and growth of (silicate) dust grains typically occur between 1100 and 900\,K, i.e.\ extending over a few stellar radii.
\emph{(ii.)} The extinction efficiencies used in the GASTRoNOoM-code represent the Fe-rich silicate MgFeSiO$_4$.
Thanks to their high absorption efficiencies at optical and near-infrared wavelength  Fe-rich silicates like MgFeSiO$_4$ (and solid Fe) are efficient wind drivers \citep{Woitke2006A&A...460L...9W}. However, other oxides or pure silicates such as Al$_2$O$_3$, SiO$_2$, Mg$_2$SiO$_4$ and MgSiO$_3$ have low absorption efficiencies at optical and near-infrared wavelengths, resulting in a negligible radiative pressure on all glassy condensates. If these latter molecules were the most abundant in the envelope of \object{IK~Tau}, the wind acceleration would be much lower. No medium resolution infrared (from the Infrared Space Observatory - Short Wavelength Spectrometer or the Spitzer - InfraRed Spectrograph) data are, however, available for IK~Tau, hence we were unable to study the circumstellar dust composition in detail.
\emph{(iii.)} 'Complete momentum coupling' is assumed. This means that the grain motion everywhere in the flow can be computed by equating the local radiative and collisional drag forces implying that virtually all of the momentum gained by a grain through the absorption of radiation from the stellar photosphere is transferred via collisions to the atmospheric gas \citep{MacGregor1992ApJ...397..644M}. For physical conditions typical of the circumstellar envelopes of oxygen-rich red giants, \citet{MacGregor1992ApJ...397..644M} find that silicate grains with initial radii smaller than about $5 \times 10^{-2}$\,$\mu$m decouple from the ambient gas near the base of the outflow. \emph{(iv.)} The momentum equation used in the GASTRoNOoM-code \citep[see Eq.~2 in ][]{Decin2006A&A...456..549D}, implicitly assumes that the mass outflow is steady in time and that the circumstellar dust is optically thin to the stellar radiation \citep{Goldreich1976ApJ...205..144G}.
Dust emission modeling by \citet{Ramstedt2008AandA...487..645R} suggests a circumstellar envelope which is slightly optically thick at 10\,$\mu$m ($\tau_{10}$\,=\,1.2).
These results suggest that the acceleration of the gaseous particles in the inner wind might be slower than deduced from solving the momentum equation ('model 1' and 'model 2'), since not all dust species take part in the momentum transfer. 

\begin{figure*}[htp]
\begin{center}
\includegraphics[height=.9\textwidth,angle=90]{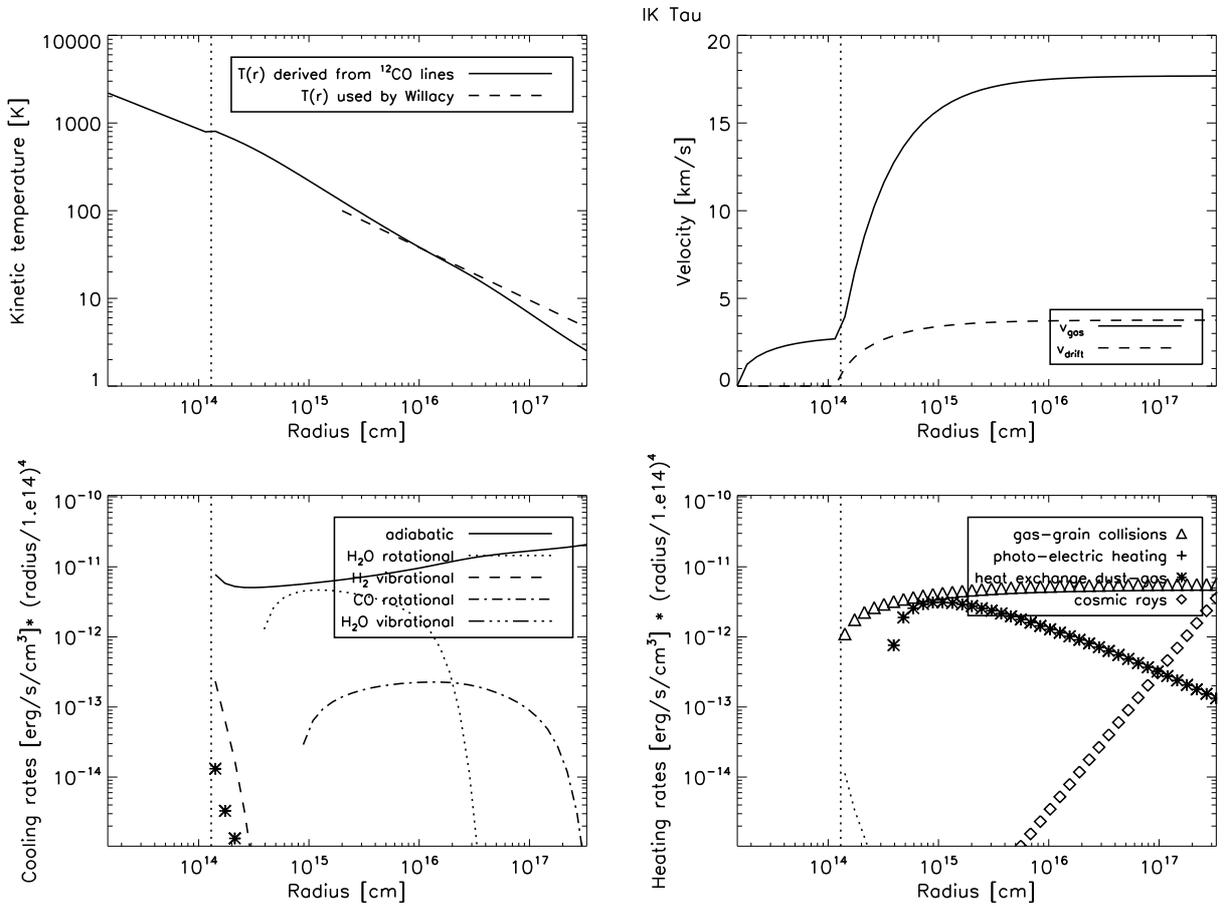}
\caption{Thermodynamic structure in the envelope of \object{IK Tau} as
  derived from the $^{12}$CO and HCN rotational line transitions for the stellar parameters of model 3 in Table~\ref{param_IKTau}. \emph{Upper
    left:} Estimated temperature profile, \emph{upper right:}
  estimated gas and drift velocity, \emph{lower left:} cooling rates,
  and \emph{lower right}: heating rates. The start of the dusty
  envelope, R$_{\rm{inner}}$, is indicated by the dotted line. The
  temperature structure as used in the chemical kinetic calculations
  by \citet{Willacy1997AandA...324..237W} is plotted in the upper left
  panel in dashed lines.}
\label{fig:structure_IKTau}
\end{center}
\end{figure*}

Solving the momentum equation and taking both the CO and HCN line profiles into account, the inner radius of the dusty envelope is shifted towards higher values, $R_{\rm inner}\,=\,1.3 \times 10^{14}$\,cm, and the mass-loss rate slightly increases ('model 2' in Table~\ref{param_IKTau}). 
Using the same stellar parameters as in `model 2', but simulating a smoother velocity law still compatible with the maser line mapping ($\beta\,=\,1$) decreases the mass-loss rate to $8 \times 10^{-6}$\,\Msun/yr (model 3 in Table~\ref{param_IKTau}), due to the condition of mass conservation (model 3 in Table~\ref{param_IKTau}). 
The narrow Gaussian line profiles of the HCN lines (Sect.~\ref{HCN}) give more support to model 3 than to model 2, for which reason the thermodynamic structure as deduced from model 3 (Fig.~\ref{fig:structure_IKTau}) will be used to model the other molecular line transitions.

\begin{table*}[htp]
  \caption{Parameters of the models with best goodness-of-fit for
    \object{IK Tau}. Details for each model are given in the text.  The numbers in italics
    are input parameters that have been kept fixed at the given
    value. }
\label{param_IKTau}
\centering
\begin{tabular}{lcccc}
\hline \hline
 \rule[0mm]{0mm}{5mm}& model 1 & model 2 & model 3 & 95\% confidence\\
 based on & \multicolumn{1}{c}{CO} & \multicolumn{1}{c}{CO \& HCN} & \multicolumn{1}{c}{CO \& HCN} & interval for\\
$v(r)$ & momentum eq. & momentum eq. & power-law $\beta\,=\,1$ & model 3\\
\hline
\rule[-3mm]{0mm}{8mm}parameter & value & value & value & \\
\hline
\rule[0mm]{0mm}{5mm}\Teff [K] & 2200 & 2200 & 2200 & 100  \\
$R_{\star}$ [$10^{13}$\,cm] & 2.5 & 1.5 & 1.5 & 1.0 \\
$[$CO/H$_2]$ [$10^{-4}$]    & \emph{2} & \emph{2} & \emph{2}& $-$  \\
distance  [pc] & 250 & 265 & 265 & 20 \\
$R_{\rm{inner}}$ [\Rstar] & 2.2 & 8.7 & 8.7 & 2 \\
$R_{\rm{outer}}$ [\Rstar] & 15170 & 27000 & 27000 & $-$ \\
$v_{\infty}$     [km\,s$^{-1}$] & \emph{17.7} & \emph{17.7}& \emph{17.7} & $-$ \\
\Mdot$(r)$ [{\Msun}/yr] & $8 \times 10^{-6}$ & $9 \times 10^{-6}$& $8 \times 10^{-6}$& $1 \times 10^{-6}$\\
$^{12}$C/$^{13}$C & 20 & 16 &  14 &  2 \\
    \hline
  \end{tabular}
\end{table*}

\subsection{Stellar parameters derived from the CO and HCN lines}

The stellar parameters for the best-fit model (model 3) are listed in Table~\ref{param_IKTau}. The outer radius of CO is computed using the results of \citet{Mamon1988ApJ...328..797M}. The derived 95\,\% confidence intervals in Table~\ref{param_IKTau} are statistical uncertainties, which should be interpreted in the light of the model assumptions of a spherically symmetric wind. As described in \citet{Decin2007A&A...475..233D}, the log-likelihood function can also be used to compare different models with a different number of parameters. In case of \object{IK~Tau}, we have assessed the likelihood preference of a model with constant mass-loss rate compared to a model with  mass-loss rate variations. The preferences pointed towards the simpler model, i.e.\ with a constant mass-loss rate (of $8 \times 10^{-6}$\,\Msun/yr). We also derived the dust-to-gas mass ratio from the amount of dust needed to drive a wind at a terminal velocity of 17.7\,km/s  for a gas mass-loss rate of $8 \times 10^{-6}$\,\Msun/yr with the deduced velocity profile. A dust-to-gas mass ratio of $1.9\times10^{-2}$ (or a dust mass-loss rate of $1.52 \times 10^{-7}$\,\Msun/yr) is obtained for model 3, with an estimated uncertainty of a factor $\sim$5.

\subsection{Fractional abundances} \label{frac_abundances}

\begin{figure*}[htp]
\begin{center}
\ \hspace{1.5cm}C-bearing molecules \hspace{5cm} Si and S bearing molecules\\
\vspace*{-3ex}
\includegraphics[height=.45\textwidth,angle=90]{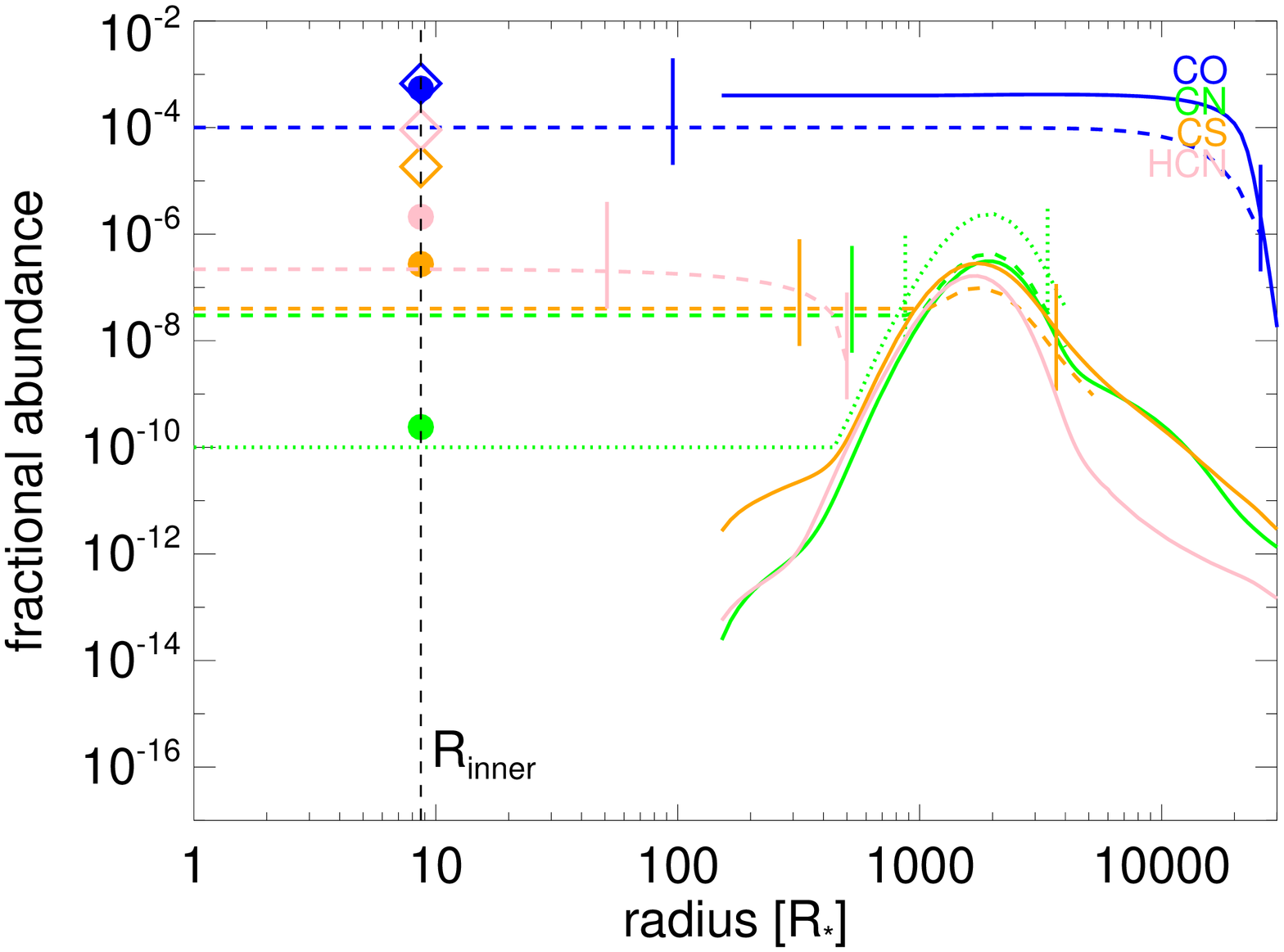}
\includegraphics[height=.45\textwidth,angle=90]{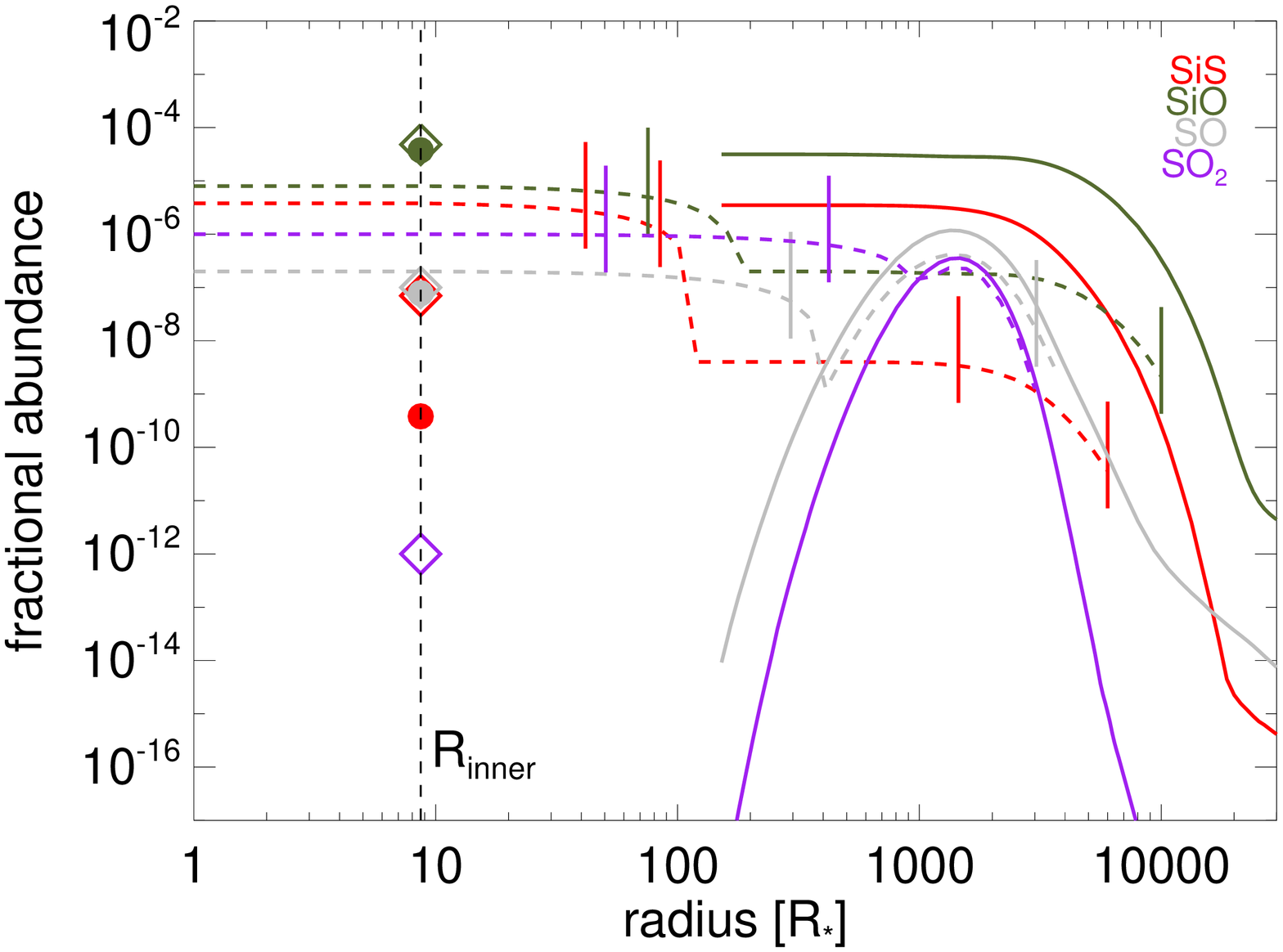}
\caption{Predicted abundance stratifications [mol/H$_{\rm tot}$] (full line, see
  Fig.~\ref{fig:Willacy_data}) are compared to the deduced abundance
  structures (dashed lines). In the left panel, the carbon bearing molecules are shown, the right panel gives the Si and S bearing molecules.
For CN, an alternative solution is given in dotted lines. For each molecule, the line formation region traced by the observed molecular lines is
indicated by the vertical dashes (see also Table~\ref{Table:abund_results}).}
\label{fig:abundances}
\end{center}
\end{figure*}

Using the thermodynamic envelope structure derived above (see Fig.~\ref{fig:structure_IKTau}), the abundance stratification of all molecules is derived. 
A comparison to the theoretical inner and outer wind predictions (as discussed in Sect.~\ref{outer}--\ref{inner}) is given in Fig.~\ref{fig:abundances}. 
The studied molecular line transitions are not sensitive to the full envelope size, but have a limited formation region. The part in the envelope we can trace by combining the different available rotational line transitions is indicated with vertical dashes in Fig.~\ref{fig:abundances} and tabulated in Table~\ref{Table:abund_results}\footnote{Estimates of these ranges are found by considering the place where $I(p) p^3$, with $I$ the intensity and $p$ the impact parameter, is at half of its maximum value.}. 

A comparison between observed and predicted line profiles and a discussion on the deduced abundance stratification are given for each molecule separately in the following subsections. We will always first briefly describe the deduced abundances, then compare the results to the theoretical inner and outer wind predictions and finally compare to other results found in the literature (see Table~\ref{table:comparison}). For the literature results, a difference is made between studies based on the assumptions of optically thin unresolved emission and a population distribution thermalized at an excitation temperature that is constant throughout the envelope, and those based on a full non-LTE  radiative transfer calculation. One should also realize that most studies make use of integrated line intensities, and do not deal with a full line profile analysis as is done here. With the exception of \citet{Omont1993AandA...267..490O}, the other literature studies listed in Table~\ref{table:comparison} assume the shell to expand at a constant velocity. As discussed in Sect.~\ref{velocity_structure}, the Gaussian line profiles of the HCN and a few of the SiO lines are the result of line formation partially in the inner wind, where the stellar wind has not yet reached its full terminal velocity. As a result, more emission is produced at velocities near the line center than would be the case for a uniform-velocity wind. Hence, observational studies assuming a constant expansion velocity, will be unable to predict the line profiles properly.

\begin{table}
\caption{Molecular fractional abundance relative to H$_{\rm{tot}}$=n(H)+2n(H$_2$) (see Fig.~\ref{fig:abundances}). }
\label{Table:abund_results}
\setlength{\tabcolsep}{1mm}
\centering
\begin{tabular}{lllll}
\hline
\hline
 \rule[-3mm]{0mm}{8mm}& $f_1$(mol) & $f_2$(mol) & $R_{2}$ [\Rstar] & ranges [\Rstar]\\
\hline
\rule[0mm]{0mm}{5mm}$^{12}$CO & $1.0\times10^{-4}$ & 1.0$\times$10$^{-4}$ & 25170 & 100--25000 \\
$^{13}$CO & 7.1$\times$10$^{-6}$ & 7.1$\times$10$^{-6}$ & 25170 & 500--15000\\
SiS & 5.5$\times$10$^{-6}$ & 4.0$\times$10$^{-9}$ & 120 & 40--80, 1500--6000\\
$^{28}$SiO & 8.0$\times$10$^{-6}$ & 2.0$\times$10$^{-7}$ & 180 & 70--10000 \\
$^{29}$SiO & 3.0$\times$10$^{-7}$ & 7.5$\times$10$^{-9}$ & 180 & 70-2000 \\
$^{30}$SiO & 1.0$\times$10$^{-7}$ & 2.5$\times$10$^{-9}$ & 180 & 70--2000 \\
CS & 4.0$\times$10$^{-8}$ & 4.0$\times$10$^{-8}$ & 1160 & 300--4000 \\
SO & 2.0$\times$10$^{-7}$ & 1.0$\times$10$^{-9}$ & 400 & 200--3000 \\ 
SO$_2$ & 1.0$\times$10$^{-6}$ & 1.0$\times$10$^{-7}$ & 1000 & 50--400 \\
HCN & 2.2$\times$10$^{-7}$ & 4.0$\times$10$^{-9}$ & 500 & 50--500 \\
CN &   3.0$\times$10$^{-8}$   & 3.0$\times$10$^{-8}$  &   1000   & 500--3500     \\
\hline
\multicolumn{5}{c}{\emph{alternative solution}}\\
CN& 1.0$\times$10$^{-10}$ & 1.0$\times$10$^{-10}$ & 450 & 900--3500 \\
\hline
\end{tabular}
\end{table}

\begin{table*}
\caption{Comparison of the deduced fractional abundances to other observational studies and theoretical predictions.  All fractional abundances are given relative to the total H-content. In cases where values found in literature were given relative to H$_2$, they were re-scaled relative to the total H-content by assuming that all hydrogen is in its molecular form H$_2$.}
\label{table:comparison}
\setlength{\tabcolsep}{1.mm}
{\scriptsize{\begin{center}
\begin{tabular}{|l|c|c|c|c|c|c|c|c|c|}
\hline
 \rule[0mm]{0mm}{5mm}& $^{12}$CO & $^{13}$CO & CS & HCN & CN & SiO & SiS & SO & SO$_2$ \\
\hline
\rule[0mm]{0mm}{5mm}\citet{Lindqvist1988AandA...205L..15L}$^a$ & $1.5 \times 10^{-4}$ & $-$ & $1.5 \times 10^{-7}$ & $3.0 \times 10^{-7}$  & $-$ & $-$ & $3.5 \times 10^{-7}$ & $-$ &$-$ \\
\citet{Omont1993AandA...267..490O}$^b$ & $-$& $-$ & $-$ & $-$ &  $-$ &$1.5 \times 10^{-6}$ & $-$ & $9 \times 10^{-7}$ & $2.05 \times 10^{-6}$ \\
\citet{Bujrrabal1994AandA...285..247B}$^c$ & $1.5 \times 10^{-4}$ & $1.6 \times 10^{-5}$ & $5.0 \times 10^{-8}$ & $4.9 \times 10^{-7}$  &  $-$ & $8.5 \times 10^{-6}$ & $2.2 \times 10^{-7}$ & $1.3 \times 10^{-6}$ & $-$ \\
\citet{Kim2009}$^d$ ('case A') & $1.5 \times 10^{-4}$ & $1.75 \times 10^{-5}$ & $3.0 \times 10^{-7}$ & $1.4 \times 10^{-6}$ &   $1.6 \times 10^{-7}$& $1.3 \times 10^{-5}$ & $1.3 \times 10^{-6}$ & $7.8 \times 10^{-7}$ & $1.4 \times 10^{-5}$ \\
 \citet{Kim2009}$^d$ ('case B') &$1.5 \times 10^{-4}$ & $1.75 \times 10^{-5}$ & $8.1 \times 10^{-8}$ & $4.3 \times 10^{-7}$ &  $5.1 \times 10^{-8}$ & $5.1 \times 10^{-6}$ & $3.1 \times 10^{-7}$ & $2.7 \times 10^{-7}$ & $4.2 \times 10^{-6}$ \\
\hline
\rule[0mm]{0mm}{5mm}Gonz\'alez Delgado$^e$  &  &  & &  &   & & &  &  \\
et al. (2003) & {\raisebox{1.5ex}[0pt]{$1.0 \times 10^{-4}$}} &{\raisebox{1.5ex}[0pt]{$-$}}  & {\raisebox{1.5ex}[0pt]{$-$}} &{\raisebox{1.5ex}[0pt]{$-$}}  &  {\raisebox{1.5ex}[0pt]{$-$}} & {\raisebox{1.5ex}[0pt]{$2.0 \times 10^{-7}$}} & {\raisebox{1.5ex}[0pt]{$-$}} & {\raisebox{1.5ex}[0pt]{$-$}} & {\raisebox{1.5ex}[0pt]{$-$}}  \\
\citet{Schoier2007AandA...473..871S}$^f$ & $1.0 \times 10^{-4}$ & $-$ &$-$  &$-$  &  $-$ &  $-$ & $5 \times 10^{-6}$& $-$ & $-$\\
 & & & & & & & -- $5.0 \times 10^{-9}$ & &  \\
this work& $1.0 \times 10^{-4}$ & $7.1 \times 10^{-6}$ & $4 \times 10^{-8}$ & $2.2 \times 10^{-7}$ &  $1.0 \times 10^{-10}$  & $8.0 \times 10^{-6}$ & $5.5 \times 10^{-6}$ & $2.0 \times 10^{-7}$ & $1.0 \times 10^{-6}$ \\
 & & & & & -- $3.0\times 10^{-8}$  & -- $2.0 \times 10^{-7}$ &--  $4.0 \times 10^{-9}$&  & \\
\hline
\rule[0mm]{0mm}{5mm}\citet{Duari1999AandA...341L..47D}$^h$ & $5.38 \times 10^{-4}$ & $-$ & $2.75 \times 10^{-7}$ & $2.12 \times 10^{-6}$ & $2.40 \times 10^{-10}$ & $3.75 \times 10^{-5}$ & $3.82 \times 10^{-10}$ & $7.79 \times 10^{-8}$ & $-$\\
\citet{Cherchneff2006AandA...456.1001C}$^i$ & $6.71 \times 10^{-4}$ & $-$ & $1.85 \times 10^{-5}$ & $9.06 \times 10^{-5}$ & $3 \times 10^{-11}$$^g$ & $4.80 \times 10^{-5}$ & $7 \times 10^{-8}$ & $1 \times 10^{-7}$ &  $1 \times 10^{-12}$ \\
\citet{Willacy1997AandA...324..237W}$^j$ & $4 \times 10^{-4}$ & $-$ & $2.9 \times 10^{-7}$ & $1.4 \times 10^{-7}$ & $3.5 \times 10^{-7}$ & $3.2 \times 10^{-5}$ & $3.5 \times 10^{-6}$ & $9.1 \times 10^{-7}$ & $2.2 \times 10^{-7}$\\
\hline
\end{tabular}\\
\end{center}
\tablefoot{In the first part of the table, observational results are listed based on the assumption of optically thin emission and a population distribution which is thermalized at one excitation temperature.  The second part gives observational results based on a non-LTE radiative transfer analysis. Theoretical predictions for either the inner envelope \citep{Duari1999AandA...341L..47D, Cherchneff2006AandA...456.1001C} or outer envelope \citep{Willacy1997AandA...324..237W} fractional abundances are given in the last part.}
\tablebib{ \newline
$^a$: no information on used distance and mass-loss rate\\
$^b$: distance is 270\,pc, \Mdot\,=\,$4.5 \times 10^{-6}$\,\Msun/yr \\
$^c$: distance is 270\,pc, \Mdot\,=\,$4.5 \times 10^{-6}$\,\Msun/yr \\
$^d$: distance is 250\,pc, assumed \Mdot\ of $4.7 \times 10^{-6}$\,\Msun/yr, LTE is assumed, 'case B' represents a solution with a larger outer radius than 'case A' \\
$^e$: $r_e$ in Gaussian distribution for SiO is $2.5 \times 10^{16}$\,cm, distance is 250\,pc and \Mdot\,=\,$3 \times 10^{-5}$\,\Msun/yr \\
$^f$: $r_e$  in Gaussian distribution for SiS, distance is 260\,pc and \Mdot\,=\,$1 \times 10^{-5}$\,\Msun/yr. For 2-component model: f$_c$ is $5.5 \times 10^{-6}$ and taken constant out to $1.0 \times 10^{15}$\,cm and  the lower abundance Gaussian component has f$_0$ of $5.0 \times 10^{-9}$ and $r_e$ of $1.6 \times 10^{16}$\,cm. Using one (Gaussian) component distribution, f$_0$ is $5 \times 10^{-8}$ and $r_e$ is $1.6 \times 10^{16}$\,cm.\\
$^g$: only value at  5\,\Rstar\ is given\\
$^h$: predicted values at 2.2\,\Rstar\ in the envelope for IK~Tau\\
$^i$: predicted values at 2\,\Rstar\ in the envelope for TX~Cam\\
$^j$: predicted peak fractional abundances in the outer envelope\\
}}}
\end{table*}

\subsubsection{CO} \label{CO}
\begin{figure*}[htp]
  \includegraphics[height=\textwidth,angle=90]{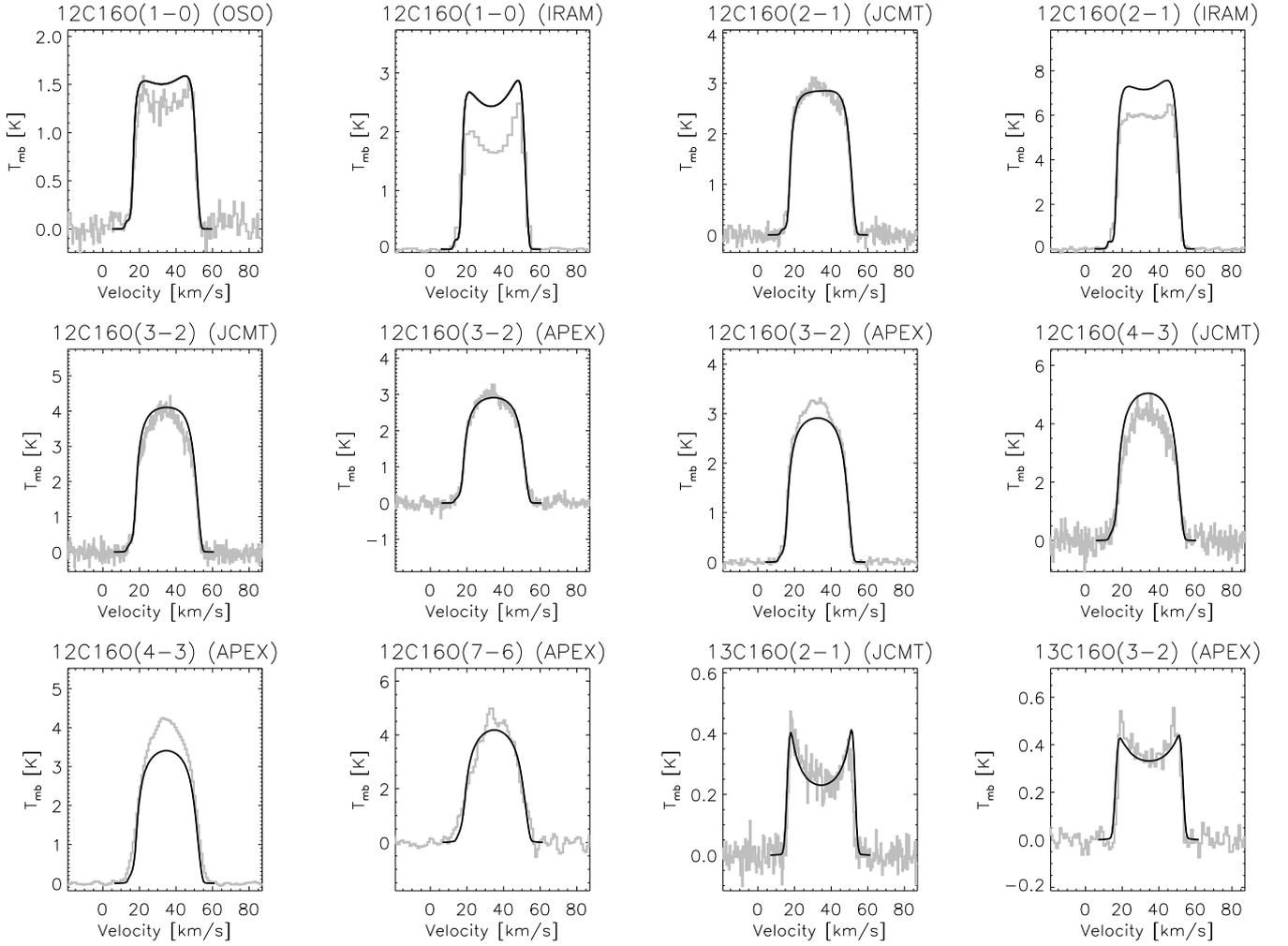}
\vspace*{2ex}
\caption{CO rotational line profiles of \object{IK Tau} (plotted in
  grey) are compared with the GASTRoNOoM non-LTE line predictions (in black) using the
  parameters of `model 3' as specified in Table~\ref{param_IKTau}. The rest frame of
  the velocity scale is the local standard of rest (LSR) velocity. }
  \label{CO_model} 
\end{figure*}

\paragraph{Results:} A comparison between the observed rotational $^{12}$CO and $^{13}$CO lines and theoretical predictions is shown in Fig.~\ref{CO_model}. The \twaalfco\ and \dertienco\ lines are very well reproduced by the GASTRoNOoM-predictions, both in integrated intensities and in line shapes. Only the IRAM \twaalfco(1--0) and \twaalfco(2--1) lines are slightly over-predicted. It is, however, not the first time that the non-compatibility of the IRAM absolute flux level is reported \citep[see, e.g.][]{Decin2008A&A...484..401D, Schoier2006A&A...454..247S}.

\paragraph{Comparison to theoretical predictions:} The CO fractional abundance assumed in all observational studies (see also Table~\ref{table:comparison}) is always lower than the deduced inner wind theoretical non-TE values of \citet{Duari1999AandA...341L..47D} and \citet{Cherchneff2006AandA...456.1001C}. The non-TE theoretical values are comparable to the TE-value of $6.95 \times 10^{-4}$ at 1\,\Rstar\ \citep{Duari1999AandA...341L..47D}. Increasing the CO fractional abundance by a factor 5, would decrease the mass-loss rate by a factor $\sim$2.8 to reproduce the observed CO rotational line profiles \citep[using the scaling laws deduced by][]{Debeck}.

\paragraph{Comparison to other observational studies:} Most observational studies listed in Table~\ref{table:comparison} assume a fractional CO abundance of [CO/H]\,=\,$1-1.5 \times 10^{-4}$. Using different CO rotational lines the mass-loss rate is then derived. The deduced $^{12}$CO/$^{13}$CO ratio ranges between 9 and 14.

Using CO rotational line transitions, other studies have also estimated the mass-loss rate (see Table~\ref{CO_others}). The results depend on the assumed or derived temperature distribution, the distance, the adopted [CO/H$_2$] abundance ratio, and the radiative transfer model or analytical approximation used. All (scaled) mass-loss rate values are in the narrow range between $6.5 \times 10^{-6}$ and $9 \times 10^{-6}$\,\Msun/yr, the exception being the result of \citet{GonzalesDelgado2003AandA...411..123G}, which is a factor $\sim$4 higher. We note that the work of these authors was not devoted to the study of CO, and it remains unclear what line intensities they used in their modeling.

\begin{table*}
 \caption{Mass-loss rate values derived from \twaalfco\ rotational line transitions for \object{IK~Tau}. The first and second columns list the distance and [CO/H$_2$] value used in the different studies, the third column gives the derived gas mass-loss rate, the fourth column contains the mass-loss rate scaled to our adopted values for the distance (D\,=\,265\,pc) and CO-abundance ([CO/H$_2$]\,=\,$2\,\times\,10^{-4}$) using the scaling laws deduced by \citet{Debeck}, the fifth column lists the rotational line transitions used, and last column gives the reference.}
\label{CO_others}
\begin{tabular}{llllcl}
\hline \hline
\multicolumn{1}{c}{\rule[0mm]{0mm}{5mm}D} & \multicolumn{1}{c}{[CO/H$_2$]} & \multicolumn{1}{c}{\Mdot} &\multicolumn{1}{c}{\Mdot}(scaled) & \multicolumn{1}{c}{Lines} & \multicolumn{1}{c}{Reference} \\
\rule[-3mm]{0mm}{3mm}[pc] & & [\Msun/yr] & [\Msun/yr] & &\\
\hline
\rule[0mm]{0mm}{5mm}350 &$2 \times 10^{-4}$ & $1.0 \times 10^{-5}$ & $6.7 \times 10^{-6}$ & \twaalfco(1--0) & \citet{Bujarrabal1989AandA...219..256B}\\
270 &$5 \times 10^{-4}$ & $4.0 \times 10^{-6}$ & $6.8 \times 10^{-6}$ & \twaalfco(1--0) & \citet{Sopka1989AandA...210...78S}\\
260 & $5 \times 10^{-4}$ & $4.4 \times 10^{-6}$ & $7.9 \times 10^{-6}$ & \twaalfco(1--0) & \citet{Loup1993AandAS...99..291L} \\
260 & $5 \times 10^{-4}$ & $3.8 \times 10^{-6}$ & $6.8 \times 10^{-6}$ & \twaalfco(1--0) & \citet{Neri1998AandAS..130....1N} \\
250 & $2 \times 10^{-4}$  & $3.0 \times 10^{-5}$ &$3.2 \times 10^{-5}$ &\twaalfco(1--0)$\rightarrow$(4--3) & \citet{GonzalesDelgado2003AandA...411..123G} \\
250 & $3 \times 10^{-4}$  & $4.7 \times 10^{-6}$ & $6.5 \times 10^{-6}$ & \twaalfco(1--0)$\rightarrow$(3--2) &\citet{Teyssier2006AandA...450..167T}\\
300 &  $2 \times 10^{-4}$  & $1.0 \times 10^{-5}$ &$7.7 \times 10^{-6}$ & \twaalfco(1--0)$\rightarrow$(4--3) & \citet{Ramstedt2008AandA...487..645R} \\ 
265 & $2 \times 10^{-4}$  & $9.0 \times 10^{-6}$ & $9.0 \times 10^{-6}$ &\twaalfco(1--0)$\rightarrow$(7--6) & this study\\
\hline
\end{tabular}
\end{table*}

\subsubsection{HCN} \label{HCN}

\begin{figure}
 \subfigure{\includegraphics[width=.24\textwidth,angle=0]{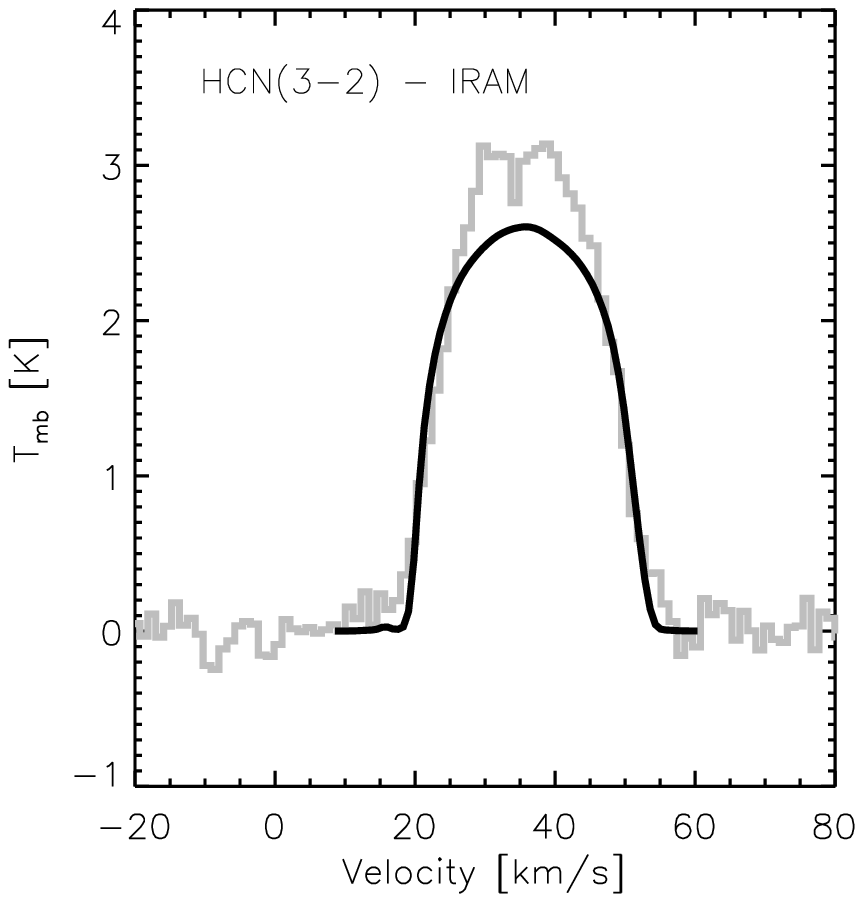}}
\subfigure{\includegraphics[width=.24\textwidth,angle=0]{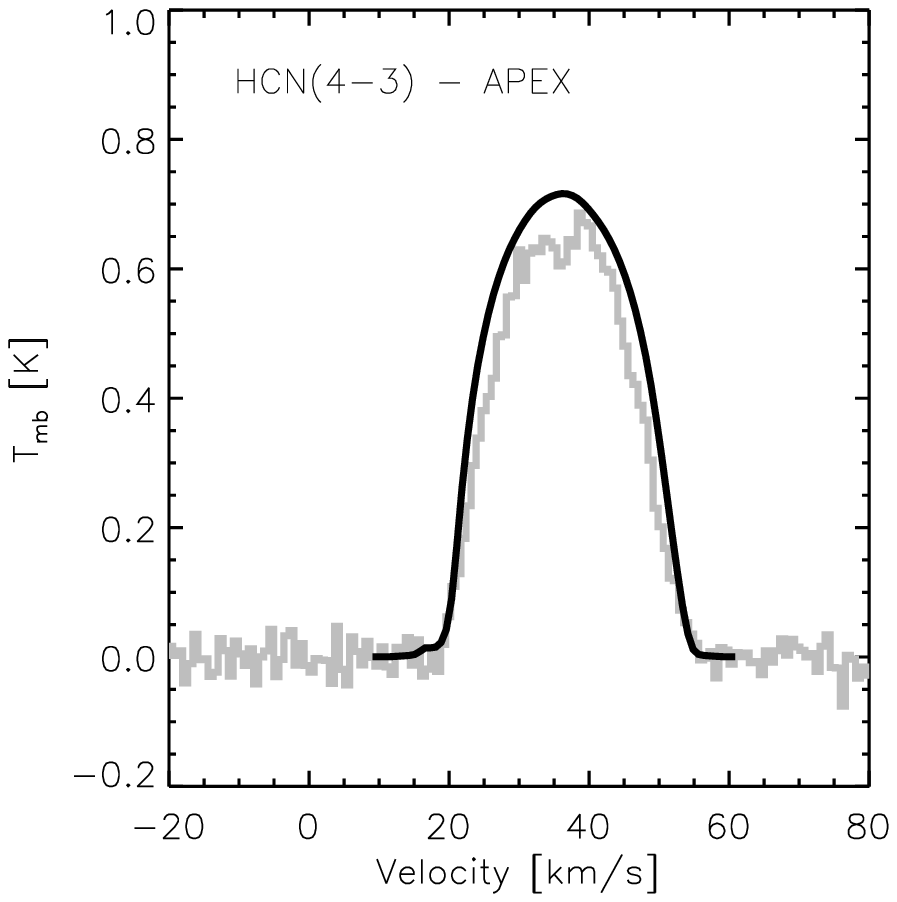}}
\caption{HCN observed spectral lines (gray) are compared to the
  spectral line predictions (black) based on the CSE model shown in
  Fig.~\ref{fig:structure_IKTau} and the abundance stratification
  displayed in Fig.~\ref{fig:abundances}.} 
\label{HCN_model}
\end{figure}

\paragraph{Results:} The narrow Gaussian line profiles clearly point toward (at least) an inner wind origin of HCN. 
As discussed above, we impose a value $R_{2}$\,=\,500\,\Rstar\ to simulate the  mapping results of \citet{Marvel2005AJ....130..261M}. 
The deduced abundance around 50\,\Rstar\ is [HCN/H]\,=\,$2 - 2.5 \times 10^{-7}$.

\paragraph{Comparison to theoretical predictions:} Theoretical predictions by \citet{Duari1999AandA...341L..47D} and \citet{Cherchneff2006AandA...456.1001C} and observational studies by, e.g., \citet{Bieging2000ApJ...543..897B} and \citet{Marvel2005AJ....130..261M}  indicate that HCN forms in the inner wind region of M-type envelopes by shock-induced non-equilibrium chemical processes. This is contrary to the photochemical models of \citet{Willacy1997AandA...324..237W}, where HCN is only produced in the outer envelope by photochemical reactions involving NH$_3$ and CH$_3$, which they assumed to be parent species originating close to the stellar photosphere and injected in the outer region. This formation route, however, leads to line shapes which are clearly non-Gaussian but are flat-topped (for optically thin unresolved emission) since the wind is already at its full velocity in this region. 

The maximum size distribution of 3.85\arcsec\ derived by \citet{Marvel2005AJ....130..261M} is suggested to be caused by photodissociation of HCN. \citet{Bieging2000ApJ...543..897B} used a parametrized formula to describe the photodissociation radius for HCN as a function of gas mass-loss rate and wind velocity (see their Eq.~2). This estimate leads to a HCN photodissociation radius of $1.2 \times 10^{16}$\,cm, in good agreement with the result of \citet{Marvel2005AJ....130..261M} of $7.6 \times 10^{15}$\,cm (at 265\,pc). Using HCN as a parent species  with an injected abundance of $1.5\times10^{-7}$, new chemical outer wind models were calculated using the code as described in \citet{Willacy1997AandA...324..237W}. The derived photodissocation radius is around $2 \times 10^{16}$\,cm, being a factor $\sim$3 higher than the observed value of \citet{Marvel2005AJ....130..261M}. 


Compared to the theoretical predictions for \object{TX~Cam} at 5\,\Rstar\ by \citet{Cherchneff2006AandA...456.1001C} or for \object{IK~Tau} at 2.2\,\Rstar\ by \citet{Duari1999AandA...341L..47D}, our deduced abundance is a factor 10 to 40 lower, respectively. There are a few possibilities for the origin of this difference.
\emph{(i.)} Contrary to what is thought \citep[e.g.][]{Duari1999AandA...341L..47D}, HCN may participate in the formation of dust grains in the inner envelope. \emph{(ii.)} The formation mechanism of HCN in the inner wind is directly linked to its radical CN by
\begin{equation}
 \mathrm{CN + H_2 \rightarrow HCN + H\,.}
\end{equation}
The destruction route is the reverse reaction. Shocks trigger CN and further HCN formation in the gas \citep{Cherchneff2006AandA...456.1001C}. The formation processes of both molecules depend critically on the physical parameters of the shocked gas, specifically on the physics in the `very fast chemistry zone', that is the narrow region after the shock front itself. The modeling of this zone is still subject to many uncertainties (e.g.\ cooling rate, velocity, shock strength), yielding uncertainties on the theoretical fractional abundances of at least one order of a magnitude.


\paragraph{Comparison to observational studies:} All observational deduced values agree within a factor $\sim$2, but are clearly lower than the inner wind non-TE theoretical predictions (see Table~\ref{table:comparison}).

\subsubsection{SiS} \label{SiS}

\begin{figure*}[htp]
\begin{center}
\subfigure{\includegraphics[width=.24\textwidth,angle=90]{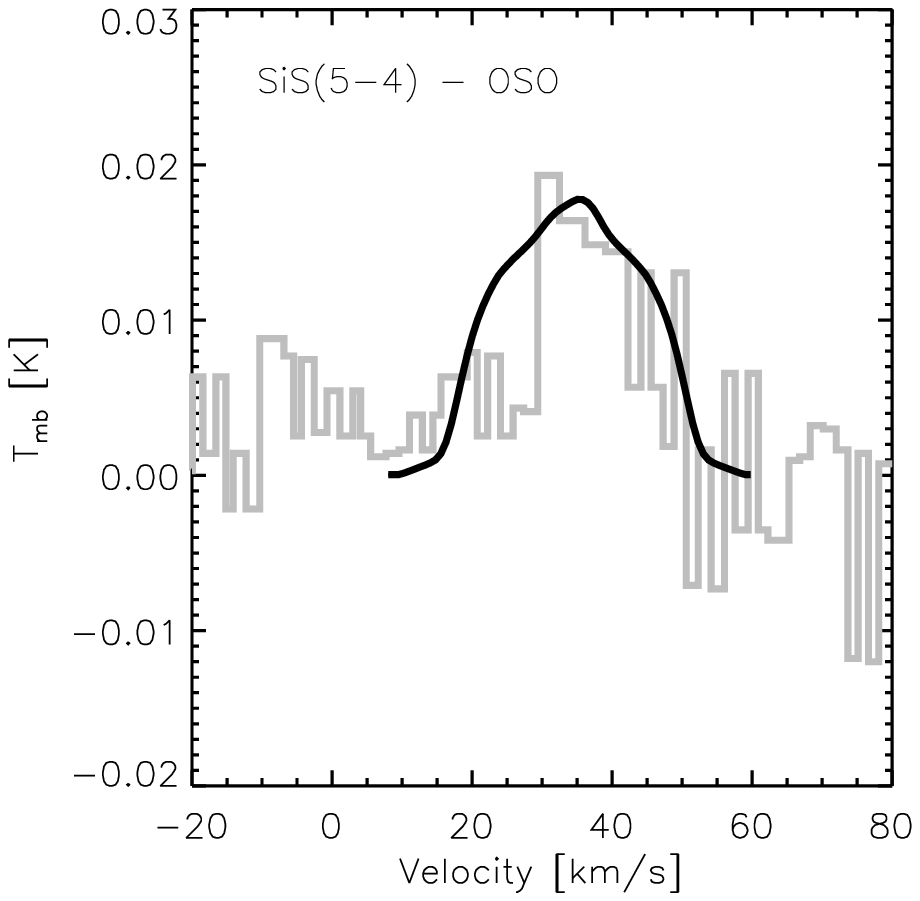}}
\subfigure{\includegraphics[width=.24\textwidth,angle=90]{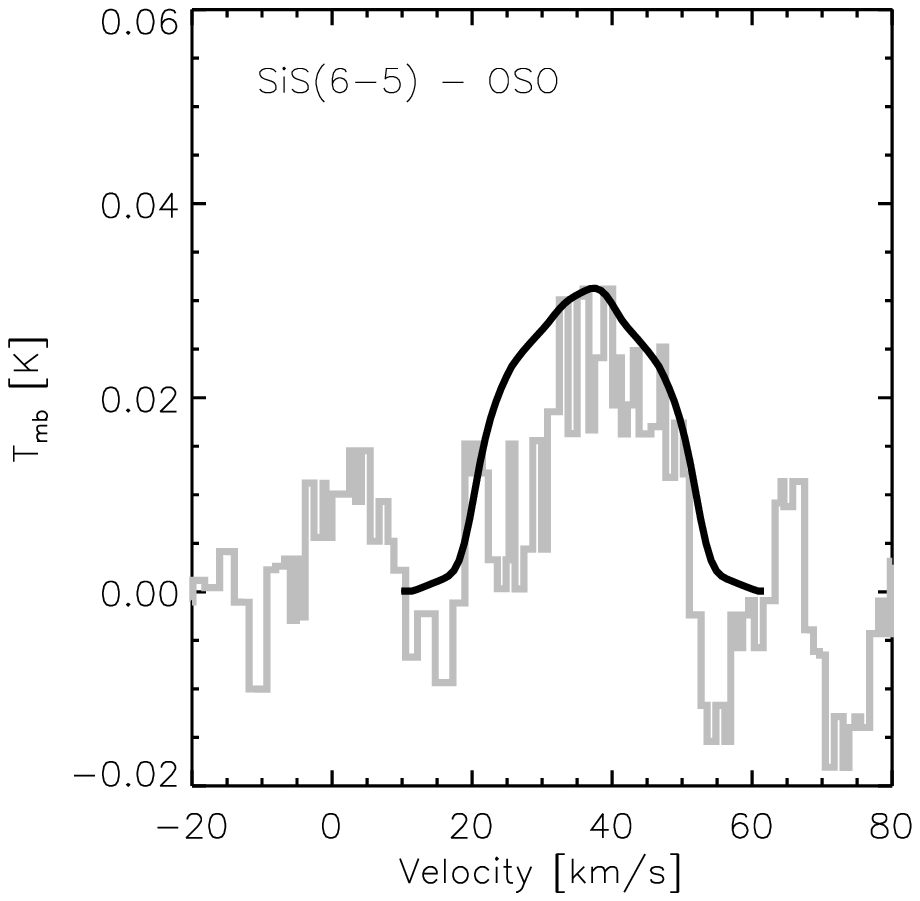}}
\subfigure{\includegraphics[width=.24\textwidth,angle=90]{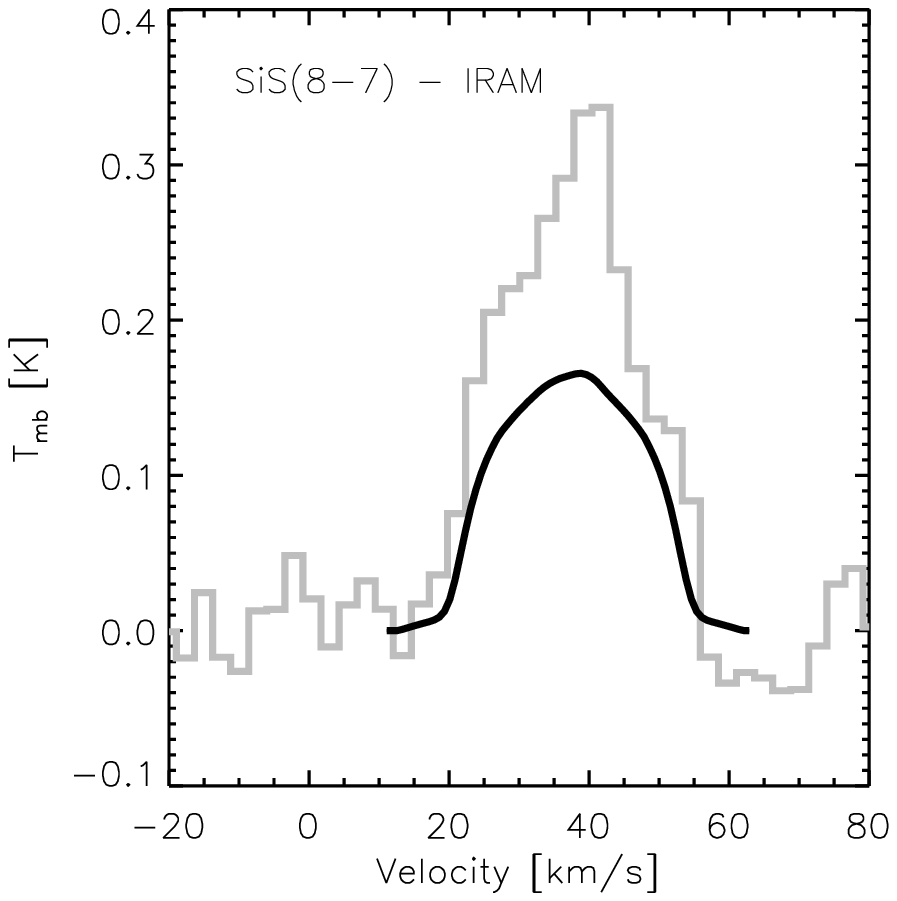}}
\subfigure{\includegraphics[width=.24\textwidth,angle=90]{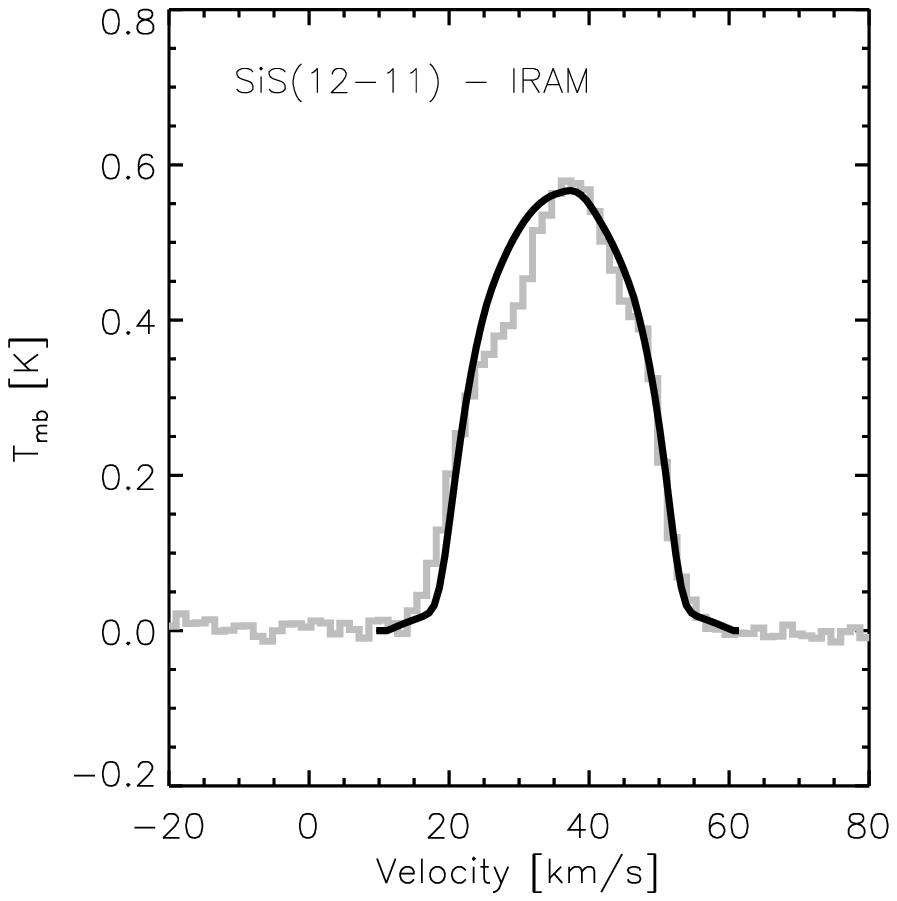}}
\subfigure{\includegraphics[width=.24\textwidth,angle=90]{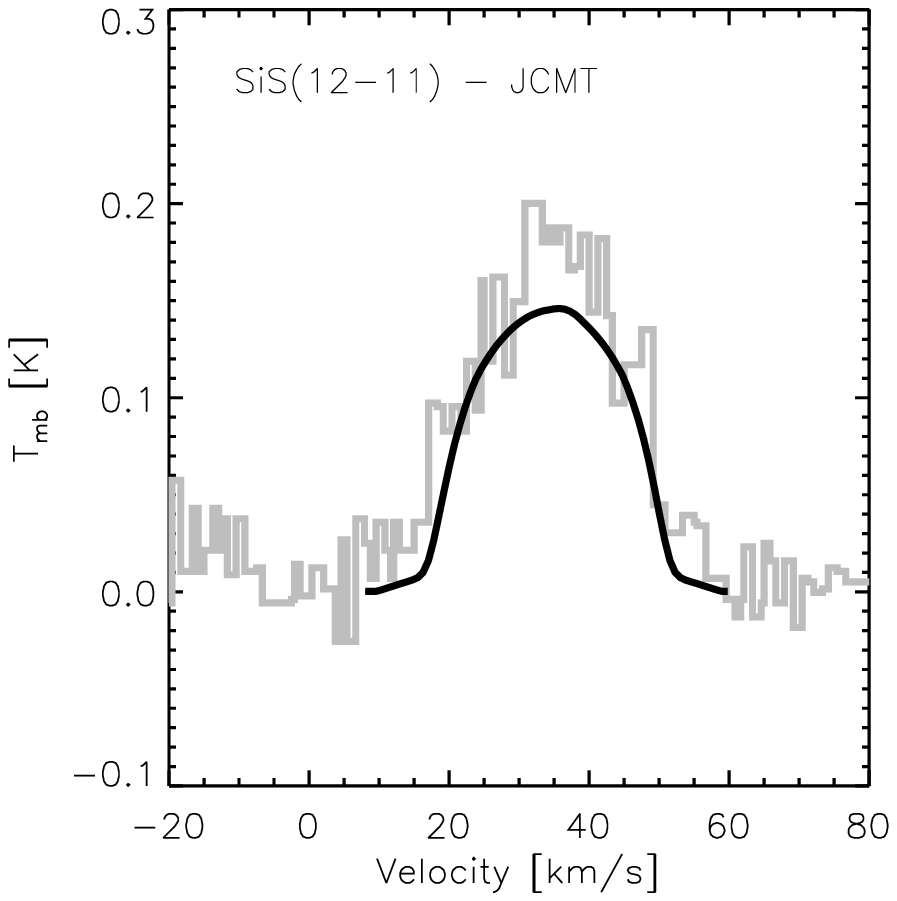}}
\subfigure{\includegraphics[width=.24\textwidth,angle=90]{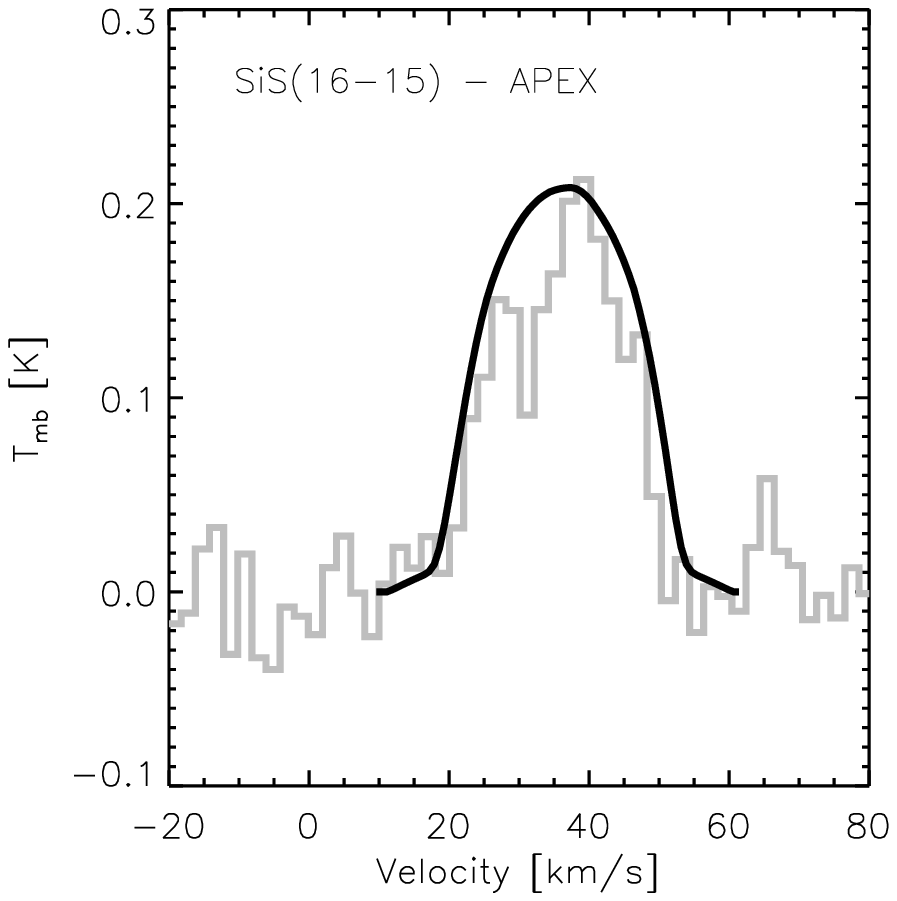}}
\subfigure{\includegraphics[width=.24\textwidth,angle=90]{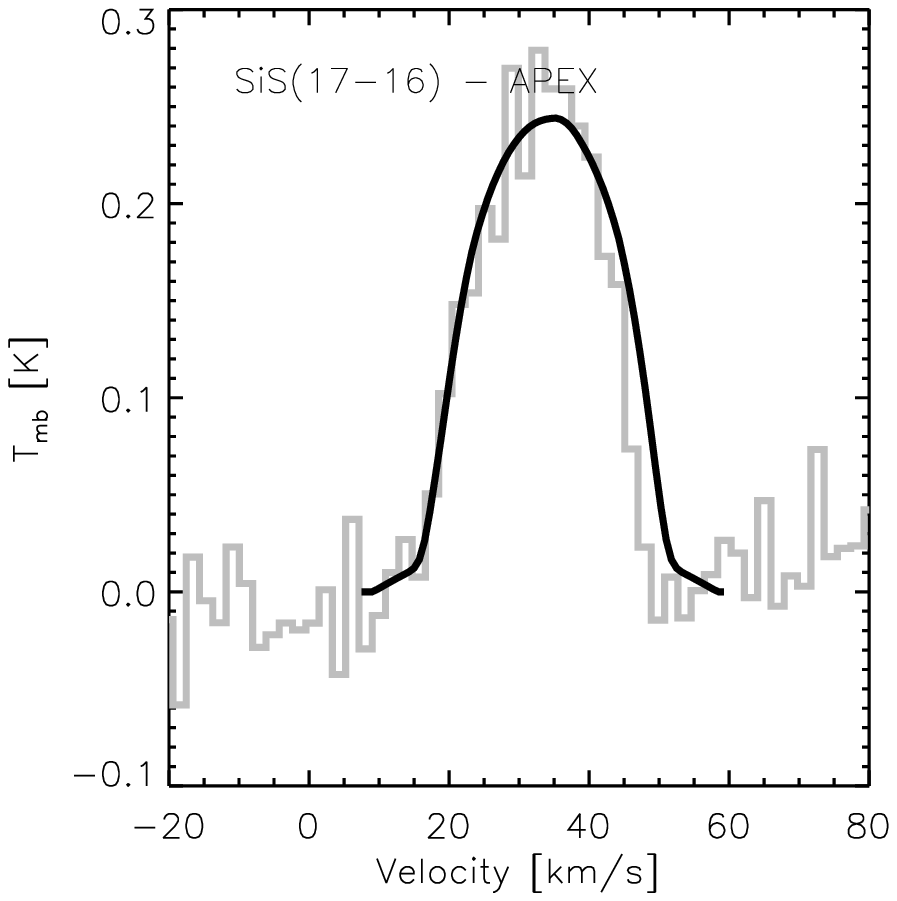}}
\subfigure{\includegraphics[width=.24\textwidth,angle=90]{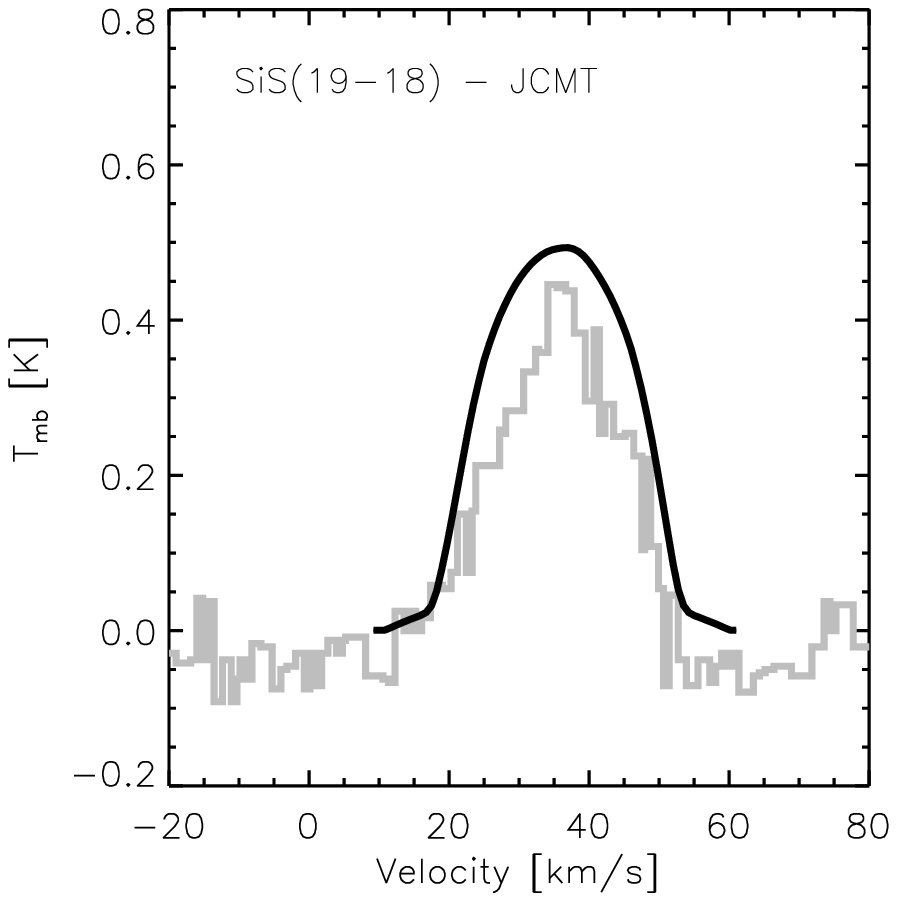}}
\subfigure{\includegraphics[width=.24\textwidth,angle=90]{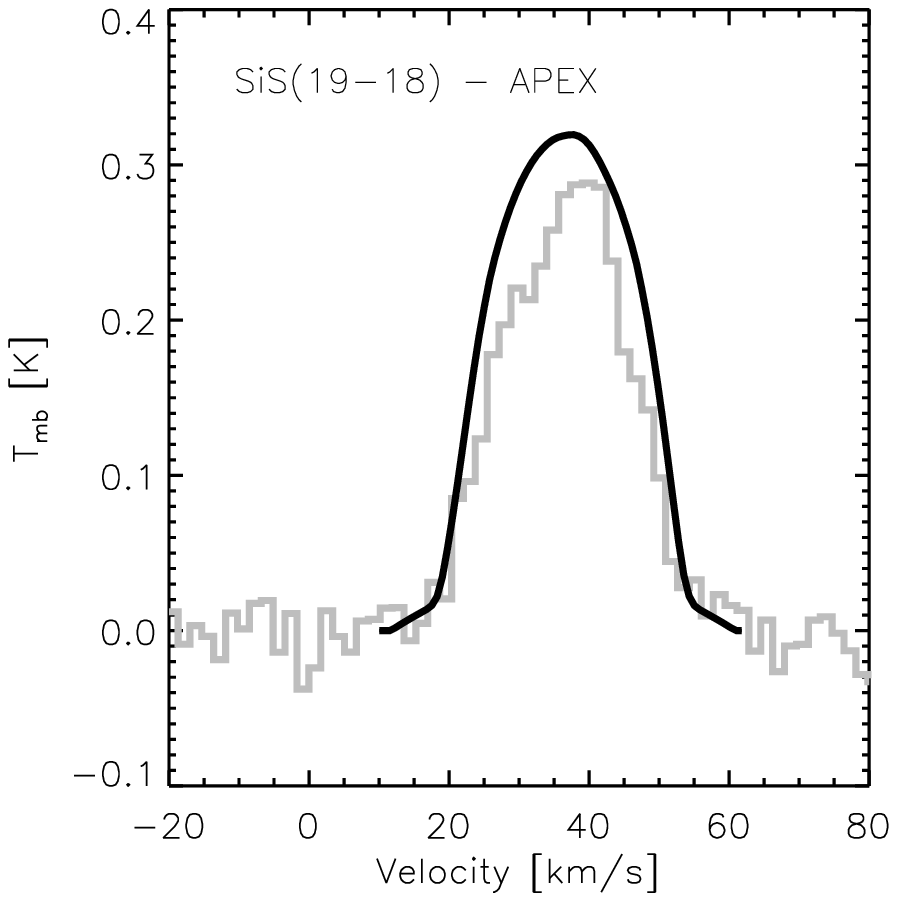}}
\subfigure{\includegraphics[width=.24\textwidth,angle=0]{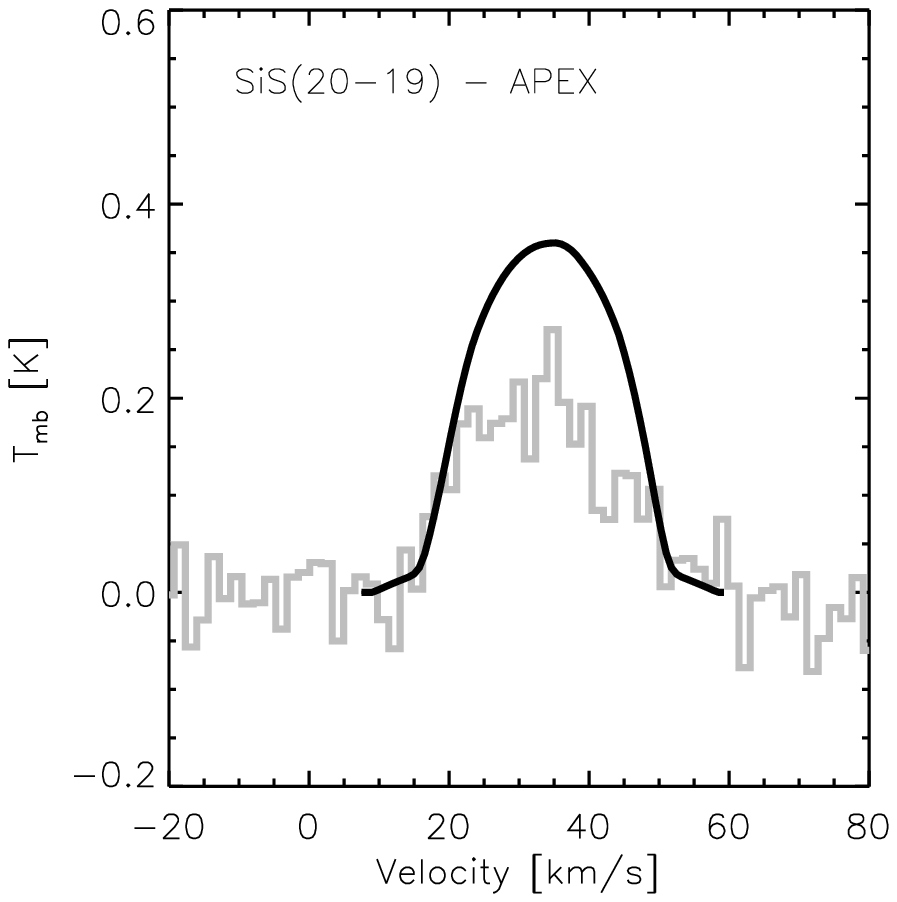}}
\caption{SiS observed spectral lines (gray) are compared to the
  spectral line predictions based on the CSE model shown in
  Fig.~\ref{fig:structure_IKTau} and the abundance stratification
  displayed in Fig.~\ref{fig:abundances}.} 
\label{SiS_model}
\end{center}
\end{figure*}

\paragraph{Results:} Next to CO, SiS is the molecule with the most molecular line transitions at our disposal. From the SiS(5--4) up to SiS(20--19), excitation temperatures from 13 to 183\,K are covered, and one can trace the envelope abundance spatial variations between 40 and 6000\,\Rstar. The lower lying transitions (SiS(5--4) and (6--4)) probe material at larger radial distances from the star; the SiS(19--18) and (20--19) line intensities are sensitive to the abundance in the inner wind region.
Modeling the strength of the low and high-excitation lines gives evidence of a depletion of SiS: the abundance at 40\,\Rstar\ is estimated to be $\sim5.5\times10^{-6}$ and decreases to $\sim4\times10^{-9}$ at 120\,\Rstar. This is also evidenced by the extra sub-structure in the high signal-to-noise high-resolution line profiles of the higher excitation lines observed with JCMT and IRAM, where the small peak traces a high abundance component at low velocity close to the star and the broader plateau the lower abundance component further away in the envelope.
SiS can be depleted due to the adsorption of SiS molecules onto dust grains, before being photodissociated much farther away. Using ISI 11\,$\mu$m interferometric data. \citet{Hale1997ApJ...490..407H} detected dust excess, with a first intensity peak around 0.1\arcsec, going out up to at least 0.7\arcsec (or $2.8 \times 10^{15}$\,cm\,$\approx$\,180\,\Rstar) around \object{IK~Tau}. SiS condensation onto dust species has been postulated already \citep[e.g.][]{Bieging1993AJ....105..576B}, but not all atomic Si or molecular SiS should condense onto dust species \citep[e.g.][]{Dominik1993A&A...277..578D}.  The different rotational lines give us the potential to constrain the compact, pre-condensation, SiS fractional abundance component quite well, but the uncertainty on the extended post-condensation component is higher.

\paragraph{Comparison to theoretical predictions:} The abundance at 40\,\Rstar\ is higher than the theoretical value predicted in the inner wind envelope by \citet{Duari1999AandA...341L..47D} and \citet{Cherchneff2006AandA...456.1001C}. This may either indicate that the destruction reaction which occurs at larger radii in the inner O-rich envelope via \citep{Cherchneff2006AandA...456.1001C}
\begin{equation}
 {\rm S + SiS \rightarrow S_2 + Si}
\end{equation}
is not as efficient, or that an extra formation route is not yet taken into account in the theoretical modeling. In general, the uncertainties on chemical reaction rates involving sulfur are still very high (I.\ Cherchneff, \emph{priv.\ comm.}).
The high abundance around 40\,\Rstar\ is in accordance with the observational suggestion by \citet{Decin2008A&A...480..431D} that SiS forms close to the star, whatever the C/O ratio of the target. 

 \paragraph{Comparison to observational studies:} Similar results concerning the depletion of SiS in the intermediate wind region are obtained by \citet{Schoier2007AandA...473..871S}. From the simplified analyses assuming optically thin emission thermalized at one excitation temperature (see Table~\ref{table:comparison}) it is not possible to derive this kind of abundance depletion pattern. 

\subsubsection{SiO} \label{SiO}
\begin{figure}[htp]
\begin{center}
\subfigure{\includegraphics[width=.24\textwidth,angle=90]{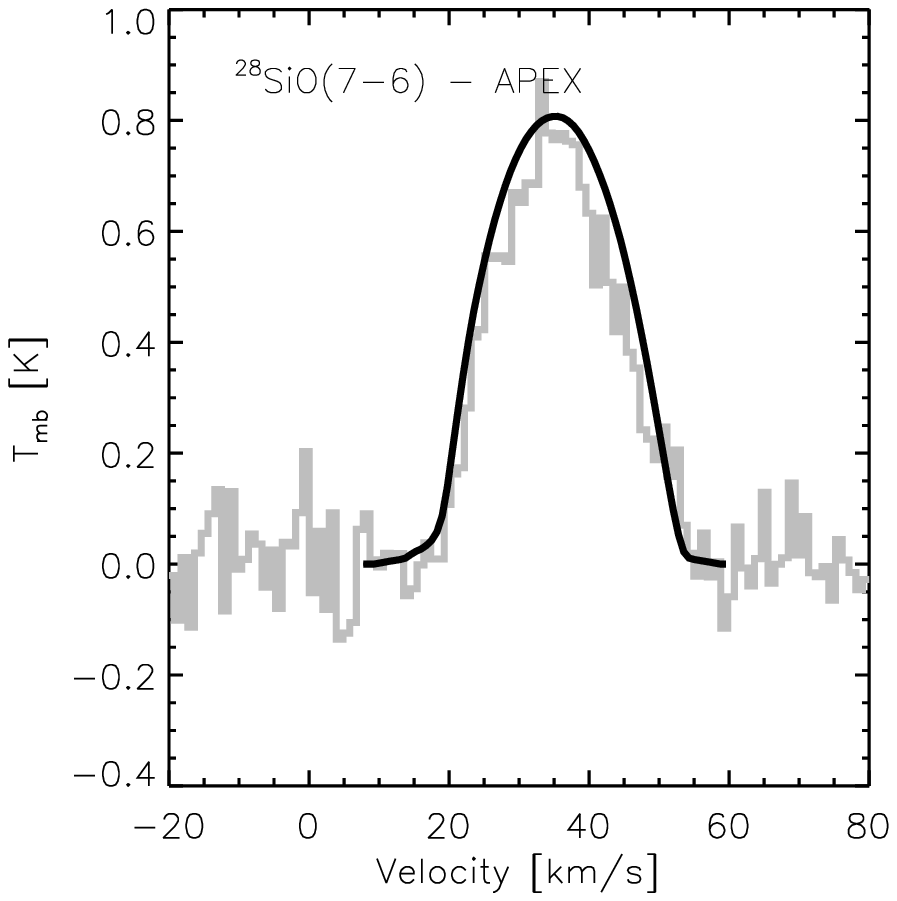}}
\subfigure{\includegraphics[width=.24\textwidth,angle=0]{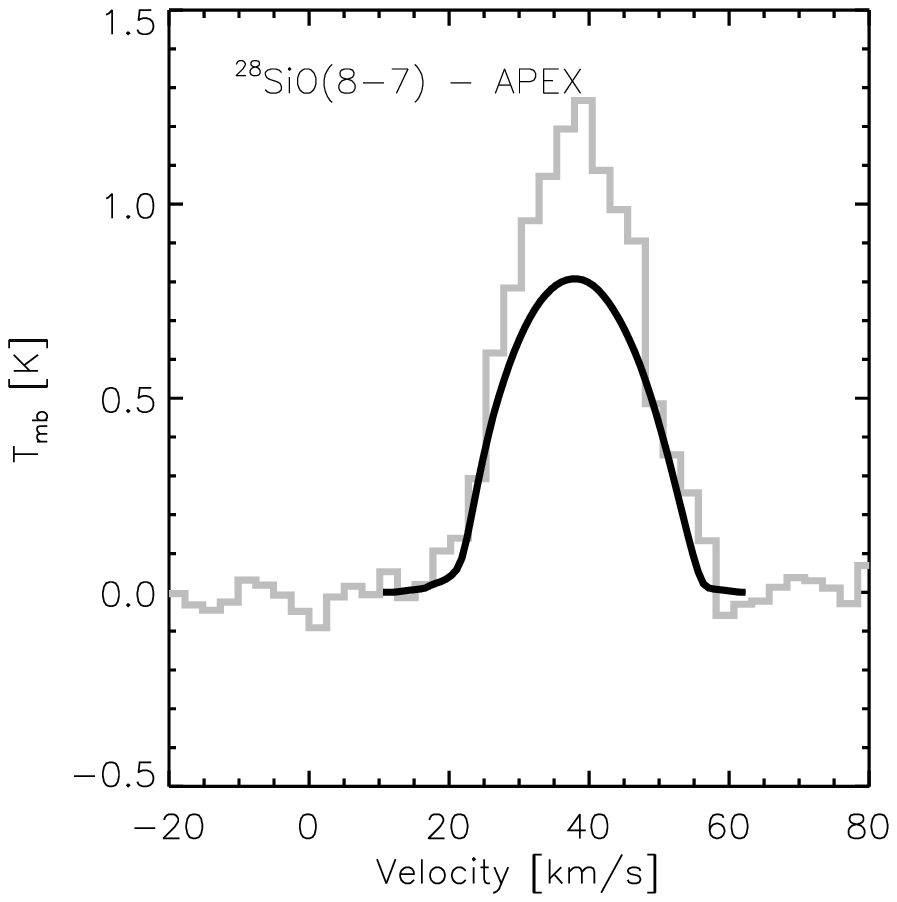}}
\caption{$^{28}$SiO observed spectral lines (gray) are compared to the
  spectral line predictions based on the CSE model shown in
  Fig.~\ref{fig:structure_IKTau} and the abundance stratification
  displayed in Fig.~\ref{fig:abundances}.}
\label{28SiO_model}
\end{center}
\end{figure}

\begin{figure}
\begin{center}
\includegraphics[height=.45\textwidth,angle=270]{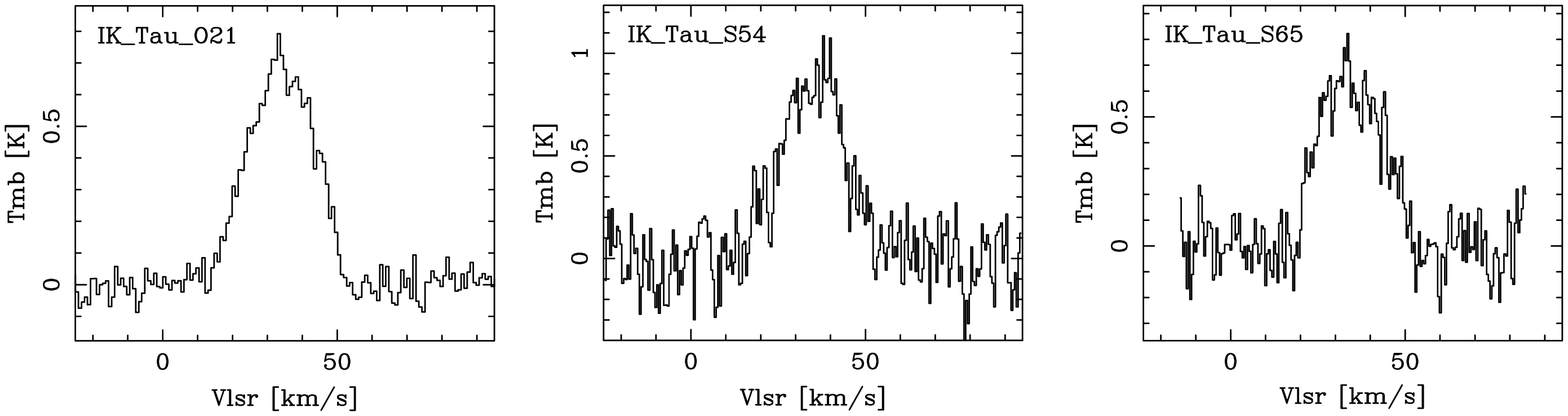}
\includegraphics[width=.45\textwidth,angle=0]{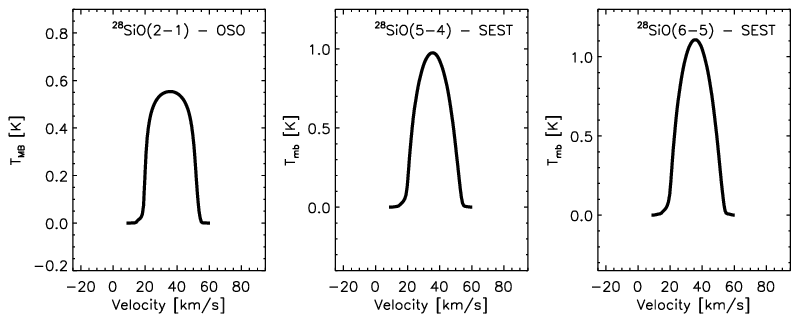}
\caption{\emph{Upper panel:} $^{28}$SiO spectra of \object{IK Tau} from
  \citet{GonzalesDelgado2003AandA...411..123G}.  \emph{Lower panel:} $^{28}$SiO spectral line predictions based
  on the CSE model shown in Fig.~\ref{fig:structure_IKTau} and the
  abundance stratification displayed in Fig.~\ref{fig:abundances}.}
\label{fig:SiO_Gonzales}
\end{center}
\end{figure}

\begin{figure}[htp]
\begin{center}
\subfigure{\includegraphics[width=.24\textwidth,angle=0]{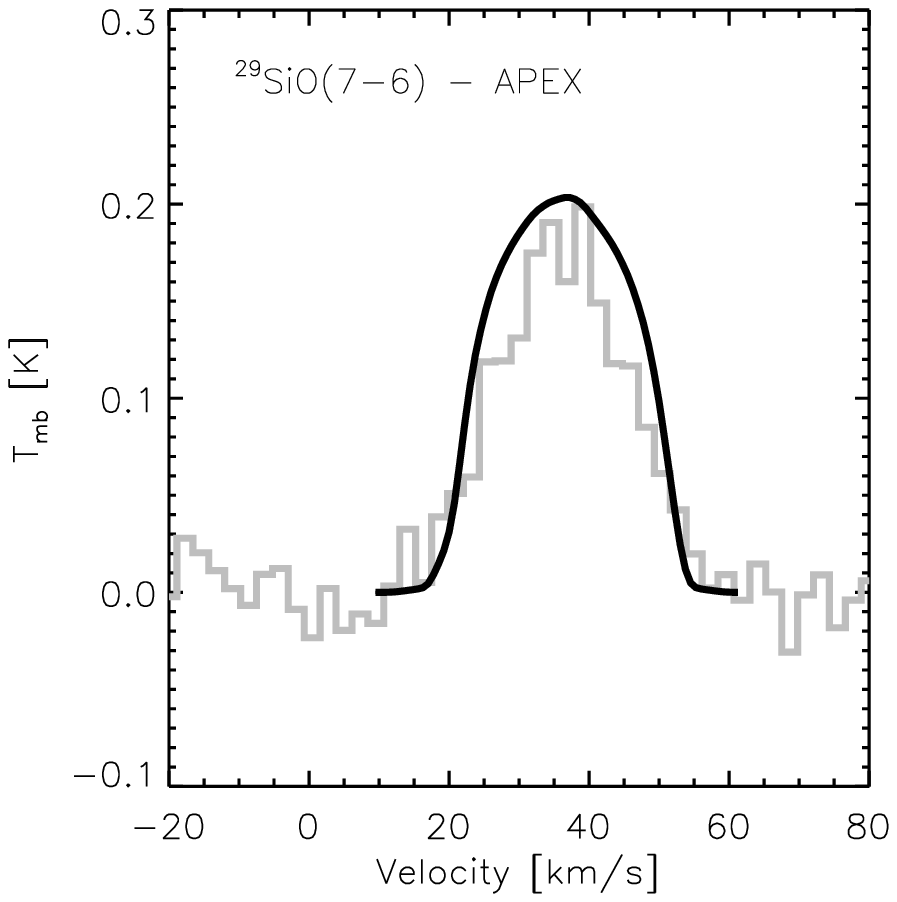}}
\subfigure{\includegraphics[width=.24\textwidth,angle=0]{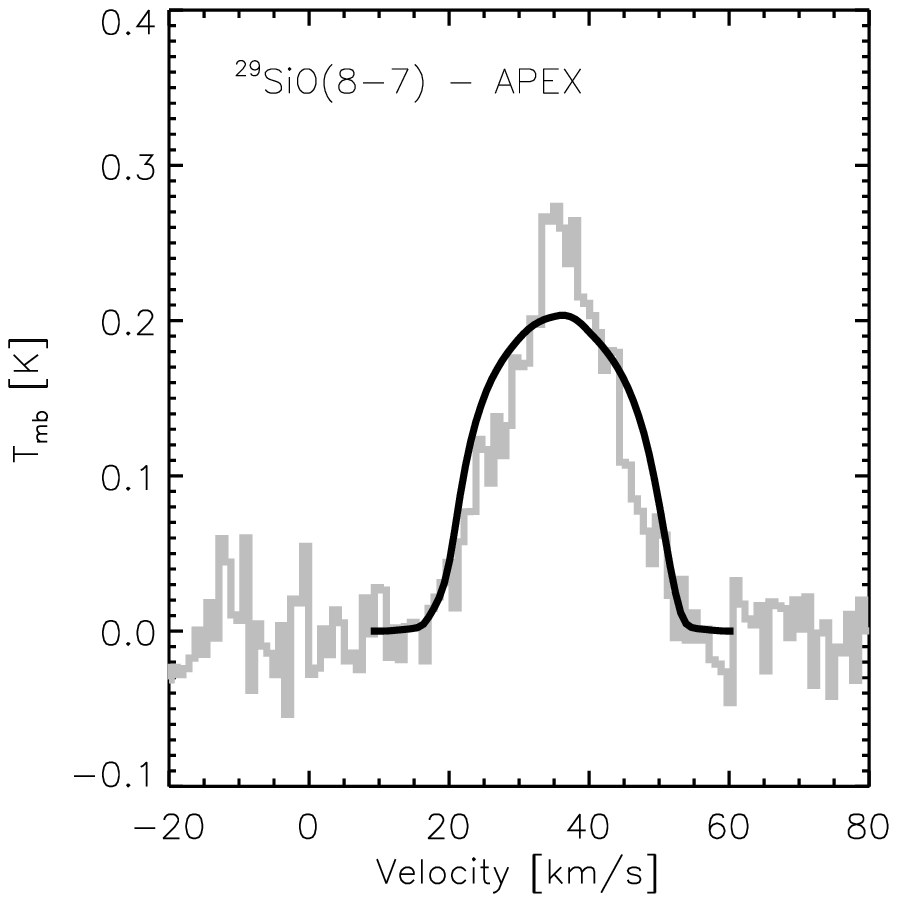}}
\caption{$^{29}$SiO observed spectral lines (gray) are compared to the
  spectral line predictions based on the CSE model shown in
  Fig.~\ref{fig:structure_IKTau} and the abundance stratification
  displayed in Fig.~\ref{fig:abundances}.}
\label{29SiO_model}
\end{center}
\end{figure}

\begin{figure}[htp]
\begin{center}
\subfigure{\includegraphics[width=.24\textwidth,angle=0]{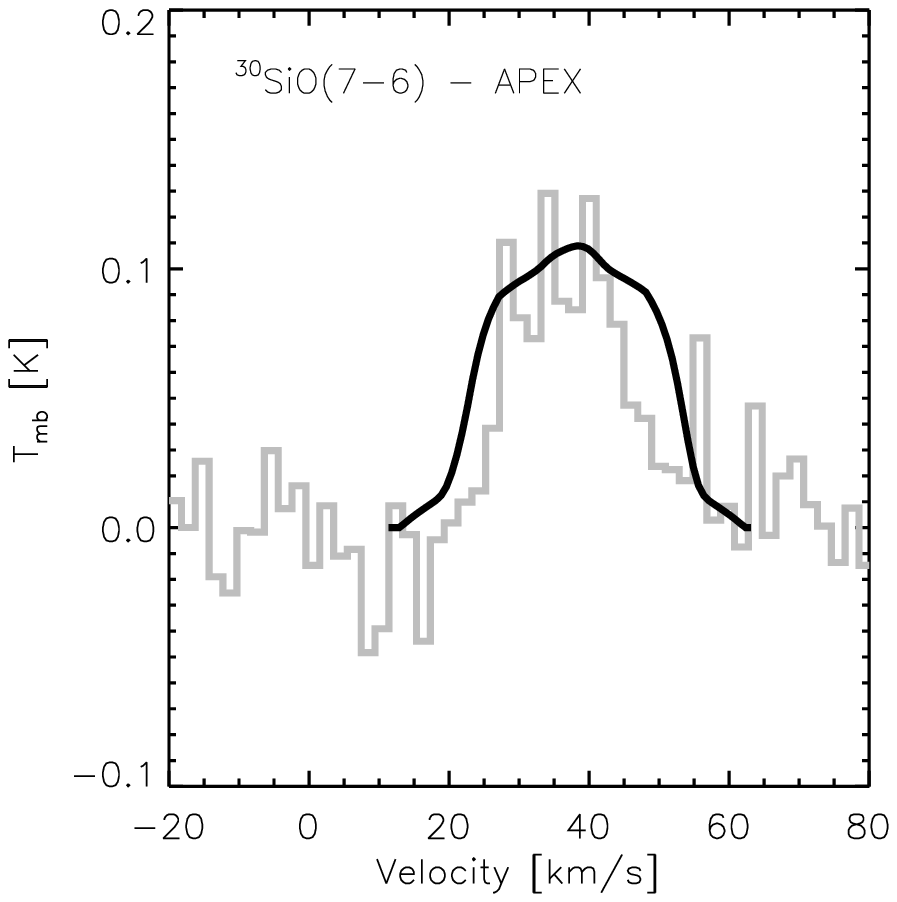}}
\subfigure{\includegraphics[width=.24\textwidth,angle=0]{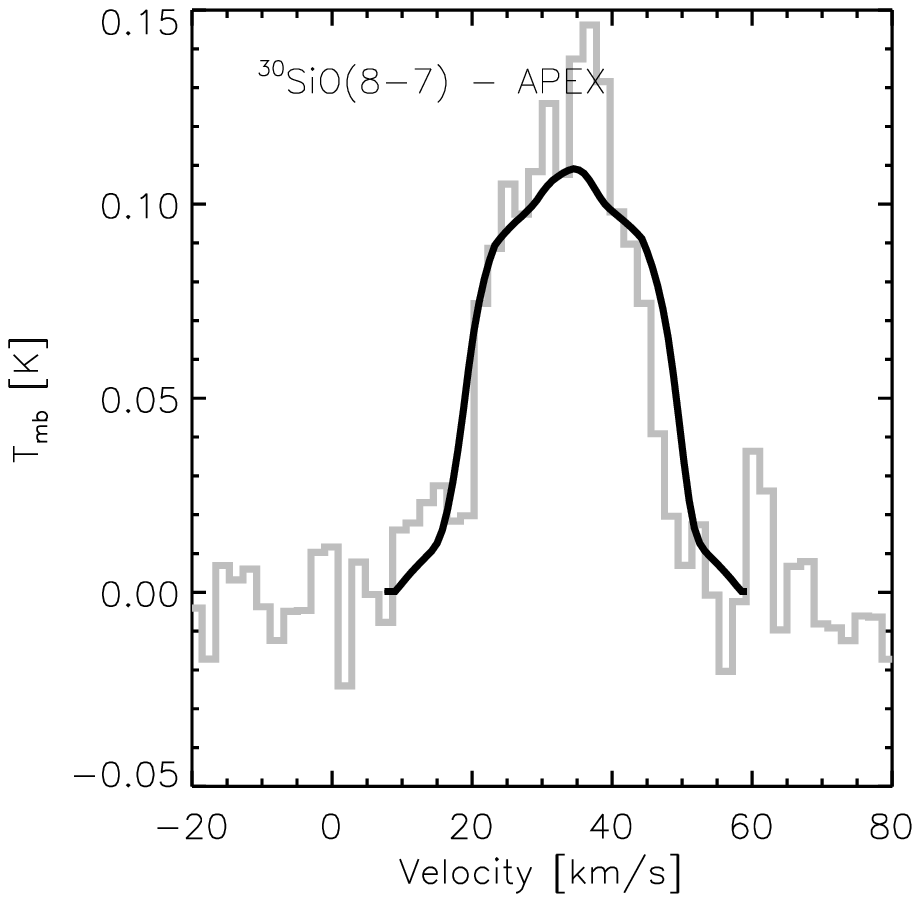}}
\caption{$^{30}$SiO observed spectral lines (gray) are compared to the
  spectral line predictions based on the CSE model shown in
  Fig.~\ref{fig:structure_IKTau} and the abundance stratification
  displayed in Fig.~\ref{fig:abundances}.}
\label{30SiO_model}
\end{center}
\end{figure}

\paragraph{Results:} For the $^{28}$SiO isotopolog, 5 transitions were observed, with excitation energies ranging between 6 and 75\,K. The line strengths and profile shapes of all $^{28}$SiO, $^{29}$SiO, and $^{30}$SiO lines are well predicted, except for the $^{28}$SiO(6--5) line as observed with the SEST by \citet{GonzalesDelgado2003AandA...411..123G} (see Fig.~\ref{fig:SiO_Gonzales}). Since the strength of both the $^{28}$SiO(5--4) and $^{28}$SiO(7--6) are well reproduced, and both of these lines share the line formation region with the $^{28}$SiO(6--5) transition, an absolute calibration uncertainty can be the cause of this discrepancy, but time variability of the lines should also be considered (see Sect.~\ref{timevar}).

Although not as pronounced as for SiS, the modeling of the different rotational transitions indicates an abundance decrease with a factor $\sim$40 around 180\,\Rstar. SiO is a parent molecule and
a strong candidate to be depleted in the wind of O-rich envelopes: at larger radii in the inner envelope, OH alters the SiO abundance via \citep{Cherchneff2006AandA...456.1001C}
\begin{equation}
 {\rm SiO + OH \rightarrow SiO_2 + H}\,.
\end{equation}
SiO$_2$ may condense as silica. It may also participate in the formation of amorphous or crystalline silicates. 

The derived isotopic ratios in the envelope are [$^{28}$SiO/$^{29}$SiO]\,$\sim$\,27 and  [$^{28}$SiO/$^{30}$SiO]\,=\,80, [$^{29}$SiO/$^{30}$SiO]\,=\,3. They are discussed in Sect.~\ref{isotopes}.

\paragraph{Comparison to theoretical predictions:} The SiO abundance at the inner dust condensation radius is slightly below the theoretical predictions of \citet{Duari1999AandA...341L..47D} and \citet{Cherchneff2006AandA...456.1001C}, which is very reasonable taking the assumptions of both the theoretical chemical kinetic calculations and our modeling into account. It possibly points toward the condensation of SiO onto dust grains in the intermediate wind zone, before 70\,\Rstar, a region where our observed lines are not very sensitive to the exact abundance distribution.  The observational study by \citet{Decin2008A&A...480..431D} indicates that SiO is formed close to the star, in support of the theoretical predictions by \citet{Duari1999AandA...341L..47D} and \citet{Cherchneff2006AandA...456.1001C}. In the outer envelope, SiO is very stable and only photodissociated around a few thousand stellar radii \citep{Willacy1997AandA...324..237W}.

\paragraph{Comparison to observational studies:}
\citet{Lucas1992A&A...262..491L} mapped the $^{28}$SiO(2-1)\,$v=0$ flux distribution, showing that it has a circular geometry. The half-peak intensity radius has a diameter of 2.2 +/- 0.1\arcsec\ (or a radius of $4.35 \times 10^{15}$\,cm at 265\,pc, in our model being 190\,\Rstar). The SiO(2-1) emission regions indicate that the final expansion velocity is not yet reached in the SiO emission region, suggesting that grain formation must still take place as far as $10^{16}$\,cm from the star. This supports our results on the velocity structure based on the study of the HCN line profiles (Sect.~\ref{velocity_structure}).

Compared to other observational studies, the deduced abundance around 70\,\Rstar\ is quite high, while the outer wind abundance agrees with the result by \citet{GonzalesDelgado2003AandA...411..123G}.

\subsubsection{CS} \label{CS}

\begin{figure}[htp]
\begin{center}
\subfigure{\includegraphics[width=.24\textwidth,angle=0]{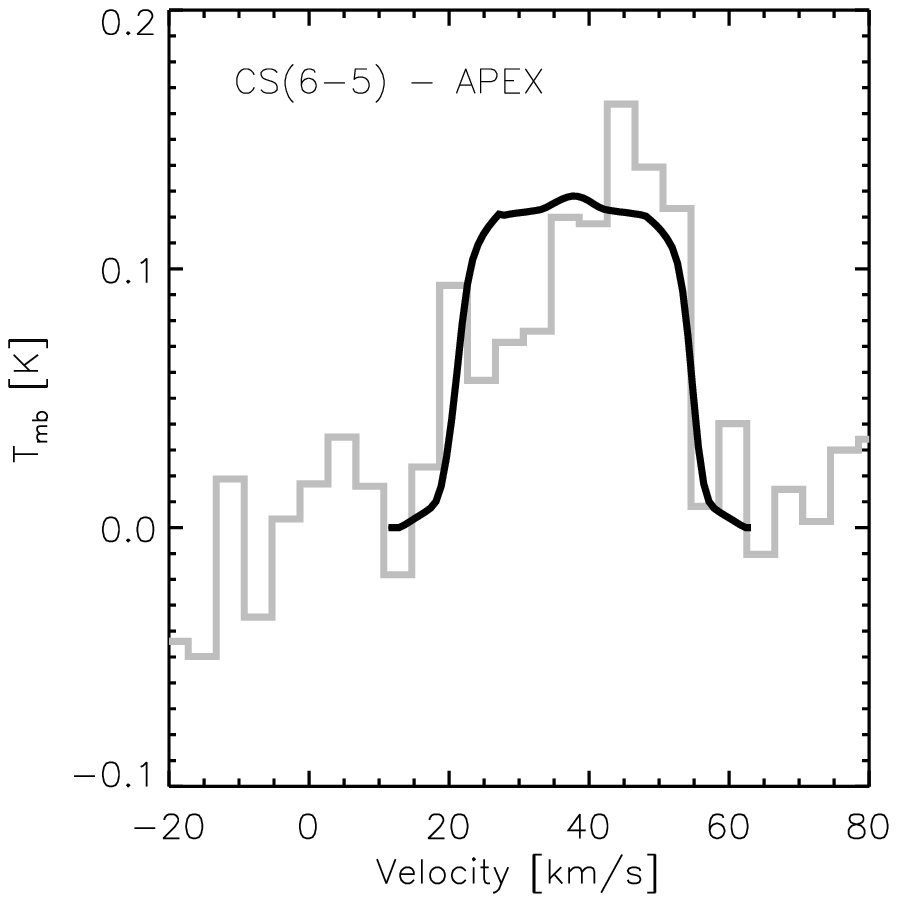}}
\subfigure{\includegraphics[width=.24\textwidth,angle=0]{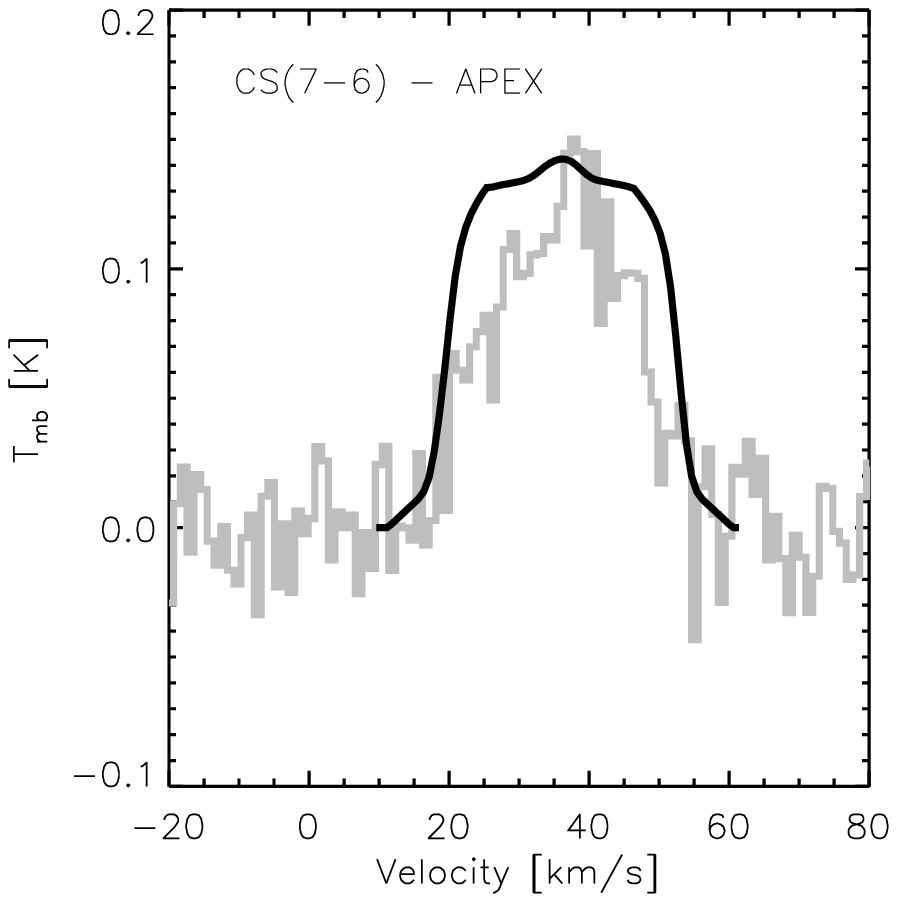}}
\caption{CS observed spectral lines (gray) are compared to the
  spectral line predictions based on the CSE model shown in
  Fig.~\ref{fig:structure_IKTau} and the abundance stratification
  displayed in Fig.~\ref{fig:abundances}.}
\label{CS_model}
\end{center}
\end{figure}

\paragraph{Results:} For CS, we only have two lines at our disposal, the  (6-5) and (7-6) rotational transitions, both with a low S/N-ratio. The fractional abundance at 300\,\Rstar\ is estimated to be $\sim 4\times10^{-8}$.

\paragraph{Comparison to theoretical predictions:} The derived abundance of $\sim4\times10^{-8}$ is much higher than the TE-abundance of $2.5\times10^{-11}$ for an oxygen-rich star. \citet{Cherchneff2006AandA...456.1001C} predicts CS to be a parent molecule, with a non-TE abundance for \object{TX~Cam} around $1.8\times10^{-5}$ at 2\,\Rstar\ and around $2\times10^{-6}$ at 5\,\Rstar\ away in the envelope, while \citet{Duari1999AandA...341L..47D} predicts a value of $2.75\times10^{-7}$ at 2.2\,\Rstar. In their theoretical modeling of the carbon-rich AGB-star \object{IRC+10216} \citet{Millar2001MNRAS.327.1173M} also argue for the need of CS as a parent species to account for the vibrationally excited CS lines detected in \object{IRC+10216}. 

As in the case of HCN (Sect.~\ref{HCN}), our derived abundances are a factor $\sim$50 lower compared to the predictions of \citet{Cherchneff2006AandA...456.1001C} and a factor $\sim$8 lower compared to  \citet{Duari1999AandA...341L..47D}.  The dominant formation pathways of both CS and HCN occur in the fast chemistry zone of the gas parcel excursion involving CN. Knowing that this zone is very difficult to model (see Sect.~\ref{HCN}) and that the sulfur reaction rates are not well known (see Sect.~\ref{SiS}), this difference is not so cumbersome. However, as suggested for HCN, it may also be the case that CS or the radical CN are involved in dust formation, altering its abundance in the intermediate wind region.
The low-resolution of the two observed CS lines do not provide enough information to firmly prove this.
In the outer envelope, CS is first formed from H$_2$S. Somewhat farther away, the reaction of atomic carbon with SO and HS forms CS, before it is photodissociated by UV radiation \citep{Willacy1997AandA...324..237W}.


\paragraph{Comparison to observational studies:}  While our deduced CS fractional abundance agrees with the values derived by \citet{Bujrrabal1994AandA...285..247B} and \citet{Kim2009}, the observational result by \citet{Lindqvist1988AandA...205L..15L} is higher by a factor $\sim$4.

\subsubsection{CN} \label{CN}

\begin{figure}[htp]
\begin{center}
\subfigure{\includegraphics[width=.24\textwidth,angle=0]{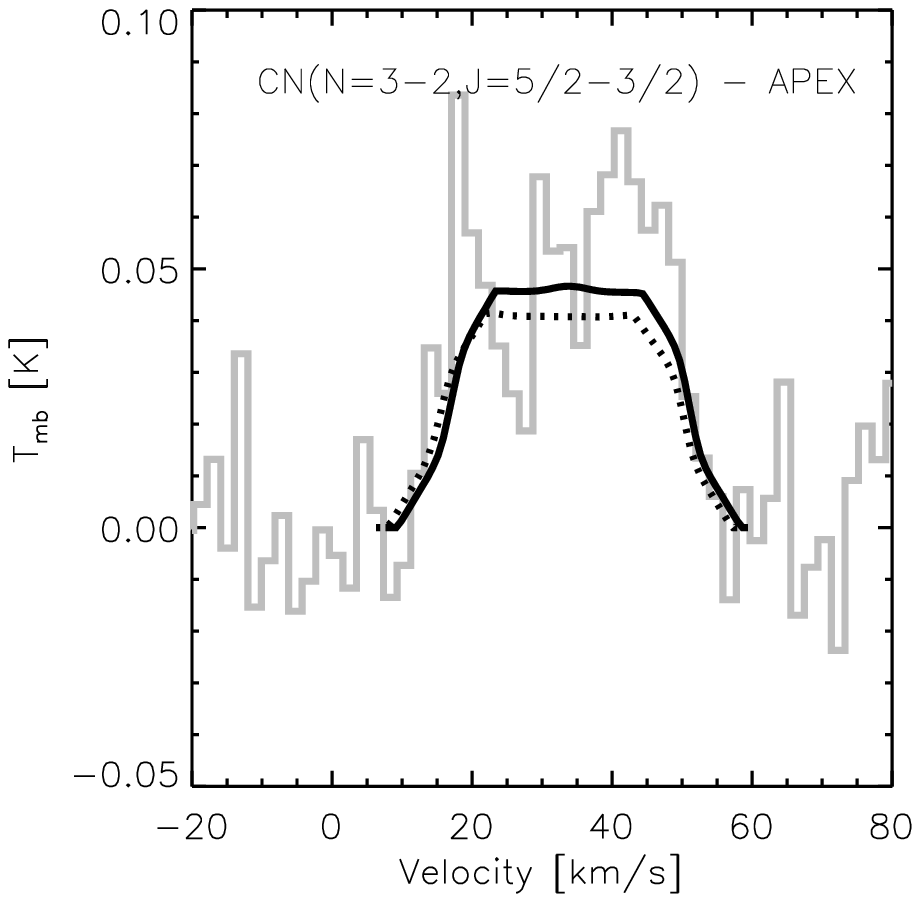}}
\subfigure{\includegraphics[width=.24\textwidth,angle=0]{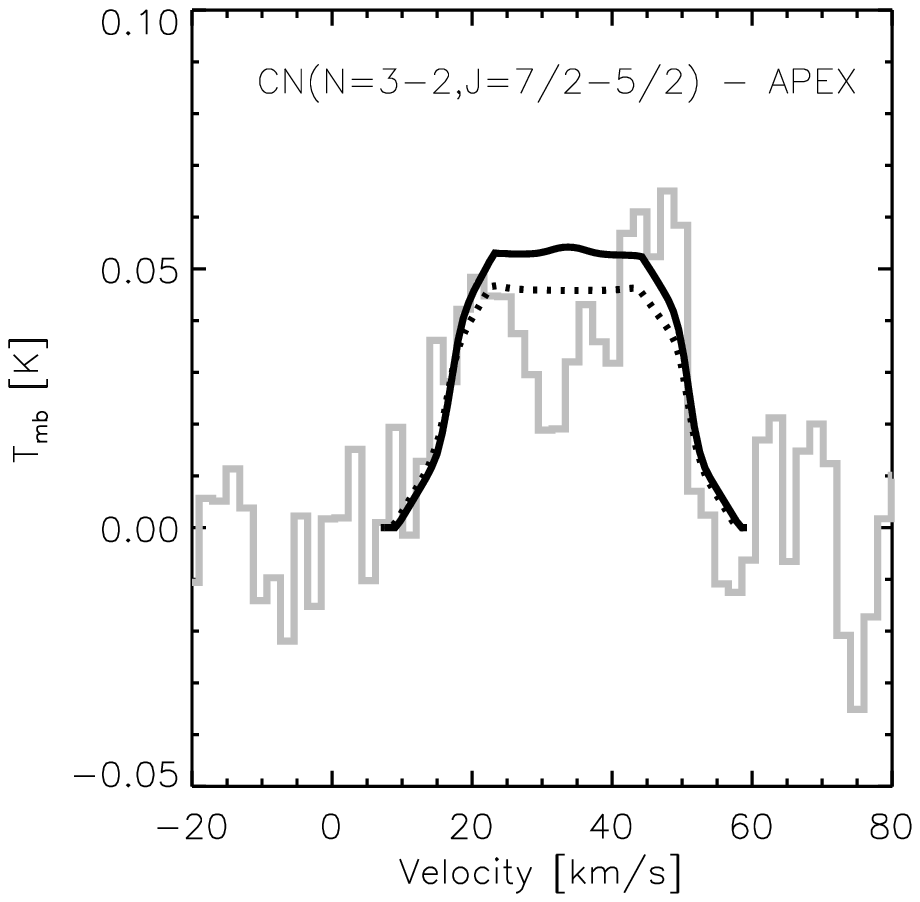}}
\caption{CN observed spectral lines (gray) are compared to the
  spectral line predictions based on the CSE model shown in
  Fig.~\ref{fig:structure_IKTau} and the abundance stratification
  displayed in Fig.~\ref{fig:abundances}. The dashed line predictions corresponds to the `alternative solution' as shown in Fig.~\ref{fig:abundances}.}
\label{CN_model}
\end{center}
\end{figure}

\paragraph{Results:} The CN lines display a peculiar profile, probably related to the hyperfine structure of the molecule. Although the signal-to-noise of the individual components is low, \citet{Kim2009} noted already that the strength of the different peaks are not in agreement with the optical thin ratio of the different hyperfine structure compoments and hint to hyperfine anomalies as already reported by \citet{Bachiller1997A&A...319..235B}. Simulations with the GASTRoNoOM-code taking all the hyperfine components into account confirm this result. We therefore opt to simulate both CN lines using the strongest component only. I.e., for the $N=3-2, J=5/2-3/2$ line we will use the $F=7/2-5/2$ component at 340031.5494\,MHz, for the $N=3-2, J=7/2-5/2$ line  the $F=9/2-7/2$ component at 340248.5440\,MHz is used.

Due to low signal-to-noise ratio of the lines and the problems with the different hyperfine componets, the derived abundance fractions are loosely constrained. To illustrate this, two model predictions are shown in Fig.~\ref{CN_model}. For one model, the inner abundance ratio is taken to be  $3 \times 10^{-8}$ and from 1000\,\Rstar\ onward, the abundance stratification follows the predictions by \citet{Willacy1997AandA...324..237W} (dashed line in Fig.~\ref{fig:abundances}). For the other model, the inner abundance ratio, $f_1$, is lowered to $1\times10^{-10}$ yielding a peak fractional abundance around 2000\,\Rstar\ of $\sim2.5\times10^{-6}$, being a factor $\sim$8 higher than the peak fractional abundance derived by \citet{Willacy1997AandA...324..237W} (see dotted line in Fig.~\ref{fig:abundances}).

\paragraph{Comparison to theoretical predictions:} The predicted inner wind abundance fractions give higher preference to the second model described in previous paragraph:
\citet{Duari1999AandA...341L..47D} predicts a fractional abundance of CN around $2.4\times10^{-10}$ at 2.2\,\Rstar\ for \object{IK~Tau}, while \citet{Cherchneff2006AandA...456.1001C} gives a value of $3\times10^{-10}$ at 5\,\Rstar\ for \object{TX~Cam}. Further out in the wind, CN is produced by neutral-neutral reactions and the photodissociation of HCN, yielding peak fractional abundances around $3\times10^{-7}$ \citep{Willacy1997AandA...324..237W}. 

If HCN is indeed photodissociated around 400 to 500\,\Rstar\ (see Sect.~\ref{HCN}), the peak fractional abundance of CN is not expected to occur around 2000\,\Rstar\ \citep[see the model predictions by][in Fig.~\ref{fig:abundances}]{Willacy1997AandA...324..237W}, but slightly beyond 500\,\Rstar\ since the photodissociation of HCN is the main formation route to CN in the outer envelope. Shifting the CN peak fractional abundance in the second model to 500\,\Rstar\ with an abundance value of $1.5\times10^{-7}$ also yields a good fit to the (noisy) data.

\paragraph{Comparison to observational studies:} This is the first time that the non-LTE CN fractional abundance for \object{IK~Tau} has been derived, although the uncertainty on the derived abundance is large.

\subsubsection{SO} \label{SO}

\begin{figure}[htp]
\begin{center}
\subfigure{\includegraphics[width=.24\textwidth,angle=0]{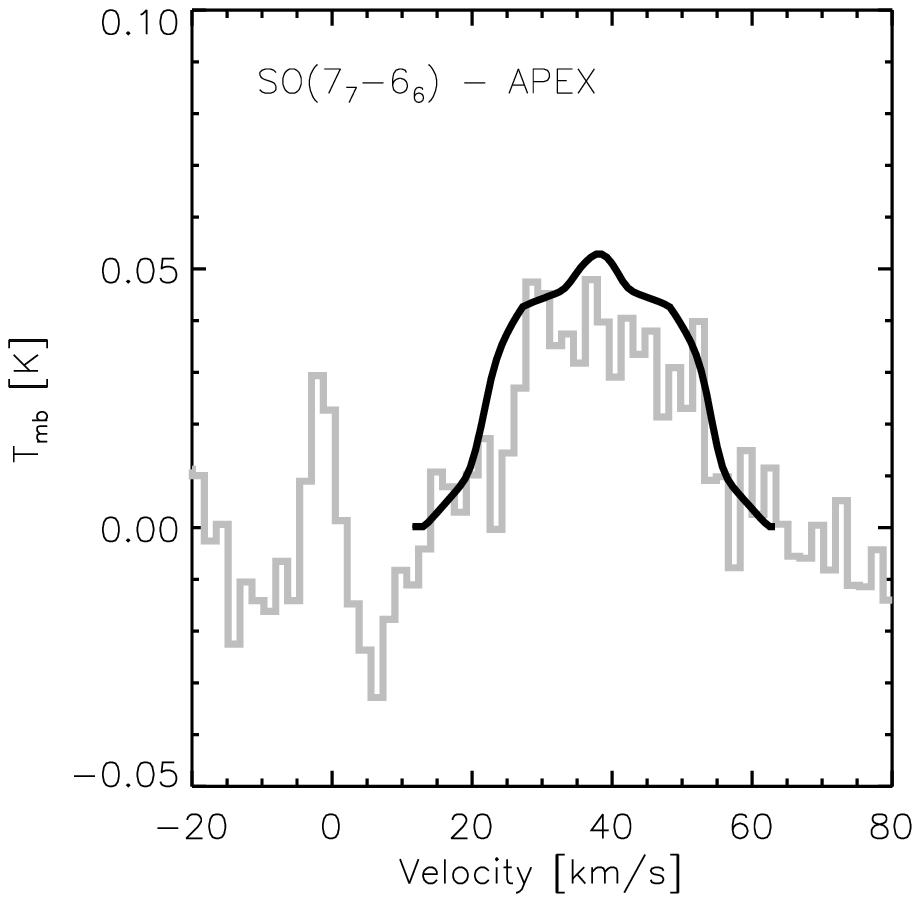}}
\subfigure{\includegraphics[width=.24\textwidth,angle=0]{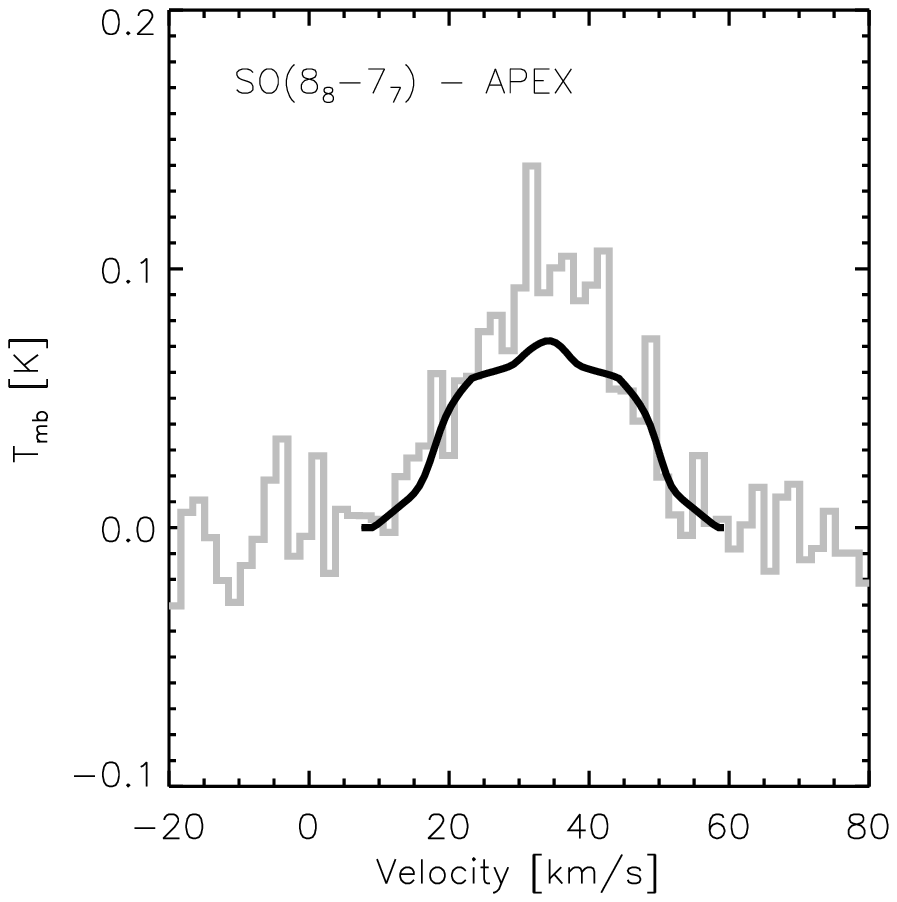}}
\caption{SO observed spectral lines (gray) are compared to the
  spectral line predictions based on the CSE model shown in
  Fig.~\ref{fig:structure_IKTau} and the abundance stratification
  displayed in Fig.~\ref{fig:abundances}.}
\label{SO_model}
\end{center}
\end{figure}

\paragraph{Results:} As in case of CN (Sect.~\ref{CN}) the noisy profiles prevent a good determination of the abundance stratification. A fractional abundance of $\sim2\times10^{-7}$ is derived around 200\,\Rstar.

\paragraph{Comparison to theoretical predictions:} An abundance stratification compatible with both the inner wind predictions of \citet{Cherchneff2006AandA...456.1001C} and the outer wind model of \citet{Willacy1997AandA...324..237W} can be derived yielding a good representation of both SO lines observed with APEX. \citet{Willacy1997AandA...324..237W} assumed no SO injection from the inner wind at large radii, but in-situ formation processes only. Assuming that SO is indeed injected to larger radii can increase the predicted peak fractional abundance computed by \citet{Willacy1997AandA...324..237W} which was somewhat too low compared to the observed value listed in their Table~6. 

\paragraph{Comparison to observational studies:} For the first time, the SO abundance fraction is derived using a non-LTE radiative transfer analysis, although the low S/N prevents an accurate abundance determination. The result agrees with the LTE analysis by \citet{Kim2009}, but is  a factor of  a few lower than \citet{Omont1993AandA...267..490O} and \citet{Bujrrabal1994AandA...285..247B}. 

\subsubsection{SO$_2$}\label{SO2}

\begin{figure*}[htp]
\begin{center}
\subfigure{\includegraphics[width=.24\textwidth,angle=0]{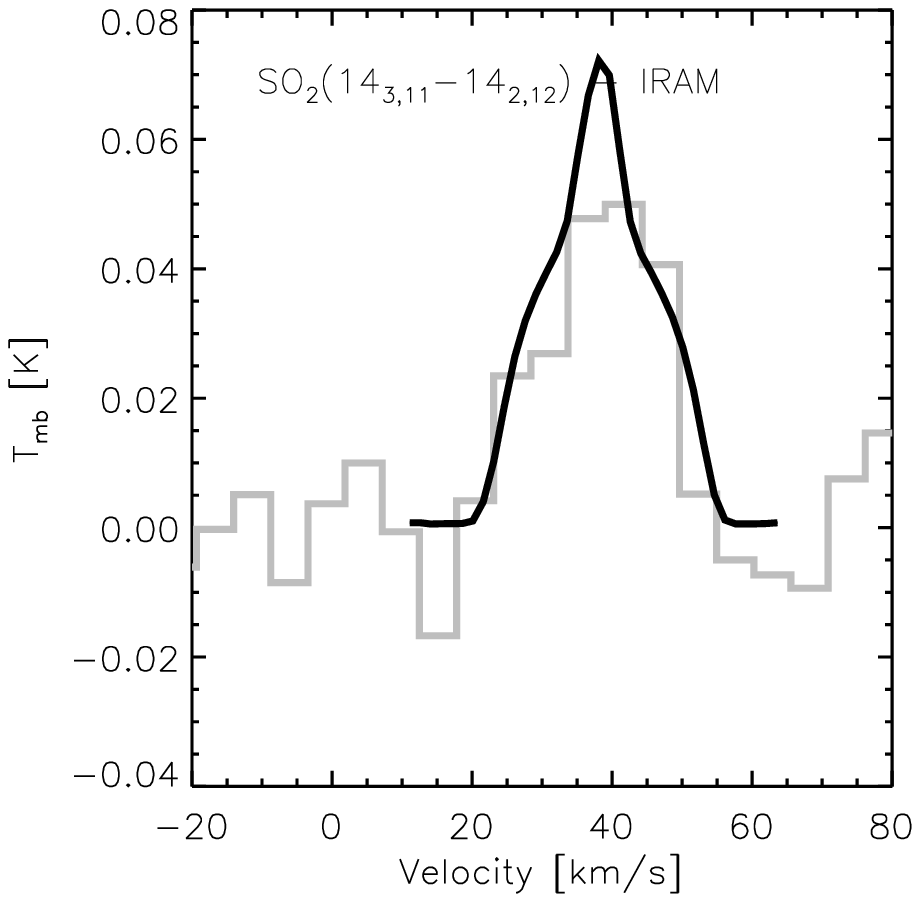}}
\subfigure{\includegraphics[width=.24\textwidth,angle=0]{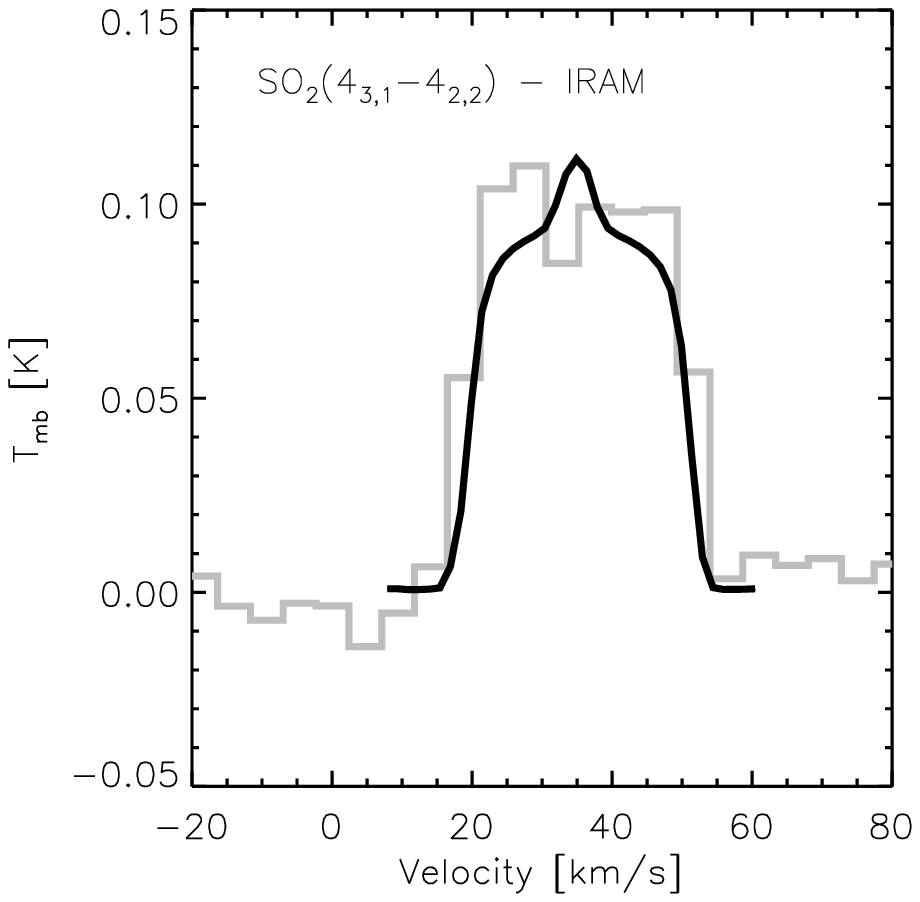}}
\subfigure{\includegraphics[width=.24\textwidth,angle=0]{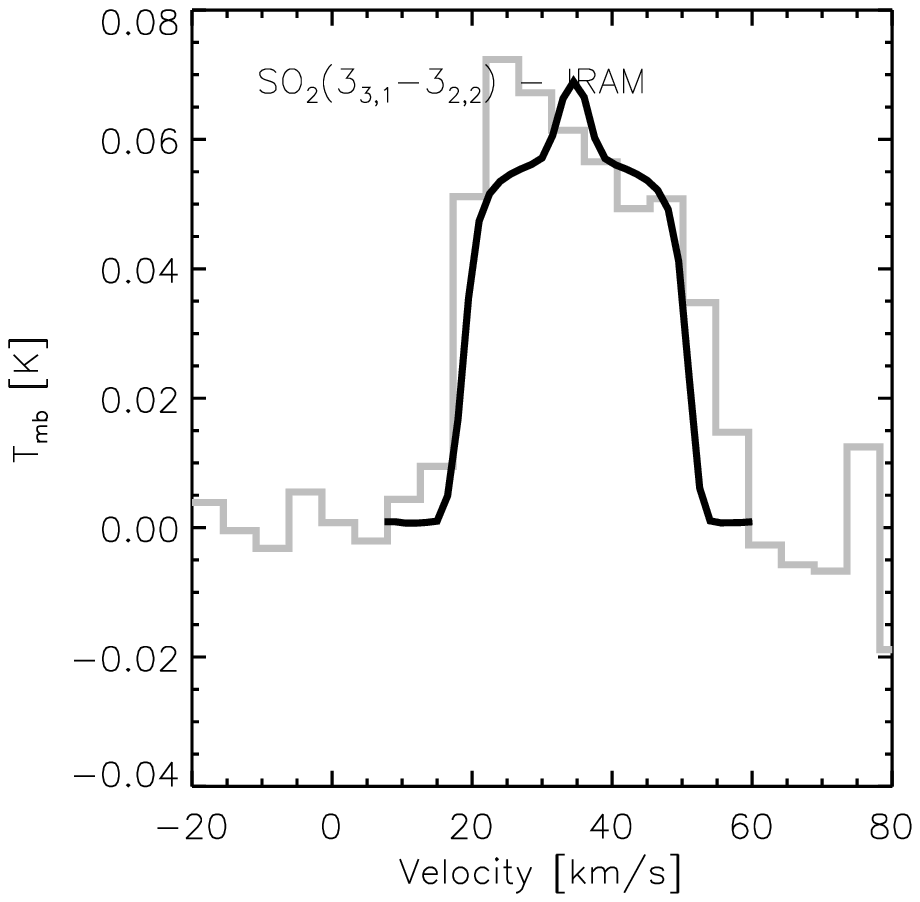}}
\subfigure{\includegraphics[width=.24\textwidth,angle=0]{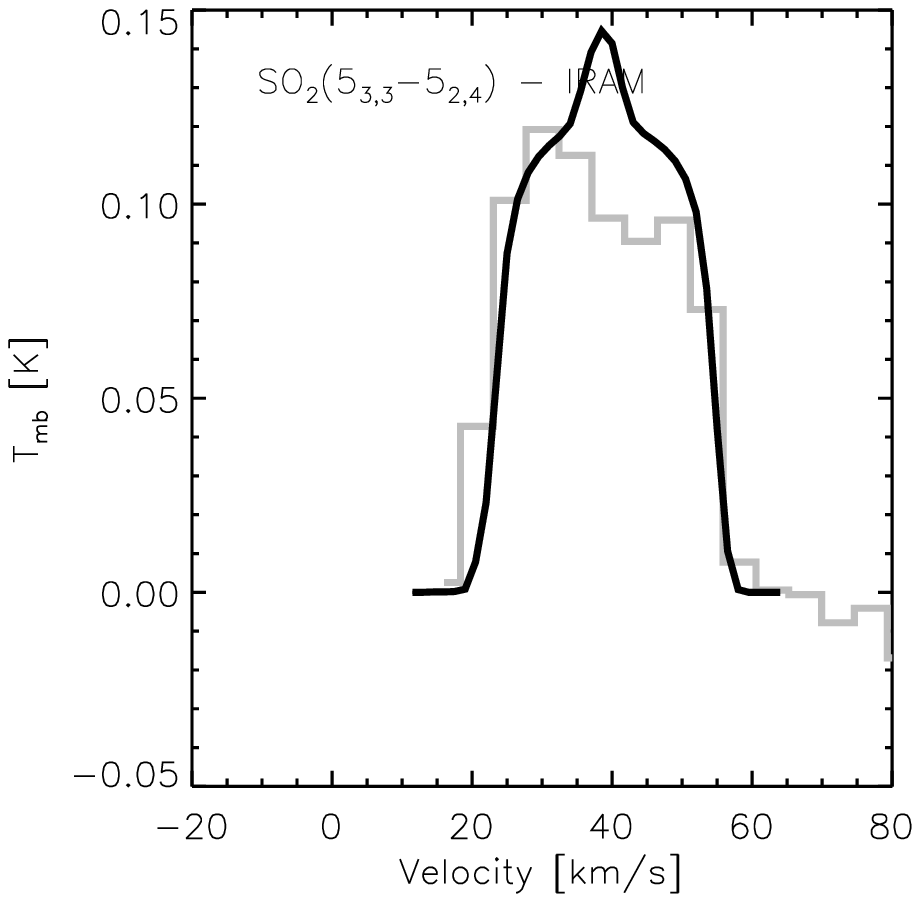}}
\subfigure{\includegraphics[width=.24\textwidth,angle=0]{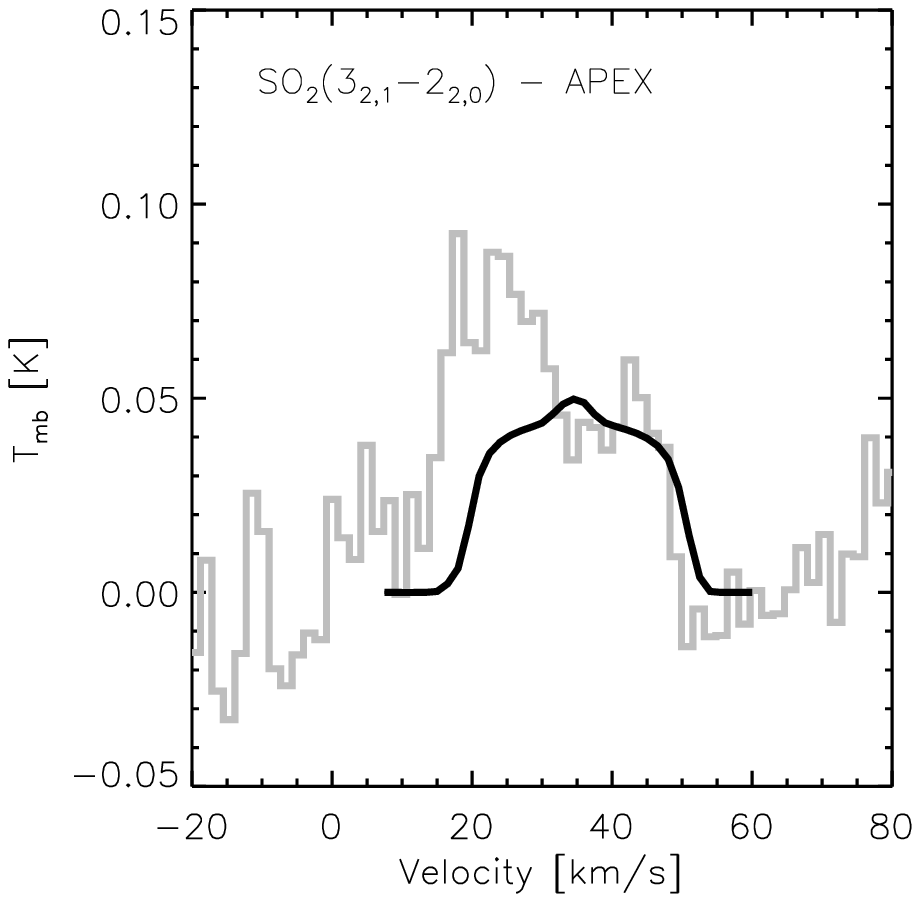}}
\subfigure{\includegraphics[width=.24\textwidth,angle=0]{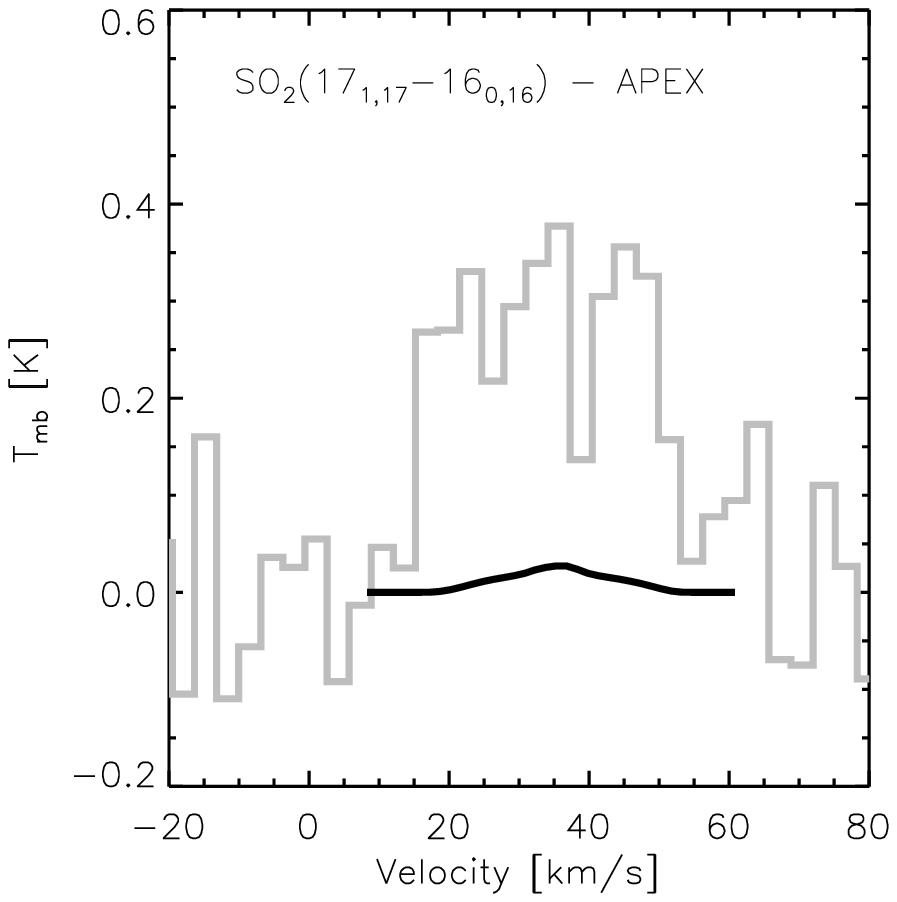}}
\subfigure{\includegraphics[width=.24\textwidth,angle=0]{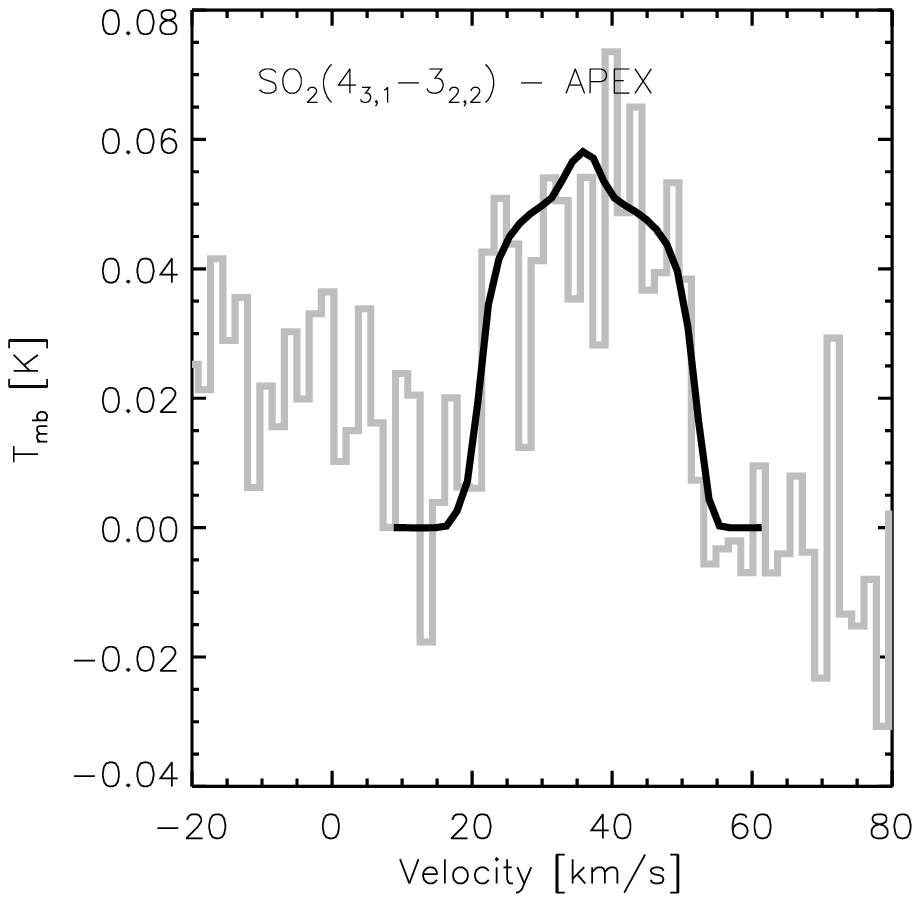}}
\subfigure{\includegraphics[width=.24\textwidth,angle=0]{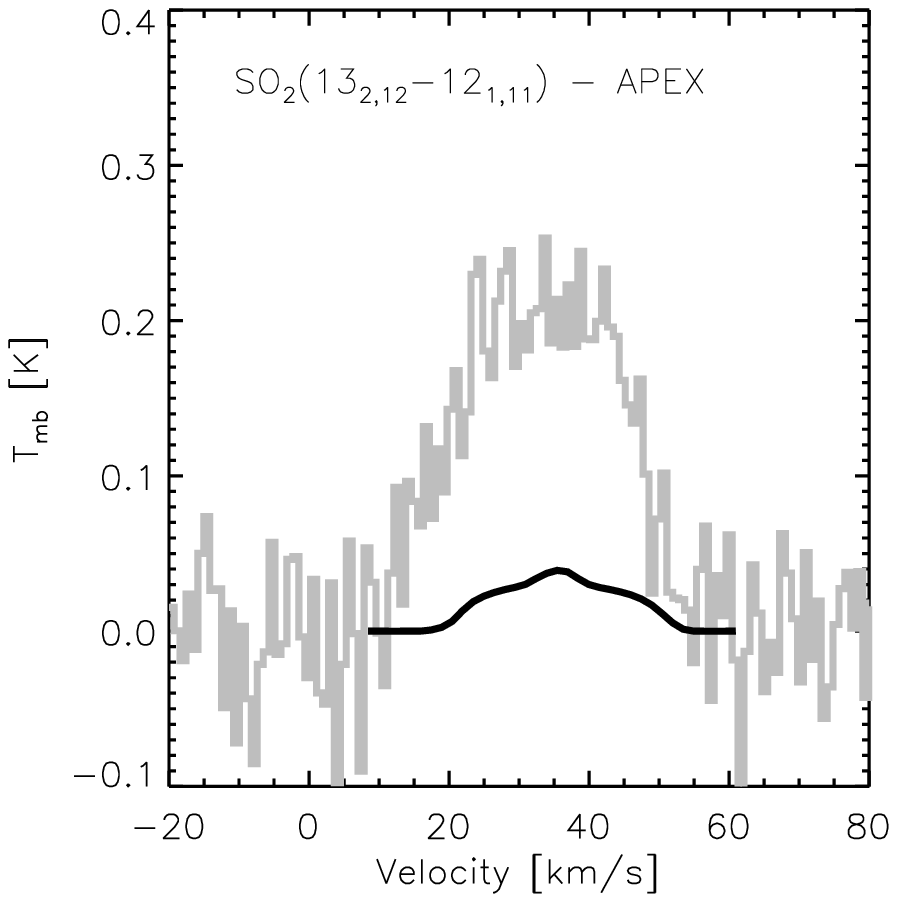}}
\subfigure{\includegraphics[width=.24\textwidth,angle=0]{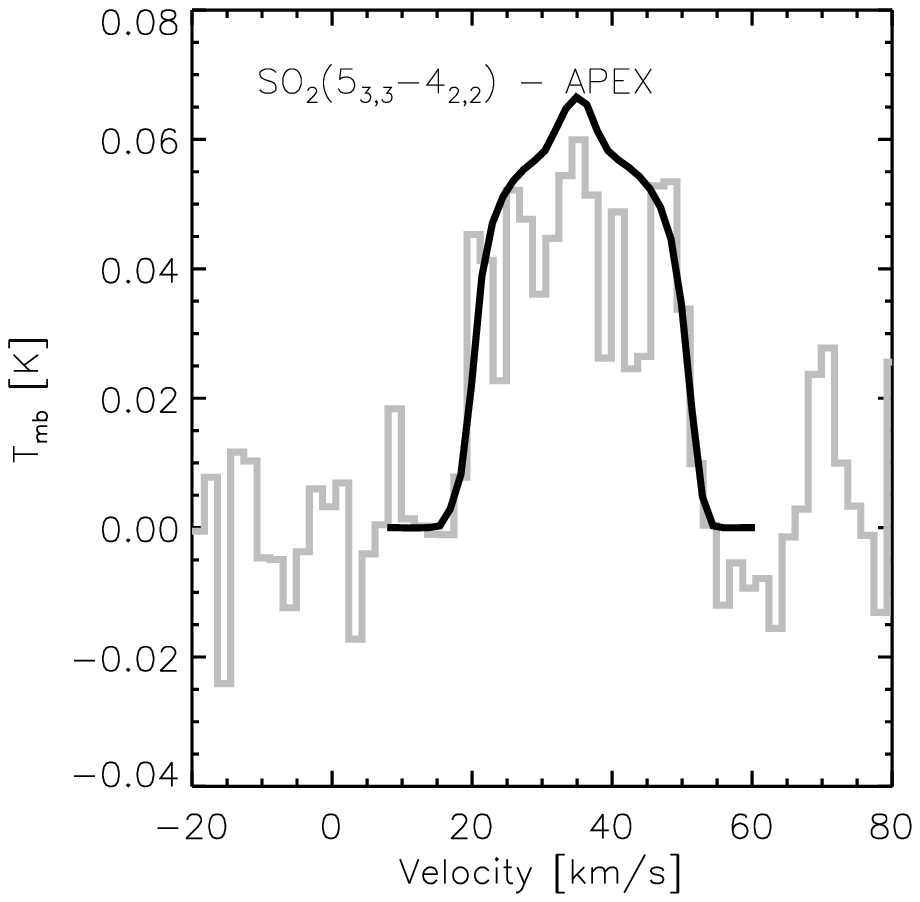}}
\subfigure{\includegraphics[width=.24\textwidth,angle=0]{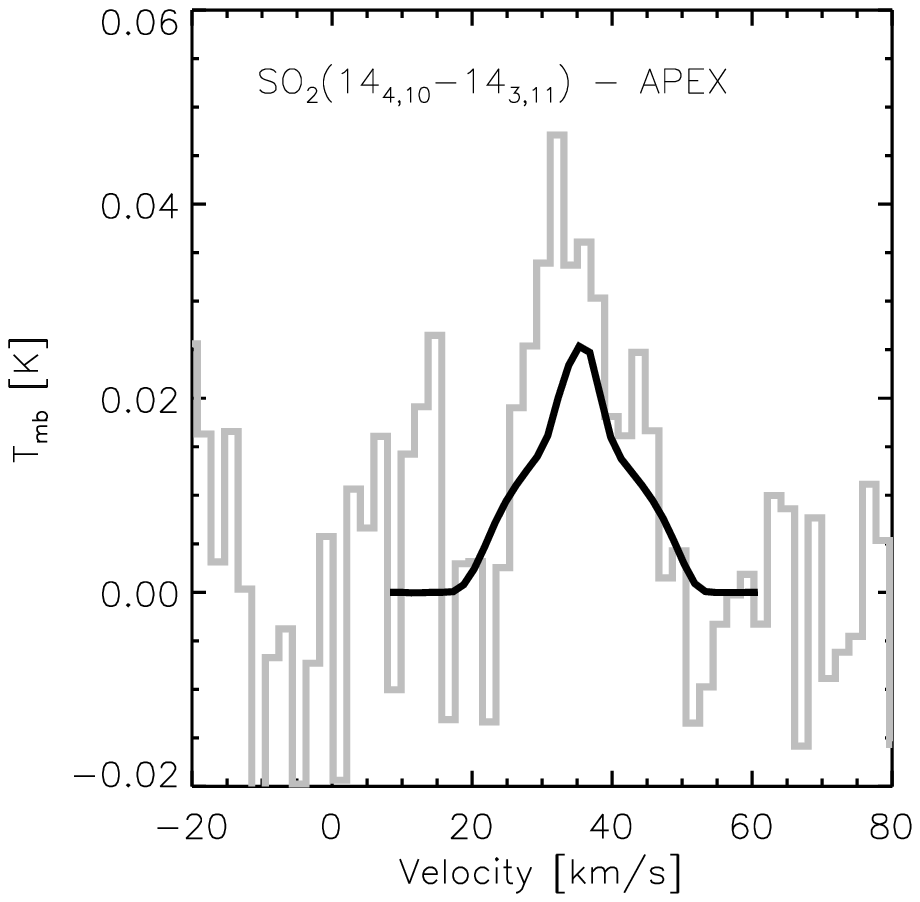}}
\caption{SO$_2$ observed spectral lines (gray) are compared to the
  spectral line predictions based on the CSE model shown in
  Fig.~\ref{fig:structure_IKTau} and the abundance stratification
  displayed in Fig.~\ref{fig:abundances}.}
\label{SO2_model}
\end{center}
\end{figure*}

As for SO$_2$ the main formation channel in both the inner and outer wind region is
\begin{equation}
 \mathrm{SO + OH \rightarrow SO_2 + H}\,.
\end{equation}
In the outer wind, SO$_2$ rapidly photodissociates back to SO \citep{Willacy1997AandA...324..237W}.
The inner wind predictions for \object{TX Cam} only yield an abundance of $\sim1.3 \times 10^{-12}$, but the detection of SO$_2$ at 7.3\,$\mu$m in several O-rich giants \citep{Yamamura1999A&A...341L...9Y} would imply a formation site close to the star and an abundance in the range $10^{-8}-10^{-7}$, slightly below the SO values. \citet{Cherchneff2006AandA...456.1001C} argues that a limited number of reactions in the involved SO$_2$ formation scheme and the lack of measured reaction rates may explain this discrepancy.

\paragraph{Results:} The availability of 10 different transitions gives some hope that we can shed light on the discussion about the inner wind fractional abundance. However, it turns out that we are unable to fit the high-excitation SO$_2$($17_{1,17}-16_{0,16}$) and SO$_2$($13_{2,12}-12_{1,11}$) lines observed with APEX (see below). 
The high-excitation SO$_2$($14_{4,10}-14_{3,11}$) and SO$_2$($14_{3,11}-14_{2,12}$) can be predicted quite well.
These two lines are narrower than the other SO$_2$ lines in the sample, indicating that their full formation region is in the inner wind region where the wind has not yet reached its full expansion velocity.

Extensive modeling efforts were done to predict the high-excitation SO$_2$ observed with APEX. While the high-excitation lines involving the $J\,=\,14-14$ levels are reasonably well predicted with the models proposed above, the $J\,=13-12$ and $J\,=\,17-16$ are far too weak. In a study of SO$_2$ in star forming regions, \citet{vanderTak2003A&A...412..133V} encountered an analogous problem, which they solved by introducing a high temperature, high-abundance component. The increase in abundance could be a factor of 100--1000. Introducing an unrealistic compact, very high-abundance component with $f_1 > 1 \times 10^{-4}$ up to 200\,\Rstar\ reproduces the APEX $J\,=13-12$ and $J\,=\,17-16$ within a factor 2, but the $J\,=\,14-14$ lines, involving similar excitation levels, is a factor $\sim$15 too strong.
A possible cause of the discrepancy could be a misidentification of the observed lines. However, different line data bases always point towards an identification as SO$_2$ transitions. But also the collision rates from the LAMDA database may be problematic. In the LAMDA database, \citet{Schoier2005A&A...432..369S} have extrapolated the collisional rates as computed by \citet{Green1995ApJS..100..213G}. \citet{Green1995ApJS..100..213G} computed the collisional rates  for temperatures in the range from 25 to 125\,K including energy levels up to 62 cm$^{-1}$; \citet{Schoier2005A&A...432..369S} extrapolated this set of collisional rates to include energy levels up to 250 cm$^{-1}$ and for a range of temperatures from 10 to 375\,K.

Neglecting the SO$_2$($17_{1,17}-16_{0,16}$) and SO$_2$($13_{2,12}-12_{1,11}$) lines, the strength of the other eight SO$_2$ lines can only be explained using a high inner abundance ratio of $1 \times 10^{-6}$, clearly pointing toward an inner wind formation region \citep[in accordance with][] {Yamamura1999A&A...341L...9Y}.

\paragraph{Comparison to observational studies:} As for SO, the SO$_2$ fractional abundance is derived for the first time using a full non-LTE radiative transfer analysis. The derived abundance value is somewhat lower than the results from classical studies assuming optically thin emission and one excitation temperature, although we have to state clearly that the SO$_2$ modeling still poses many problems.

\section{Discussion} \label{discussion}

The derived fractional abundances are already discussed in Sect.~\ref{frac_abundances}. In this section, we focus on the possible time variability of the emission lines and on the derived SiO isotopic ratios. In the last part, H$_2$O line profile predictions for the Herschel/HIFI 
mission are performed.

\subsection{Time variability} \label{timevar}

The observed molecular emission lines could be time variable. Unfortunately, no dedicated study has yet been performed to study the time variability of the molecular lines in \object{IK~Tau}. \citet{Carlstrom1990fmpn.coll..170C} reported on a monitoring programme of the SiS($J$\,=\,4--3, 5--4, and 6--5) emission from the Mira-type carbon-rich AGB star \object{IRC~+10216}. It was found that the circumstellar $J$\,=\,5--4 and $J$\,=\,6--5 line emission toward \object{IRC~+10216} varies both in line intensity and in line shape. A clear correlation between the variations and the infrared flux (as measured using the $K$-band magnitude) is found for the $J$\,=\,5--4 and $J$\,=\,6--5 lines \citep{Bieging1993AJ....105..576B}, but not for the $J$\,=\,4--3 line. This indicates that, at least, the population of a few levels vary in phase with the stellar flux. A change in the pumping mechanism of the infrared vibrationally excited levels of a molecule will modify the excitation in the ground vibrational state. A change in dust emission may also alter the excitation of a molecule, since dust emission has
the potential of affecting the level populations in the ground vibrational state, in particular through pumping via excited vibrational states. 

 \citet{Cernicharo2000A&AS..142..181C} have looked for time-related intensity variations in a line survey at 2\,mm of \object{IRC+10216}. Among the 2-mm lines, the most likely lines to be affected are: \emph{(i)} CS, HC$_3$N, SiO and SiS, four species whose IR lines are known to be optically thick, as well as \emph{(ii)} the vibrationally excited lines of C$_4$H and HCN. During the 10-year long run, these lines were observed at several occasions. The ground-state mm lines were found to have stable shapes and intensities (within 20\,\% which is consistent with the calibration uncertainty). Only the strong HCN, $\nu_2 = 1$, $J=2-1$ line, which is known to be masering, showed a factor of 2 intensity variation with time.

Currently the effect of time variability on circumstellar line emission is unknown for AGB stars in general. 

\subsection{SiO isotopic ratios} \label{isotopes}
The SiO abundance isotopic ratios derived for \object{IK~Tau} are $^{28}$SiO/$^{29}$SiO\,=\,27, $^{28}$SiO/$^{30}$SiO\,=\,80, and $^{29}$SiO/$^{30}$SiO\,=\,3, with an uncertainty of a factor of $\sim$2 due to the low signal-to-noise ratio of the $^{29}$SiO and $^{30}$SiO lines. The $^{29}$SiO/$^{30}$SiO is similar to the simple ratio of integrated intensities corrected for a difference in transition strength and beam filling factor (the combined effect is a frequency factor of $\nu^{-2}$), but the $^{28}$SiO/$^{29}$SiO and $^{28}$SiO/$^{30}$SiO ratios are a factor $\sim$7 and $\sim$10 larger, respectively,  due to neglect of optical depth effects of the $^{28}$SiO lines in case the simplified intensity ratio is used. Compared to the solar isotopic ratios of ($^{28}$SiO/$^{29}$SiO)$_\odot$\,=\,19.6, ($^{28}$SiO/$^{30}$SiO)$_\odot$\,=\,29.8 and ($^{29}$SiO/$^{30}$SiO)$_\odot$\,=\,1.52, \object{IK Tau} is underabundant in neutron-rich isotopologs or overabundant in $^{28}$Si.

Due to the weakness of the $^{29}$SiO and $^{30}$SiO lines, only few results on the silicon isotopic ratios in the circumstellar envelopes around AGB stars are reported in literature. By fitting the SiO maser emissivity \citet{Jiang2003ASSL..283..323J} estimated the $^{28}$SiO/$^{29}$SiO isotopic ratio in two oxygen-rich Miras, \object{R Cas} and \object{NV Aur}, as being 29 and 32, respectively. \citet{Ukita1988LNP...305...51U} measured the relative intensities of the $^{29}$SiO/$^{30}$SiO lines ($J=2-1,v=0$), being 2.4 for the S-type Mira \object{$\chi$ Cyg} (with C/O ratio slightly lower than 1), 1.5 for \object{IK~Tau}, and 2.9 for the oxygen-rich Mira \object{V1111 Oph}.
For the carbon-rich AGB star \object{IRC+~10216}, \citet{Cernicharo2000A&AS..142..181C} and \citet{He2008ApJS..177..275H}  derived respectively $^{28}$SiO/$^{29}$SiO\,=\,$15.4 \pm 1.1$ ($17.2 \pm 1.1$) and $^{28}$SiO/$^{30}$SiO\,=\,$20.3 \pm 2.0$ ($24.7 \pm 1.8$) and  $^{29}$SiO/$^{30}$SiO\,=\,$1.45 \pm 0.13$ ($1.46 \pm 0.11$) from the ratios of integrated line intensities, being close to the solar values. The results on IRC+10216 are, however, lower limits, since no correction for opacity effects has been done. 

\citet{Lambert1987Ap&SS.133..369L} analysed high-resolution spectra of the SiO first overtone band around 4\,$\mu$m. They obtained estimates of the atmospheric $^{28}$SiO/$^{29}$SiO abundance ratios for four red giants. For the M-type $\beta$ Peg and the S-type star HR~1105, the $^{28}$SiO/$^{29}$SiO ratio is close to the solar ratio. $^{29}$SiO appears to be underabundant in the MS star $o^1$ Ori ($^{28}$SiO/$^{29}$SiO\,=\,40) and the M-type star 10~Dra ($^{28}$SiO/$^{29}$SiO$\sim$53). The $^{30}$SiO isotope appears to be underabundant by a factor of $\sim$2 in all four red giants.

\citet{Tsuji1994A&A...289..469T} reported on high spectral resolution observations of the 4\,$\mu$m SiO first overtone band in six late-type M giants and two M supergiants. The atmospheric $^{28}$Si/$^{29}$Si  and $^{28}$Si/$^{30}$Si  and $^{29}$Si/$^{30}$Si ratios in the M giants are always slightly lower than the terrestrial values, i.e.\ more neutron-rich nuclei tend to be more abundant. This is opposite to the results of \citet{Lambert1987Ap&SS.133..369L}, and assuming that the isotopic ratios are not modified in the circumstellar envelope, the result of \citet{Tsuji1994A&A...289..469T} is also in contrast to the circumstellar isotopic ratios of M-type giants listed above.

The silicon isotopic ratios are not obviously correlated with other stellar properties. 
 In the literature, it is conventional to express the silicon (and other element) isotope ratios in parts per thousand deviation from the solar silicon isotope ratio: 
\begin{eqnarray}
 \delta_\odot(^{29}{\rm Si}) \equiv \delta_\odot\left(\frac{^{29}{\rm Si}}{^{28}{\rm Si}}\right) \equiv 1000\left[\left(\frac{^{29}{\rm Si}}{^{28}{\rm Si}}\right) / \left(\frac{^{29}{\rm Si}}{^{28}{\rm Si}}\right)_\odot -1 \right]\,, \nonumber \\
\delta_\odot(^{30}{\rm Si}) \equiv \delta_\odot\left(\frac{^{30}{\rm Si}}{^{28}{\rm Si}}\right) \equiv 1000\left[\left(\frac{^{30}{\rm Si}}{^{28}{\rm Si}}\right) / \left(\frac{^{30}{\rm Si}}{^{28}{\rm Si}}\right)_\odot -1 \right]\,, \\
\end{eqnarray}
yielding values for \object{IK~Tau} of $-274$ and $-627$, respectively. These values reflect both the initial isotopic composition of the star and possible effects due to nucleosynthesis. 

For an AGB star, the  initial Si isotopic composition in the envelope is altered by slow neutron capture reactions ($s$-process) in the He intershell and subsequent third dredge-up (TDU) events, increasing the $^{29}$Si and $^{30}$Si abundance fractions. A low-metallicity star is expected to have initial $^{29}$Si/$^{28}$Si and $^{30}$Si/$^{28}$Si ratios that are smaller than the solar ratios. The inferred isotopic shifts of the Si isotopes are smaller for an O-rich than for a C-rich AGB star since a C-rich star goes through more dredge-up events increasing the $^{12}$C and $s$-processed material \citep{Zinner2006ApJ...650..350Z, Vollmer2008ApJ...684..611V}.  Using two different stellar evolution codes, \citet{Zinner2006ApJ...650..350Z} studied the change in silicon isotopic ratios in AGB stars: the shift in Si isotopic ratios and the increase of the $^{12}$C/$^{13}$C in the envelope during third dredge-up are higher for higher stellar mass, lower metallicity, and lower mass-loss rate, but their predicted silicon isotopic shifts are always much higher than the observational values derived for \object{IK~Tau}. The mimimum values plotted in their Fig.~6 correspond to the value $\delta_\odot(^{29}{\rm Si}) $ and $\delta_\odot(^{30}{\rm Si}) $ at C/O=1, and are larger than $-200$. They find that no noticeable changes in the Si isotopes occur when the star is still O-rich. Consequently, the isotopic anomalies in silicon found for several M-type giants probably reflect those of the interstellar medium out of which stars were formed.

For the solar system material, the silicon isotopic ratios are thought to be understood by a mixture of the nuclear products by types I and II supernovae \citep{Tsuji1994A&A...289..469T}. The predicted silicon isotopic ratios by type II supernovae are around $^{28}$Si/$^{29}$Si$\sim$15 and $^{28}$Si/$^{30}$Si$\sim$35 according to \cite{Hoppe2009ApJ...691L..20H},
while \citet{Hashimoto1989A&A...210L...5H} arrive at lower values being 8.9 and 12.6 respectively. Type I supernovae produce mostly $^{28}$Si with little $^{29}$Si and $^{30}$Si \citep{Thielemann1986A&A...158...17T}. Non-terrestrial silicon isotopic ratios can then be reasonably explained in the same way as for the solar system but by assuming a different contribution of types I and II supernovae.

The silicon isotopic shifts reported here for \object{IK~Tau} are much lower than values deduced from presolar silicate grains  \citep{Vollmer2008ApJ...684..611V, Mostefaoui2004ApJ...613L.149M}, of which the origin spans the range from red giant branch (RGB) and AGB stars up to supernovae. 
Looking to silicon isotopic ratios derived from presolar SiC grains \citep[Fig.~2 in][]{Zinner2006ApJ...650..350Z}, the silicon isotope ratios of \object{IK~Tau} correspond to the X-grains, which are thought to originate in Type II supernovae  \citep[however major discrepancies between model predictions and observed isotopic ratios still exist;][]{Nittler1995ApJ...453L..25N}. The $^{12}$C/$^{13}$C ratio inferred for \object{IK~Tau} (=14) is at the lower limit of the values derived for X-type grains \citep[see Fig.~1 in][]{Zinner2006ApJ...650..350Z}.

Hence, if the atmospheric (and circumstellar) silicon isotopic ratio is indeed not changed due to nucleosynthesis and subsequent dredge-ups, the above arguments seem to suggest that the interstellar medium out of which IK Tau was born has a mixture analogous to  X-type grains of which supernovae type II are thought to be the main contributors. The measurement of other isotopic ratios can shed new light on this discussion.

\subsection{H$_2$O line profile predictions}\label{H2O}

\begin{figure*}[!hpt]
 \includegraphics[height=\textwidth,angle=90]{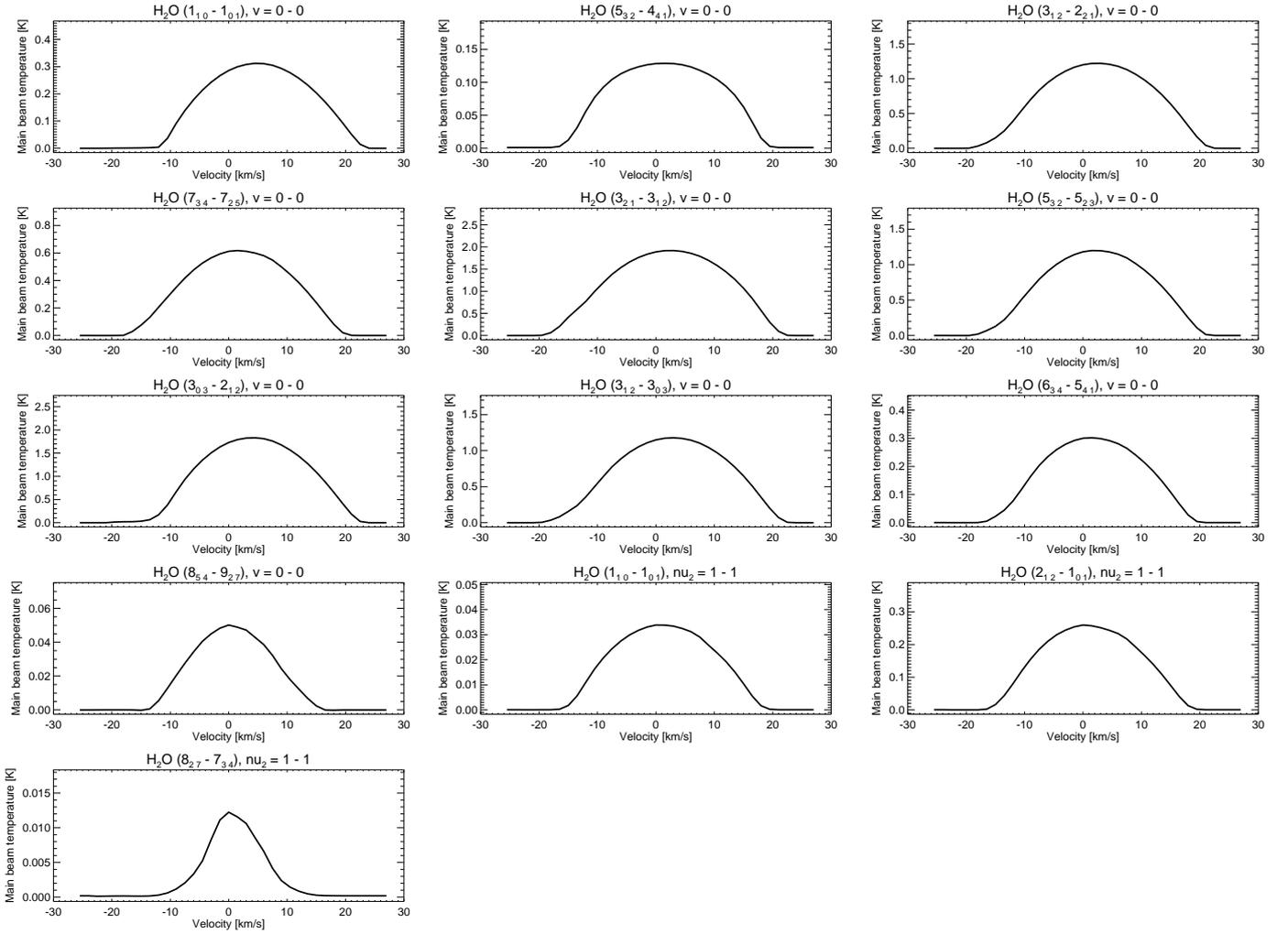}
\caption{Snapshot of a few ortho-H$_2$O lines which will be observed with Herschel/HIFI.}
\label{HIFI_H2O}
\end{figure*}

Line profile predictions are performed for a few water lines which will be observed by Herschel/HIFI in the framework of the Guaranteed Time Key Programme HIFISTARS (P.I.\ V.\ Bujarrabal) (see Table~\ref{H2O_frequencies}). This key programme focusses on the observations of CO, H$_2$O and HCN lines in a well-selected sample of evolved stars in order to gain deeper insight into the structure, thermodynamics, kinematics and chemistry of CSEs and into the mass-loss history of evolved stars. The inner wind abundance fraction is assumed to be [H$_2$O/H$_2$]=$3.5 \times 10^{-4}$ \citep{Cherchneff2006AandA...456.1001C}. The photodissociation radius is taken from the modeling of \citet{Willacy1997AandA...324..237W}, being around 1600\,\Rstar. Using the analytical formula from \citet{Groenewegen1994A&A...290..531G} deduced from the results of \citet{Netzer1987ApJ...323..734N}, a photodissociation radius of $2.8 \times 10^{16}$\,cm or 1870\,\Rstar\ would be obtained. As standard set-up, the Barber H$_2$O line list (see Appendix~\ref{appa}) is used, including 45 levels in both the ground state and first excited vibrational state (the bending mode $\nu_2=1$ at 6.3\,$\mu$m).

\begin{table}[htp]
\caption{Line frequencies, upper energy levels and Einstein A-coefficients for the ortho-H$_2$O lines which will be observed with Herschel/HIFI.}
\label{H2O_frequencies}
\begin{center}
 \begin{tabular}{ccccc}
 \hline
Vibrational & Transition & Frequency & E$_{\rm upper}$ & A \\
state        &                & [GHz]   & [cm$^{-1}$] & [s$^{-1}$]\\
\hline \hline
\rule[3mm]{0mm}{0mm}$v=0$ & $1_{1,0}-1_{0,1}$ & 556.933 & 42.371 & $3.497 \times 10^{-3}$\\ 
$v=0$ & $5_{3,2}-4_{4,1}$ & 620.882 & 50.880 & $1.106 \times10^{-4}$\\
$v=0$ & $3_{1,2}-3_{0,3}$ & 1097.488 & 173.365 & $1.664 \times10^{-2}$\\
$v=0$ & $3_{1,2}-2_{2,1}$ & 1153.219 & 173.365 & $2.693 \times10^{-3}$ \\ 
$v=0$ & $6_{3,4}-5_{4,1}$ & 1158.391 & 648.967 & $1.417 \times10^{-3}$ \\ 
$v=0$ & $3_{2,1}-3_{1,2}$ & 1162.849 & 212.154 & $2.307 \times10^{-2}$\\
$v=0$ & $8_{5,4}-9_{2,7}$ & 1595.961 & 1255.144 & $2.947 \times10^{-4}$ \\
$v=0$ & $3_{0,3}-2_{1,2}$ & 1716.731 & 136.757 & $5.109 \times10^{-2}$ \\
$v=0$ & $7_{3,4}-7_{2,5}$ & 1797.168 & 842.355 & $9.143\times10^{-2}$ \\ 
$v=0$ & $5_{3,2}-5_{2,3}$ & 1867.562 & 508.806 & $8.198 \times10^{-2}$ \\
\hline
\rule[3mm]{0mm}{0mm}$v=1$ &  $1_{1,0}-1_{0,1}$ & 658.081 & 1618.683 & $5.575 \times10^{-3}$\\ 
$v=1$ &  $8_{2,7}-7_{3,4}$ & 967.403 & 2495.278 & $5.800 \times10^{-4}$ \\
$v=1$ &  $2_{1,2}-1_{0,1}$ & 1753.922 & 1677.187  & $6.335 \times10^{-2}$\\
\hline
\end{tabular}
\end{center}
\end{table}

\paragraph{Description of the line profiles:}  Most H$_2$O lines displayed in Fig.~\ref{HIFI_H2O} have a parabolic shape, characteristic for optically thick unresolved emission.
A few lines suffer from self-absorption in the blue wing (e.g., the $8_{5,4} - 9_{2,7}$ line in the ground-state).
Particularly for lines where the optical depths at the line centre can be up to $\sim$100, effective self-absorption on the blue-shifted side can be seen (e.g., the $1_{1,0} - 1_{0,1}$ and $3_{0,3} - 2_{1,2}$ lines in the ground-state). Lines in the first vibrational state  allow one to trace the wind acceleration zone. These lines are considerably narrower than $2\,v_\infty$ (e.g., the $8_{2,7} - 7_{3,4}$ line in the $\nu_2=1$-state).



\begin{figure}
 \includegraphics[height=.5\textwidth,angle=90]{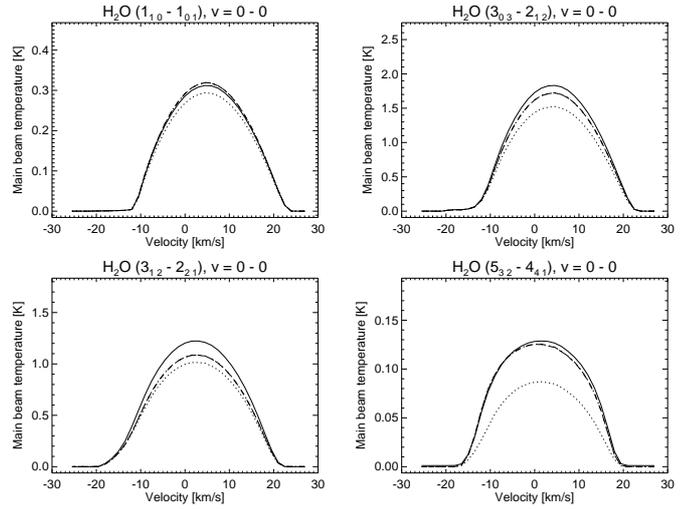}
\caption{Comparison between H$_2$O line profile predictions using \emph{(1)} the Barber line list, including the ground-state and $\nu_2=1$-state (full line, see also Fig.~\ref{HIFI_H2O}), \emph{(2)} the LAMDA line list, including only the ground-state (dotted line), \emph{(3)} the LAMDA line list, including the ground-state and $\nu_2=1$-state (dashed line), and \emph{(3)} the LAMDA line list, including the ground-state,  the $\nu_2=1$- state and $\nu_3=1$-state (dash-dotted line). One can barely see the difference between including or omitting the $\nu_3=1$-state in the predictions for \object{IK Tau}, the only exception to this is the center of the optically thin $8_{5 4} - 9_{2 7}$ line in the ground state.}
\label{HIFI_H2O_nu}
\end{figure}

\paragraph{Comparison to other line lists, including or omitting the $\nu_2$ and $\nu_3$ vibrational state} (see Fig.~\ref{HIFI_H2O_nu}): Using the Barber H$_2$O or the LAMDA linelist yields comparable results for the lines displayed here.
Omitting the first vibrational state of the bending mode ($\nu_2=1$), yields a decrease in line flux of a few percent up to 60 percent at maximum, depending on the transition. This discrepancy will increase for lower mass-loss rate objects \citep{Maercker2008A&A...479..779M}. The inclusion of excitation to the first excited vibrational state of the asymmetric stretching mode ($\nu_3=1$) yields a change in line flux of 20\,\% at maximum. This is in agreement with the recent results by \citet{Gonzales-Alfonso2007ApJ...669..412G} and \citet{Maercker2009A&A...494..243M}, who found that the inclusion of the $\nu_3=1$-state is particularly important for low mass-loss rate objects. 
Since the Einstein A-coefficients of the symmetric stretching mode ($\nu_1\,=\,1$) are an order of magnitude lower than for the asymmetric stretching, this state is not likely to affect the models. 

\begin{figure}
 \includegraphics[height=.5\textwidth,angle=90]{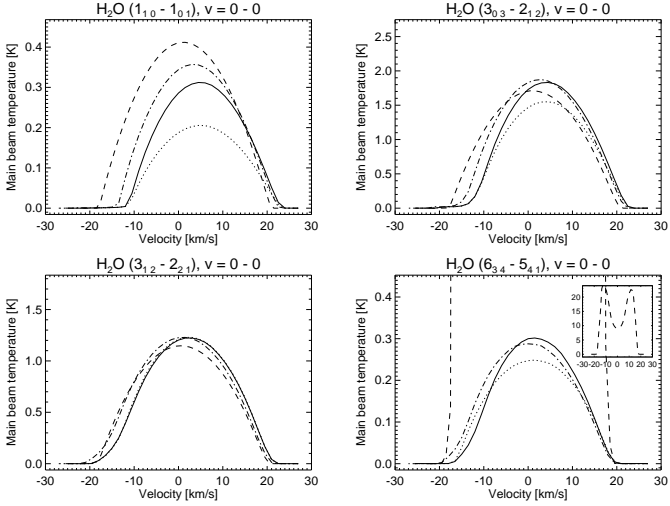}
\caption{Comparison between H$_2$O line profile predictions using the Barber line list, including the ground-state and $\nu_2=1$- state \emph{(1)}  full line: predictions using temperature and velocity structure as shown in Fig.~\ref{fig:structure_IKTau} (see also Fig.~\ref{HIFI_H2O}), \emph{(2)} dotted line: assuming a power law temperature structure $T(r) \sim r^{-0.7}$,  \emph{(3)} dashed-dotted line: assuming a velocity structure consistent with the momentum equation (see discussion in Sect.~\ref{velocity_structure}),  \emph{(5)} dashed line: assuming the same stellar parameters as used by \citet{Maercker2008A&A...479..779M}: $T_{\rm eff}$\,=\,2600\,K, \Rstar\,=\,$3.53 \times10^{13}$\,cm, D\,=\,300\,pc and \Mdot\,=\,$1 \times 10^{-5}$\,\Msun/yr. The inset in the lower right panel shows the full maser line profile of the $6_{3 4}-5_{4 1}$ transition using these parameters. }
\label{HIFI_H2O_param}
\end{figure}

\begin{figure}
 \includegraphics[height=.5\textwidth,angle=90]{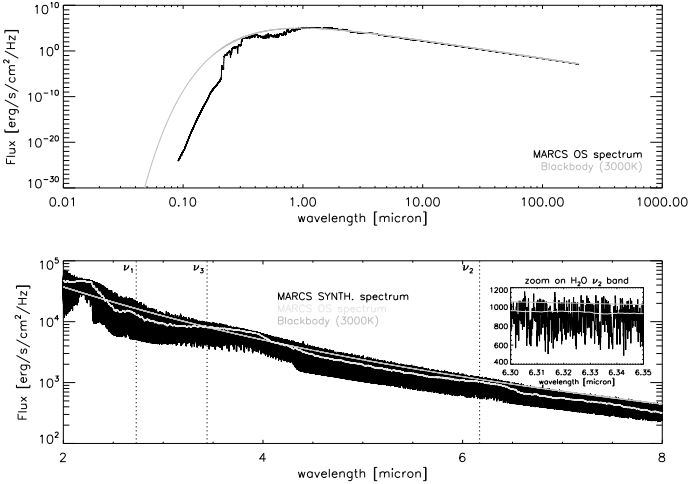}
\caption{Representation of the stellar radiation field. \emph{Upper panel:} Comparison between a blackbody at 3000\,K (gray) and a {\sc marcs} flux-sampled spectrum at a stellar temperature of 3000\,K and a logarithm of the gravity of 1.5\,dex (black). \emph{Lower panel:} Comparison between a blackbody at 3000\,K (gray), a {\sc marcs} flux-sampled spectrum (light gray) and a high-resolution theoretical spectrum generated from the {\sc marcs} model with a resolution of $\Delta \lambda$\,=\,0.5\AA\ in the wavelength region between 2 and 8\,$\mu$m. The bandheads of the different water vibrational modes (the symmetric stretching mode $\nu_1$, the bending mode $\nu_2$, and the asymmetric stretching mode $\nu_3$) are indicated by the vertical dotted lines. The inset shows a zoom around 6.3\,$\mu$m where water absorption determines the spectral appearance.}
\label{BB_MARCS}
\end{figure}

\paragraph{Dependence on parameters:} The H$_2$O line fluxes which will be observed with HIFI are very sensitive to certain (stellar) parameters, and their observations will give us a handle to constrain the circumstellar structure to even higher accuracy. To illustrate this, some simulations are shown in Fig.~\ref{HIFI_H2O_param}. 
\begin{itemize}
 \item In case the temperature structure is approximated using a power law $T(r) \sim r^{-0.7}$ (dotted line in Fig.~\ref{HIFI_H2O_param}), a differential change is seen for the predicted line fluxes. Observing a few water lines with different excitation levels will pin down the temperature structure in the CSE. 
\item A second simulation shows the effect of using a velocity structure which is computed from solving the momentum equation (dashed-dotted line in Fig.~\ref{HIFI_H2O_param}). Using another velocity law results in another gas number density and slightly different temperature structure. Using the momentum equation, a steeper velocity gradient is obtained (see, e.g., Fig.~\ref{velocity}), and the velocity reaches the terminal velocity at shorter distances from the star. This results in slightly broader line profiles and a flux  enhancement in the blue wing since that part of the CSE contributing to the line profile at a certain velocity $v$ will be shifted somewhat inward, hence attaining a higher source function.
\item Using a blackbody to represent the stellar radiation instead of a high-resolution theoretical spectrum calculated from a {\sc marcs} model atmosphere (see Sect.~\ref{rad_model}), only induces a change in the predicted line fluxes smaller than 2\,\% (not shown in Fig.~\ref{HIFI_H2O_param}). For wavelengths beyond 200\,$\mu$m the stellar flux is always represented by a blackbody in the GASTRoNOoM-calculations (note that the ground-state of ortho-water is at 23.794\,cm$^{-1}$ or around 420\,$\mu$m), the flux difference between the blackbody and the theoretical high-resolution spectrum around the $\nu_2$ bending mode  is shown Fig.~\ref{BB_MARCS}. The reason for this negligible difference is the fact that the stellar radiation field is not important (in this case) for the H$_2$O excitation. Excluding the stellar radiation field only yields a reduction of the line emission by 2\,\% at maximum.
\item The dashed line in Fig.~\ref{HIFI_H2O_param} shows the model predictions using the same stellar and envelope parameters as in \citet{Maercker2008A&A...479..779M}: $T_{\rm eff}$\,=\,2600\,K, \Rstar\,=\,$3.53 \times10^{13}$\,cm, D\,=\,300\,pc and \Mdot\,=\,$1 \times 10^{-5}$\,\Msun/yr. This example shows how maser action in the $5_{3 2}-4_{4 1}$ transition at 621\,GHz is very sensitive to the structural parameters. 

\end{itemize}

\section{Conclusions} \label{conclusion}

In this paper, we have for the first time performed a self-consistent, non-LTE radiative transfer analysis on 11 different molecules and isotopologs ($^{12}$CO, $^{13}$CO, SiS, $^{28}$SiO, $^{29}$SiO, $^{30}$SiO, HCN, CN, CS, SO, SO$_2$) excited in the circumstellar envelope around the oxygen-rich AGB star \object{IK~Tau}. In contrast to previous studies, the temperature and velocity structure in the envelope are computed self-consistently, the circumstellar fractional abundances are linked to theoretical outer wind non-chemical equilibrium studies and the full line profiles are used as criteria to deduce the abundance structure. The Gaussian line profiles of HCN and SiO clearly point toward  formation partially in the region where the wind has not yet reached its full velocity. Using the HCN line profiles as criterion, we can deduce that the wind acceleration is slower than deduced from classical theories \citep[e.g.][]{Goldreich1976ApJ...205..144G}. For a few molecules, a significantly different result is obtained compared to previous, more simplified, studies. SiO and SiS seem to be depleted in the intermediate wind region due to adsorption onto dust grains. The HCN and CS intermediate wind abundance around 50-300\,\Rstar\ is clearly below the inner wind theoretical predictions by \citet{Duari1999AandA...341L..47D} and \citet{Cherchneff2006AandA...456.1001C}, which may either indicate a problem in the theoretical shock-induced modeling or possibly that, contrary to what is thought, HCN and CS do participate in the dust formation, maybe via the radical CN through which both molecules are formed. The lack of high signal-to-noise data for CN and SO prevent us from accurately determining the circumstellar abundance stratification. It turned out to be impossible to model all the SO$_2$ line profiles, particularly a few of the high-excitation lines. This may be due to a misidentification of the lines or to problems with the collisional rates. The SiO isotopic fractions point toward high $^{28}$SiO/$^{29}$SiO and $^{28}$SiO/$^{30}$SiO ratios, which are currently not understood in the framework of nucleosynthesis altering the AGB isotopic fractions, but seem to reflect the chemical composition of the interstellar cloud out of which the star is born.
Finally, in Sect.~\ref{H2O}, we present H$_2$O line profile predictions for a few lines which will be observed with the Herschel/HIFI instrument (launched on May, 14 2009).

\begin{appendix}
 \section{Molecular line data}\label{appa}
For each of the treated
molecules, we briefly describe the molecular line data used in this paper. Quite
often, data from the \emph{Leiden Atomic and Molecular Database}
(LAMDA) are used \citep{Schoier2005A&A...432..369S}. When appropriate,
transition probabilities are compared to the relevant data in this
database. 

\begin{figure}[htp]
\begin{center}
\includegraphics[height=.45\textwidth,angle=90]{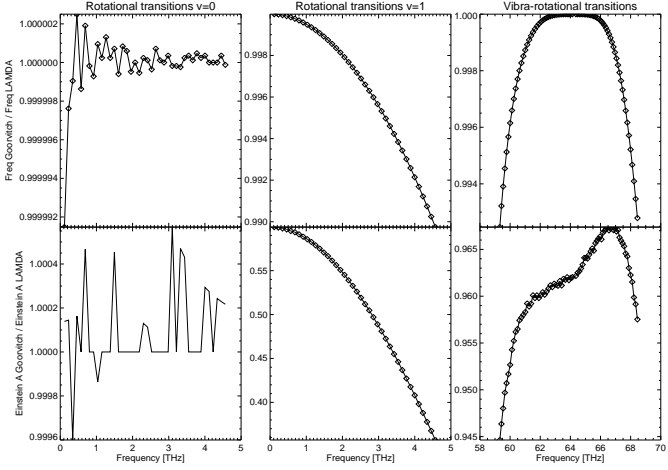}
\caption{Comparison between the transition frequencies and Einstein A
  coefficients of CO as listed in the LAMDA database and as computed by
  \citet{Goorvitch1994ApJS...91..483G}. }
\label{fig:EinsteinCO}
\end{center}
\end{figure}

\paragraph{CO -- carbon monoxide.} Both for $^{12}$CO and $^{13}$CO,
energy levels, transition frequencies and Einstein A coefficients were
taken from \citet{Goorvitch1994ApJS...91..483G}. Transitions in the
ground and first vibrational state were included up to $J = 40$. The
CO-H$_2$ collisional rate coefficients at kinetic temperatures from 10
to 4000\,K are from
\citet{Larsson2002A&A...386.1055L}. Figure~\ref{fig:EinsteinCO} shows
a good match for the transition frequencies (better than 1\,\%). The Einstein A coefficients for the rotational transitions in the ground-state and the vibra-rotational transitions correspond to better than 0.5\,\%, but  the Einstein A
coefficients for the rotational transitions in the $v=1$ state may
differ by up to a factor 3 for the high-lying rotational transitions,
i.e.\ the ones with the largest $J$ quantum number. For the CO lines of
interest to this study, i.e.\ rotational transitions in the $v=0$ state
with $J_{\rm up} \le 7$, the effect on the predicted line fluxes is small:
calculating the temperature stratification self-consistently (see Sect.~\ref{rad_model}) using the 
Einstein A coefficients of the LAMDA-data base, the largest deviation (of 6\,\%) occurs for the $^{12}$CO(1--0) line.

\paragraph{SiO --  silicon monoxide.} The SiO linelist of 
\citet{Langhoff1993} was used to extract the frequencies, energy
  levels and (vibra-)rotational radiative rates for $^{28}$SiO, $^{29}$SiO, and
$^{30}$SiO. Both ground and first vibrational state were included,
with rotational quantum number $J$ up to 40. The SiO-H$_2$
collisional rates in the ground state are taken from the
LAMDA-database. For the rotational transitions in the first
vibrational state, it is assumed that the collisional rates are equal
to the ones in the ground state. The vibra-rotational collisional
rates are assumed to be zero. 

\begin{figure}[htp]
\begin{center}
\includegraphics[height=.45\textwidth,angle=90]{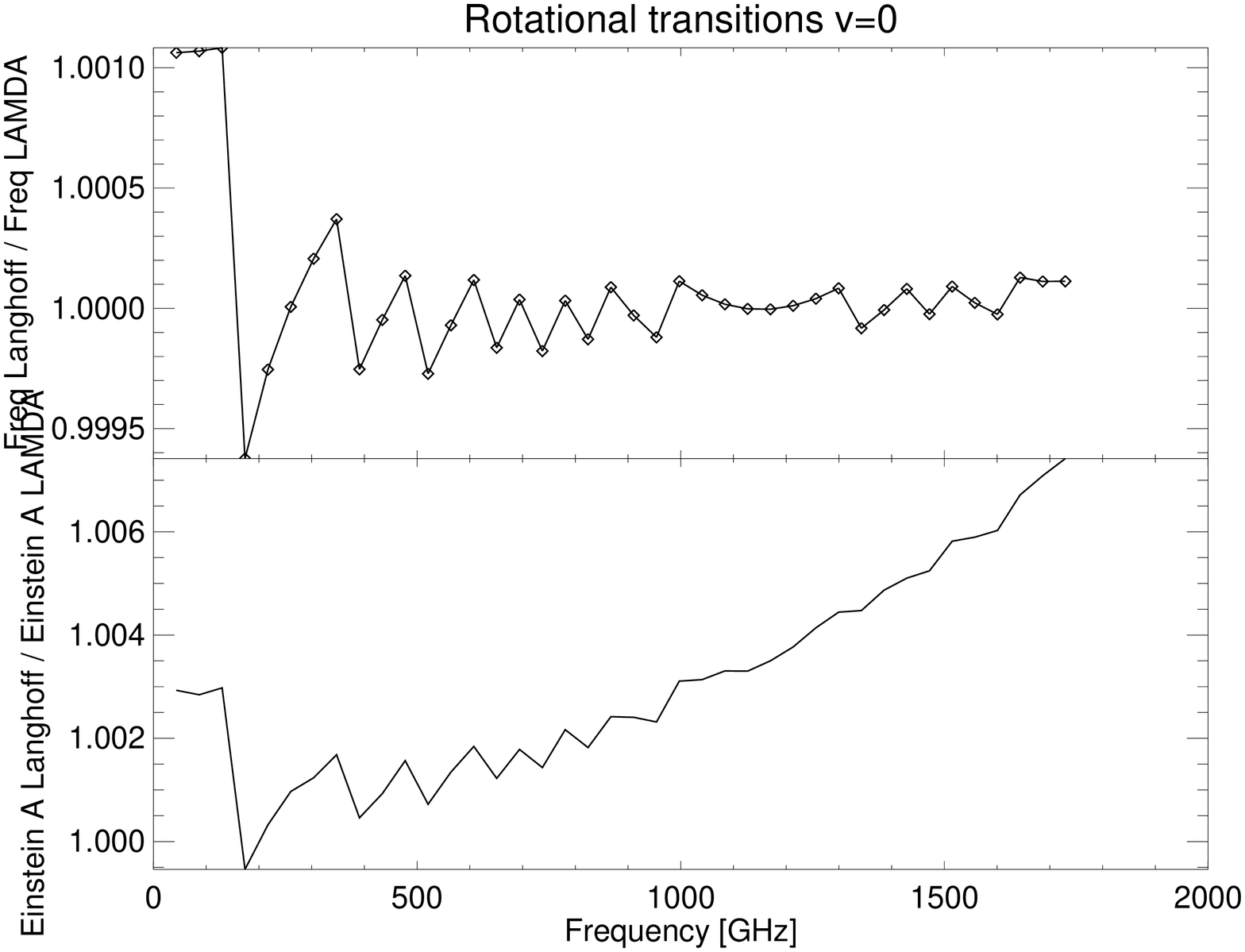}
\caption{Comparison between the transition frequencies and Einstein A
  coefficients of the rotational transitions in the ground state of SiO as
  listed in the LAMDA database and computed by \citet{Langhoff1993}. }
\label{fig:EinsteinSiO}
\end{center}
\end{figure}

The LAMDA database only lists the frequencies and transition
probabilities for the first 40 levels in the ground state. Comparison
with the line list of \citet{Langhoff1993} shows that both databases
nicely agree for the rotational transitions in the ground state (see
Fig.~\ref{fig:EinsteinSiO}). 

\begin{figure}[htp]
\begin{center}
\includegraphics[height=.45\textwidth,angle=90]{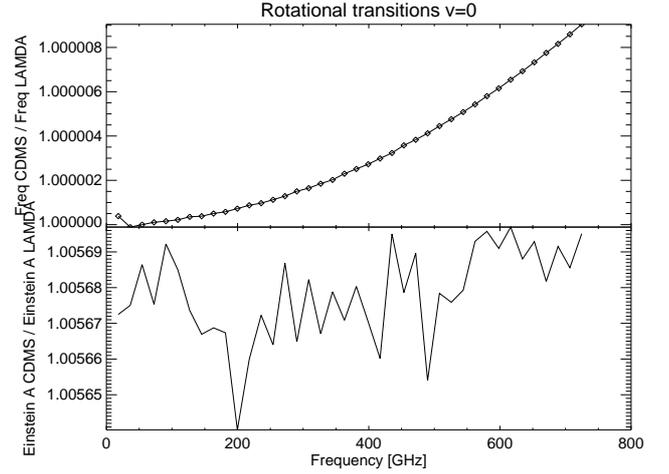}
\caption{Comparison between the transition frequencies and Einstein A
  coefficients of the rotational transitions in the ground state of SiS as
  listed in the LAMDA database and in the CDMS database. }
\label{fig:EinsteinSiS}
\end{center}
\end{figure}

\paragraph{SiS -- silicon monosulfide.}
Frequencies, transition probabilities and energy levels were taken
from the Cologne Database for Molecular Spectroscopy
\citep[CDMS,][]{Muller2005JMoSt.742..215M}. Transitions in the ground
and first vibrational state were included up to $J = 40$.  The pure
rotational transitions of silicon monosulfide ($^{28}$Si$^{32}$S) and
its rare isotopic species have been observed in their ground state as well
as vibrationally excited states by employing Fourier transform
microwave (FTMW) spectroscopy by
\citet{Muller2007PCCP....9.1579M}. The LAMDA database only lists the
lowest 40 levels in the ground vibrational state, for which the energy
levels, transition frequencies and Einstein A coefficients were taken
from the JPL catalog \citep{Pickett1998JQSRT..60..883P}. The values
listed in the JPL and CDMS database show good correspondence (see
Fig.~\ref{fig:EinsteinSiS}).  As for SiO, the collisional rates for
the rotational transitions in the ground state are taken from LAMDA,
the collisional rates in the first vibrational state are assumed to be
equal to the ones in the ground state, while the vibra-rotational
collisional rates are assumed to be zero.

\paragraph{CS -- carbon monosulfide.}
Since the LAMDA database only lists the 41 lowest levels in the ground
state (up to $J\,=\,40$, as extracted from the CDMS database), the
level energies, frequencies and Einstein A coefficients for 
the first vibrational state up to rotational quantum number $J=40$ and the vibra-rotational transitions between $v\,=\,1$ and $v\,=\,0$ 
were extracted from CDMS. The collisional rates are taken from the
LAMDA database and are treated as in the case of SiO and SiS.

\paragraph{CN -- cyanogen, cyanide radical, $^2\Sigma^+$.}
Transition rates for rotational transitions in the ground vibrational state of the cyanide radical are extracted from the CDMS catalog. The CN energy levels are indicated by three rotational quantum numbers: $N$ being the total rotational quantum numbers excluding electron and nuclear spin, $J$ the total rotational angular momentum including electron spin, and $F$ designating the spin quanta. All energy levels with $N_{\rm up} \le 39$ are included, yielding 235 energy levels and 508 transitions. For lack of anything better, we have applied the H$_2$-CS collisional deexcitation rates from the LAMDA database \citep{Black1991ApJ...369L...9B, Hogerheijde2000A&A...362..697H}. The typical deexcitation rates are estimated not to be in error by more than factors 2--3 \citep{Black1991ApJ...369L...9B}.

\paragraph{HCN -- hydrogen cyanide.}
A plethora of vibrational states of the HCN molecule are of relevance
for astronomical observations. The states are designated by
($\nu_1 \nu_2 \nu_3$). The (100) mode is the CH stretching mode at 3311.5
cm$^{-1}$; the (010) mode is the doubly degenerate bending mode at
712.0 cm$^{-1}$; and the (001) mode is the CN stretching mode at
2096.8 cm$^{-1}$. The excitation analysis includes radiative
excitation through the stretching mode at 3\,$\mu$m and the bending
mode at 14\,$\mu$m. The stretching mode at 5\,$\mu$m includes
transitions that are about 300 times weaker and is therefore not
included in the analysis. In each of the vibrational levels, we
include rotational levels up to $J\,=\,29$. Hyperfine splitting of the
rotational levels were included only in the $J\,=\,1$ levels, where
the splitting is larger than the local turbulent width. $l$-type
doubling in the 14\,$\mu$m tranistions was included. Data used are
from \citet{Schoier2007ApJ...670..766S}.

\paragraph{SO -- sulfur monoxide, $X ^3\Sigma^-$.}
While both the JPL and CDMS database (and hence also the LAMDA
database for which the values were extracted from the JPL catalog)
list the rotational transitions in the $v\,=\,0$ and $v\,=\,1$
vibrational state, the transition probabilities for the
vibra-rotational transitions could not be found. We therefore have
restricted the excitation analysis to the first 70 levels in the
ground vibrational state as given by the LAMDA database. We note that
the collisional rates as given by the LAMDA database are only listed
in the range between 50 and 350\,K, and no extrapolations to higher
temperatures are provided. In case of temperatures in the envelope
being higher than 350\,K, the rotational rates at $T\,=\,350\,$\,K
were used.

\paragraph{SO$_2$ --  sulfur dioxide.}
Energy levels, frequencies, Einstein A coefficients and collisional
rates are taken from the LAMDA database (where they were extracted from the JPL database). The first 198 levels in the
ground state are included in the analysis. We note that the collisional
rates are taken from \citet{Green1995ApJS..100..213G} and were
calculated for temperatures in the range from 25 to 125\,K including
energy levels up to 62\,cm$^{-1}$ for collisions with He. This set of
collisional rate coefficients, multiplied by 1.4 to represent
collisions with H$_2$, was extrapolated in the LAMDA database to
include energy levels up to 250\,cm$^{-1}$ and for a range of
temperatures from 10 to 375\,K.

\begin{figure}[htp]
\begin{center}
\includegraphics[height=.45\textwidth,angle=90]{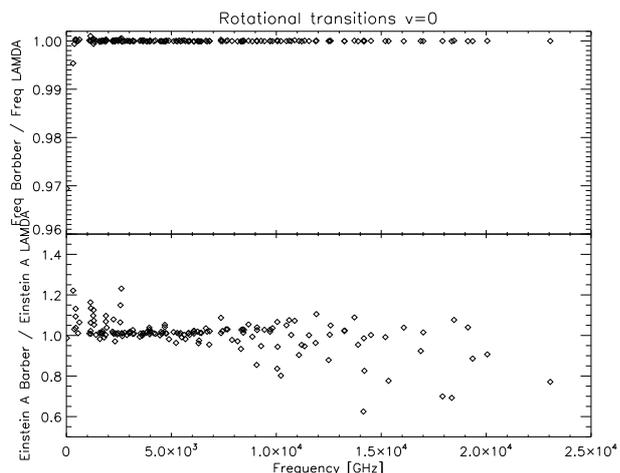}
\caption{Comparison between the transition frequencies and Einstein A
  coefficients of the rotational transitions in the ground state of H$_2$O as
  listed in the LAMDA database and as computed by
  \citet{Barber2006MNRAS.368.1087B}. }
\label{fig:EinsteinH2O}
\end{center}
\end{figure}

\paragraph{H$_2$O -- water.}
The radiative transfer modeling includes the 45 lowest levels in the
ground state and first vibrational state (i.e.\ the bending mode
$\nu_2\,=\,1$ at 6.3\,$\mu$m). Level energies, frequencies and
Einstein A coefficients are extracted from the high-accuracy computed
water line list by \citet{Barber2006MNRAS.368.1087B}, currently being
the most complete water line list. E.g.\ while the LAMDA database lists 158 transitions
between the 45 lowest levels in the ground state, the line list of
\citet{Barber2006MNRAS.368.1087B} contains 164 transitions. The
Einstein A coefficients for the common transitions agree within
$\sim$40\,\% (see Fig.~\ref{fig:EinsteinH2O}).

The H$_2$O-H$_2$ collisional rates in the ground state are taken from
the H$_2$O-He rates by \citet{Green1993ApJS...85..181G}, corrected by
a factor 1.348 to account for collisions with H$_2$. Rotational
collision rates within the first excited state are taken to be the
same as for the ground state, while collisions between the ground
$\nu_0$ and first excited state $\nu_2$ are based on the ground state
rotational collision rate coefficients scaled by a factor 0.01
\citep{Deguchi1990ApJ...360L..27D}. In their analysis,
\citet{Deguchi1990ApJ...360L..27D} show that the uncertainties in the
vibrational collisional rates have no effect on the calculated H$_2$O
lines.

In Sect.~\ref{H2O}, we compare the H$_2$O line profile predictions using the Barber and LAMDA line list and the effect of excluding the $\nu_2 = 1$ bending mode and including the asymmetric stretching mode $\nu_3 = 1$ is discussed.

\end{appendix}

\begin{acknowledgements}
We thank I.\ Cherchneff for useful discussion on the circumstellar non-TE chemistry, and F.\ Sch\"oier for providing us with an updated HCN linelist in the LAMDA database.
 LD  acknowledges financial support from the Fund for Scientific
  Research - Flanders (FWO). EDB acknowledges support from the FWO under  grant number G.0470.07. H.S.P.M.\ is very grateful to the Bundesministerium f\"ur Bildung und Forschung (BMBF) for financial support aimed at maintaining the Cologne Database for Molecular Spectroscopy, CDMS. This support has been administered by the Deutsches Zentrum fur Luft- und Raumfahrt (DLR).The computations for this research have
  been done on the VIC HPC Cluster of the KULeuven. We are grateful to
  the LUDIT HPC team for their support. 

\end{acknowledgements} 

\bibliographystyle{aa}
\bibliography{decinl}
\end{document}